\documentclass{emulateapj}

\newcommand{\cm}[1]{\, {\rm cm^{#1}}}
\newcommand\pks{PKS1302--102}
\newcommand\fuse{\emph{FUSE}}
\newcommand\hst{\emph{HST}}
\newcommand\stis{STIS}
\newcommand\kms{\,\mathrm{km\,s}^{-1}}
\newcommand\mA{\,\mbox{m\AA}}
\newcommand\Ang{\,\mbox{\AA}}
\newcommand\hinv{\,h_{75}^{-1}}
\newcommand\kpc{\,\mathrm{kpc}}
\newcommand\K{\,\mathrm{K}}
\newcommand\Lya{Ly$\alpha$} 
\newcommand\Lyb{Ly$\beta$} 
\newcommand\Lyg{Ly$\gamma$} 
\newcommand\Lyd{Ly$\delta$} 
\newcommand\Lye{Ly$\epsilon$} 
\newcommand\Lyz{Ly$\zeta$} 
\newcommand\HH{H$_{2}$}
\newcommand\zabs{z_{abs}}
\newcommand\zqso{z_{QSO}}
\newcommand\EW{$W_{\lambda}$}
\newcommand\EWr{$W_{r}$}
\newcommand\EWo{$W_{obs}$}
\newcommand\logU{$\log U$}
\newcommand\Z{[M/H]}
\newcommand\Dopb{$b$}
\newcommand\ImPar{$\rho$}
\newcommand\logOVI{$\log \mathrm{N}(\mathrm{O}^{+5})$}
\newcommand\logHI{$\log \mathrm{N}_{\rm HI}$}
\newcommand\nhi{$\mathrm{N}_{\rm HI}$}
\newcommand\mnhi{\rm N_{\rm HI}}
\newcommand\logCIII{$\log \mathrm{N}(\mathrm{C}^{++})$}
\newcommand\Dz{$\Delta z$}
\newcommand\dNOVIdz{$dN_{\rm OVI}/dz$}
\newcommand\dNCIIIdz{$dN_{\rm CIII}/dz$}
\newcommand\dNLyadz{$dN_{\mathrm{Ly\alpha}}/dz$}
\newcommand\dvg{\delta\mathrm{v}_{\mathrm{gal}}}
\newcommand\dva{\delta\mathrm{v}_{\mathrm{abs}}}
\newcommand\ie{\emph{i.e.},\ }
\newcommand\eg{\emph{e.g.},\ }

\usepackage[]{graphicx}
\usepackage[]{comment}

\begin{document}


\title{Characterizing the Low-Redshift Intergalactic Medium towards \pks}

\author{Kathy L. Cooksey\altaffilmark{1}, Jason
  X. Prochaska\altaffilmark{2}, Hsiao-Wen Chen\altaffilmark{3}, John
  S. Mulchaey\altaffilmark{4}, and Benjamin J. Weiner\altaffilmark{5}}

\altaffiltext{1}{Department of Astronomy; University of California;
  1156 High St., Santa Cruz, CA 95064; kcooksey@ucolick.org} 
\altaffiltext{2}{UCO/Lick Observatory; University of California; 1156
  High St., Santa Cruz, CA 95064; xavier@ucolick.org} 
\altaffiltext{3}{Department of Astronomy; University of Chicago; 5640
  S. Ellis Ave., Chicago, IL 60637; hchen@oddjob.uchicago.edu} 
\altaffiltext{4}{Observatories of the Carnegie Institution of
  Washington; 213 Santa Barbara St., Pasadena, CA 91101; mulchaey@ociw.edu} 
\altaffiltext{5}{Steward Observatory, University of Arizona, 933
  N. Cherry Ave., Tucson, AZ 85721; bjw@as.arizona.edu}

\shorttitle{IGM towards \pks}
\shortauthors{Cooksey et al.}

\slugcomment{Draft 3: \today}


\begin{abstract}
We present a detailed analysis of the intergalactic metal-line
absorption systems in the archival \hst/\stis\ and \fuse\ ultraviolet
spectra of the low-redshift quasar \pks\ ($\zqso=0.2784$). We
supplement the archive data with CLOUDY ionization models and a survey
of galaxies in the quasar field. There are 15 strong \Lya\ absorbers
with column densities \logHI\ $>14$. Of these, six are associated with
at least \ion{C}{3} $\lambda977$ absorption (\logCIII\ $>13$); this
implies a redshift density \dNCIIIdz\ $=36^{+13}_{-9}$ (68\%
confidence limits) for the five detections with rest equivalent width \EWr\
$>50\mA$. Two systems show \ion{O}{6} $\lambda\lambda1031,1037$
absorption in addition to \ion{C}{3} (\logOVI\ $>14$). One is a
partial Lyman limit system (\logHI\ $=17$) with associated \ion{C}{3},
\ion{O}{6}, and \ion{Si}{3} $\lambda1206$ absorption. There are three
tentative \ion{O}{6} systems that do not have \ion{C}{3} detected. For
one \ion{O}{6} doublet with both lines detected at $3\sigma$ with
\EWr\ $>50\mA$, \dNOVIdz\ $=7^{+9}_{-4}$. We also search for
\ion{O}{6} doublets without \Lya\ absorption but identify none. From
CLOUDY modeling, these metal-line systems have metallicities spanning
the range $-4\lesssim$ \Z\ $\lesssim-0.3$. The two \ion{O}{6} systems
with associated \ion{C}{3} absorption cannot be single-phase,
collisionally-ionized media based on the relative abundances of the
metals and kinematic arguments. From the galaxy survey, we discover
that the absorption systems are in a diverse set of galactic
environments.  Each metal-line system has at least one galaxy within
$500\kms$ and $600\hinv\kpc$ with $L > 0.1L_{\ast}$.
\end{abstract}



\section{Introduction}\label{sec.intro}

The baryonic content of the Universe is well constrained by Big Bang
nucleosynthesis models, the cosmic microwave background, and the
high-redshift Lyman-$\alpha$ forest
\citep[\eg][]{omearaetal06,spergeletal06ph}. However, surveys of the
nearby Universe reveal a dearth of baryons in stars, galaxies, and
clusters \citep{fukugitaandpeebles04}. Recent cosmological simulations
have placed the most likely reservoir of baryons at low redshift in
moderately overdense, collisionally-ionized gas, called the warm-hot
intergalactic medium
\citep[WHIM;][]{daveetal01,fangandbryan01,cenetal01}. With
temperatures in the range $10^{5}$--$10^{7}\K$, the most sensitive
tracer with current observational facilities is the \ion{O}{6} doublet
$\lambda\lambda1031, 1037\Ang$, which dominates collisionally-ionized
gas at $T\approx3\times10^5\K$, as discussed below.

The \ion{O}{6} doublet is a valuable absorption feature
observationally because it has a characteristic separation and rest
equivalent width (\EWr) ratio for unsaturated features ($2:1$ for the
$\lambda1031.93:\lambda1037.62$ pair).  Furthermore, oxygen is the
most abundant metal, and the O$^{+5}$ ion is an effective tracer of the
low temperature WHIM \citep{trippetal06conf}. Assuming collisional
ionization equilibrium, other ion species (e.g. \ion{O}{8},
\ion{Mg}{10}, \ion{Ne}{8}) have greater abundances in the higher WHIM
temperature range, where it is predicted there are more baryons;
however, these other ions are extremely difficult to detect at low
redshifts.  Current X-ray telescopes are not up to the task but for a
few systems \citep{wangetal05, williamsmathurandnicastro06,
nicastroetal05}.

Cosmological simulations make four important predictions about the
content, temperature, ionization mechanism, and density of the WHIM.
The WHIM contains $\sim\!40\%$ of the baryons in the low-redshift
Universe \citep{daveetal01,cenetal01}. It has characteristic
overdensity $10\lesssim\delta\lesssim30$ and is shock heated to
$T\approx10^{5}$--$10^{7}\K$ as it collapses onto large-scale
structure \citep[\eg filaments;][]{daveetal01,fangandbryan01}.  The
WHIM thermally emits soft X-rays; \citet{daveetal01} argue that the
WHIM must be in a filamentary structure to agree with the soft X-ray
background.  Collisional ionization dominates in high-temperature,
high-density regions (\eg WHIM), and photoionization dominates in
low-temperature, low-density regions \citep[\eg local \Lya\
forest;][]{fangandbryan01}. \citet{cenandostriker06} and
\citet{cenandfang06} include new and improved prescriptions for
galactic super-winds and collisional non-equilibrium. Their recent
results substantiate previous simulations which argue for a large
contribution of WHIM gas to the baryonic census as well as demonstrate
the importance of galactic super-winds in dispersing metals to large
distances from the galaxies, with impact parameters \ImPar\ $\approx
1\,\mathrm{Mpc}$.

Several observational papers propose that \ion{O}{6} absorption occurs
in a multi-phase medium, with hot collisionally-ionized components
($10^{5}\lesssim T\lesssim 10^{7}\K$) and warm photoionized components
($T \approx 10^4\K$) \citep[e.g.][]{trippetal00, simcoeetal02,
shulletal03, sembachetal04, danforthetal06}.  Other papers suggest
that the \ion{O}{6} absorbers are in collisional ionization
equilibrium \citep[CIE;][]{richteretal04}, not in equilibrium
\citep{trippandsavage00}, or photoionized and therefore not part of
the WHIM \citep{prochaskaetal04}. \citet{richteretal04} argue that
broad \Lya\ features can be used to trace the WHIM if there are no
\ion{O}{6} lines, and the simulations of \citet{richterfangandbryan06}
indeed find broad \Lya\ features at the redshift of \ion{O}{6}
absorption.

Recent observations have argued that \ion{O}{6} absorbers are often
correlated with galaxies or galaxy groups \citep[\eg][]{richteretal04,
prochaskaetal04,sembachetal04}.  Typically, \ion{O}{6} absorbers are
identified because \Lya\ absorbers were first identified at the
corresponding redshifts. If \ion{O}{6} absorption were truly tracing
the WHIM, the hydrogen should be predominantly ionized at
$T\approx10^{5}\K$ and therefore very broad and shallow due to thermal
broadening, precluding easy detection \citep{richteretal04}.  This may
explain the tendency to detect \ion{O}{6} absorbers near galaxies and
to model the absorbers as a multi-phase medium. For this reason,
\citet{trippetal07ph} and the current study perform searches for
\ion{O}{6} doublets without first detecting \Lya.

\citet{danforthandshull05} and \citet[hereafter,
DSRS06]{danforthetal06} surveyed the \emph{Hubble Space Telescope}
Space Telescope Imaging Spectrograph (\stis) and \emph{Far Ultraviolet
Spectroscopic Explorer} (\fuse) archives for \Lya, \ion{O}{6}, and
\ion{C}{3} $\lambda977$.  Of 45 \Lya\ absorbers with statistics for
both metal lines, 12 (27\%) have both \ion{O}{6} and \ion{C}{3}
absorption, 8 (18\%) have \ion{O}{6} without \ion{C}{3}, and 4 (9\%)
have \ion{C}{3} without \ion{O}{6}. The prevalence of low-ionization
absorption \Lya\ and, often, \ion{C}{3} associated with highly-ionized
\ion{O}{6} absorption supports a multi-phase model of the IGM. DSRS06
did not perform a blind search for \ion{O}{6} without \Lya\ absorbers.

\citet{trippetal07ph} searched for \ion{H}{1} and \ion{O}{6}
absorption in archival \stis\ spectra of 16 low-redshift quasars. The
spectra were supplemented with \fuse\ data for sightlines with
published complete line lists. \citet{trippetal07ph} compared and
contrasted 14 associated (those within $5000\kms$ of the quasar) and
53 intervening \ion{H}{1} and \ion{O}{6} absorption systems. From this
survey, almost half of the intervening systems are multi-phase
absorbers that may host, at least in part, the WHIM, and more than a
third of the systems are cool, single-phase absorbers.

Presented here is a detailed analysis of far-ultraviolet \stis\ and
\fuse\ spectra of the quasar \pks\ \citep[$\zqso =
0.2784\pm0.0005$,][]{corbettetal98}. Column densities are measured for
\ion{H}{1} Lyman systems and metal lines. For systems with at least
\Lya\ and \Lyb, the \ion{H}{1} Doppler parameter was measured. The
redshift density of \Lya, \ion{O}{6}, and \ion{C}{3} absorbers is
determined. A direct comparison with the results of DSRS06 is
given. In addition, the UV spectra are complemented by a galaxy survey
of the field surrounding \pks, made at Las Campanas Observatory. These
observations are used to characterize the \ion{O}{6} absorbers and
other metal-line systems with respect to galaxies. This is the first
in a series of papers on the chemical enrichment of the low-redshift
IGM (\hst\ proposal 10679; PI: J. X. Prochaska).  The paper is
organized as follows: the data and reduction procedures are discussed
in \S\ \ref{sec.data}; the identification of absorption-line systems
in \S\ \ref{sec.als}; metal-line systems in \S\ \ref{sec.mls}; strong
\Lya\ absorbers in \S\ \ref{sec.lya}; previous analysis in \S\
\ref{sec.comp}; galaxy survey and results in \S\ \ref{sec.gal}; and
final discussion and conclusions in \S\ \ref{sec.disc}.

\section{Data and Reduction}\label{sec.data}

\subsection {Space Telescope Imaging Spectrograph}\label{subsec.stis}

\begin{deluxetable*}{lrrcrrl}
\tablewidth{0pt}
\tabletypesize{\scriptsize}
\tablecaption{OBSERVATIONS SUMMARY\label{tab.obssum}  }
\tablehead{
\colhead{Instrument} & \colhead{Obs. Date} & \colhead{Data ID} & 
\colhead{T$_{\mathrm{obs}}$\tablenotemark{a}} &
\colhead{N$_{\mathrm{exp}}$\tablenotemark{b}} & 
\colhead{Aperture/Grating} & \colhead{S/N}
\\
\colhead{} & \colhead{} & \colhead{} & \colhead{(ks)} &
\colhead{} & \colhead{} & \colhead{(per pixel)}
}
\startdata

FUSE & 2000-05-21 & P108020 & 66.0 & 31 & LWRS
& 2.7\tablenotemark{c} \\  
FUSE & 2001-01-20 & P108020 & 83.3 & 34 & LWRS
& 3.7\tablenotemark{c} \\  
STIS & 2001-08-21 & 8306 & 22.1 & 2 & 0.2X0.2/E140M & 6.0 
\enddata

\tablenotetext{a}{Total observation time for coadded spectrum}
\tablenotetext{b}{Total number of exposures per observation}
\tablenotetext{c}{Best signal-to-noise ratio measured for LiF 1B}

\end{deluxetable*}

\pks\ was observed for a total of $22\,$ks by the \emph{Hubble Space
  Telescope}/Space Telescope Imaging Spectrograph (\hst/\stis) in
  August 2001 (Program 8306; PI: M. Lemoine). The observations are
  summarized in Table\ \ref{tab.obssum}. \stis\ observations were
  taken with the medium echelle grating E140M that covers
  $1140\lesssim\lambda\lesssim1730\Ang$.  The 44 orders were coadded
  individually before being coadded into a one-dimensional spectrum,
  which was used for further analysis. There are gaps between the
  orders for $\lambda \gtrsim 1600\Ang$. \stis\ E140M has a resolution
  of $R \approx 45,000$, or FWHM $\approx7\kms$.  More information
  about \stis\ can be found in \citet{mobasher02}.

The data were retrieved from the Multimission Archive of Space
Telescope (MAST)\footnote{http://archive.stsci.edu/} and were reduced
with CalSTIS v2.15b with On-the-Fly-Reprocessing. The multiple
exposures were coadded with the IDL routine COADSTIS from the XIDL
library,\footnote{http://www.ucolick.org/$\sim$xavier/IDL/} which is
described below.  Each order of each exposure was rebinned to the same
logarithmic wavelength solution. Regions with bad data quality flags
and a small neighboring buffer to the bad regions were excluded in the
coadding of the observations.  The STIS Data Handbook defines many
data quality flags \citep{mobasher02}, and all but three non-zero
flags were rejected (16, 32, and 1024).  These three accepted flags
indicated abnormally high dark rate and mild CCD blemishes.

The spectra were scaled to the spectrum with the highest
signal-to-noise ratio S/N, measured across all orders. The orders were
coadded with the XIDL routine X\_COMBSPEC, which weights by S/N.  To
coadd the orders into one spectrum, overlapping regions of the orders
were combined by taking the weighted mean of the flux.

\subsection {\emph{Far Ultraviolet Spectroscopic Explorer}}\label{subsec.fuse}

The \emph{Far Ultraviolet Spectroscopic Explorer} (\fuse) complements
the \stis\ wavelength range and enables the identification of
important absorption lines at lower redshifts. \fuse\ covers
$905\lesssim\lambda\lesssim1190\Ang$ with $R\approx20,000$, or FWHM
$\approx15\kms$. In \fuse, \Lyb\ and \ion{O}{6} absorption can be
detected at $\zabs\lesssim0.15$ and \ion{C}{3} at
$\zabs\lesssim0.22$. For \pks, the \ion{H}{1} column density is better
constrained when \Lya\ absorption from \stis\ is supplemented with
detections and upper limits of higher-order \ion{H}{1} Lyman lines
from \fuse. The \ion{O}{6} absorption that may trace the WHIM should
be more prevalent at lower redshifts, and it is important to search
for \ion{O}{6} at $\zabs<0.15$. \ion{C}{3} absorption is a common
metal line from photoionized gas and, typically, indicates a
multi-phase medium when detected in a system with the highly-ionized
\ion{O}{6} \citep{prochaskaetal04}.

The four gratings of \fuse\ disperse onto two detectors resulting in
eight spectra per exposure. More details about the \fuse\ instrument
and mission can be found in \citet{moosetal00} and
\citet{sahnowetal00}.  All \pks\ \fuse\ observations were taken in
photon address mode (\ie time-tag mode) with the low-resolution
aperture (LWRS).

\pks\ was observed for a total of $149\,\mathrm{ks}$ with \fuse\
  between May 2000 and January 2001 (Program P108; PI: K. Sembach).
  The raw \fuse\ files, also downloaded from MAST, were completely
  reduced with a modified
  CalFUSE\footnote{ftp://fuse.pha.jhu.edu/fuseftp/calfuse/} v3.0.7
  pipeline and coadded with Don Lindler's IDL tool FUSE\_REGISTER.

The two separate observations of \pks\ were coadded into one set of
eight spectra in order to increase S/N.  The CalFUSE pipeline has
procedures for this purpose. After each exposure has been processed,
the intermediate data files (IDFs), which contain all information from
the raw photon-event list to the wavelength solution, are combined
into one IDF for each channel; in a similar manner, the bad pixel
masks (BPMs) are also combined.  From these two files, the CalFUSE
pipeline extracts the final, calibrated spectra.

By default, the combined IDF has its aperture centroid defined by the
first, single-exposure IDF in the list to be combined. In addition, to
save space, the combined BPM is only defined for the regions of the
two-dimensional spectra used in the final extraction (\ie aperture and
background windows).  Each single \fuse\ exposure of \pks\ has low S/N
and a poorly measured centroid, and CalFUSE was not able to optimally
extract the final spectra.  Even if they were optimally-extracted, the
spectra would exclude good data since the centroid used was not
measured for the combined IDF.

To properly calculate the centroid for a combined IDF, several
subroutines were copied from the CalFUSE program CF\_EDIT, which is an
IDL GUI used to modify IDFs, into a customized IDL routine that
calculates the centroid and modifies the IDF header accordingly. This
new centroid was written to the headers of the individual IDFs so that
the BPMs generated with the standard CalFUSE pipeline would
automatically span the desired regions of the spectra.  These BPMs
were combined, as mentioned previously, and used with the combined
IDFs to extract the calibrated spectra.  In this manner, the \pks\
\fuse\ spectra were optimally extracted.

The calibrated spectra from each observation were coadded with
FUSE\_REGISTER.  The eight segments were not combined into one
spectrum. This allowed for the identification of the same feature in
different segments for confidence and avoids the issue that the \fuse\
channels have slightly different wavelength solutions.  Values quoted
in this paper primarily come from the detection in the channel with
the highest S/N.

According to the \fuse\ white paper about wavelength
calibration,\footnote{http://fuse.pha.jhu.edu/analysis/calfuse\_wp1.html}
the two main sources of uncertainty in the absolute wavelength
solution are the detector distortions and zero-point offsets, which,
at worst, cause uncertainties of $\delta v \approx 13\kms$ and
$\approx66\kms$, respectively.  The \stis\ wavelength solution is
accurate to $\delta v \approx 4\kms$ \citep{mobasher02}.  The partial
Lyman-limit system at $\zabs=0.09487$ spanned all the \fuse\ channels,
save SiC 1B (see \S\ \ref{subsec.z09486}), and was used to shift the
\fuse\ spectra onto the \stis\ wavelength solution. The alignment of
the Galactic features is secondary evidence that the shifts are
reasonable. For SiC 1B, the Galactic \Lyg\ emission was used. The
spectra were shifted by the following amounts: $-56\kms$ (SiC 1B);
$35\kms$ (SiC 2A); $9\kms$ (LiF 2B); $21\kms$ (LiF 1A); $-22\kms$ (SiC
1A); $18\kms$ (LiF 1B); and $2\kms$ (LiF 2A).

The SiC 2B and LiF 2B segments were not used in the analysis due to
their poor sensitivity. The other three SiC channels have poor flux
zero points, resulting in negative flux and uncertain \EWr; however,
line identification was possible. SiC 1A was excluded from analysis
because LiF 1A covered the same wavelength range; SiC 1B and 2A were
included to cover the lower wavelengths.  There are two segments
covering most wavelengths: $905\lesssim\lambda\lesssim1005\Ang$ (SiC
2A and 1B); $990\lesssim\lambda\lesssim1090\Ang$ (LiF 1A); and
$1088\lesssim\lambda\lesssim1188\Ang$ (LiF 1B and 2A).

\subsection{Continuum Fitting}\label{subsec.conti}

The spectra were normalized with a parameterized b-spline
continuum-fitting program. Once an initial breakpoint spacing was
chosen ($\approx\!6\Ang$ for \stis\ and $\approx4$--$5\Ang$ for
\fuse), the spectrum was iteratively fit with a b-spline. In each
iteration, pixels that lay outside the high/low sigma clips (\eg
2.5/2) were masked out to prevent absorption features, cosmic rays, or
other bad pixels from skewing the fit. This process was repeated until
the fit changed less than a set tolerance compared to the previous
iteration. The breakpoint spacing was automatically decreased in
regions of great change (\eg quasar \Lya\ emission) and increased in
regions of relatively little variation.  More specifically, the
spacing is made coarser in regions where the binned flux $f_{i}$
varied by $\leq\!10$\% compared to the error-weighted flux $\bar{f}$;
the spacing is refined where $f_{i}$ varied by greater than one
standard deviation $\sigma_{f_{i}}$ of $\bar{f}$. The value $f_{i}$ is
the median flux in bins defined by the initial breakpoint spacing. The
spectrum and its error were divided by the continuum to generate the
normalized spectrum used in the analysis.

The program may loosely be considered ``automatic:'' it will converge
on the best fit for the spectrum based on a given set of parameters.
However, a fit based on a random set of parameters may not be a good
fit to the continuum as judged visually by the authors.  To estimate
the errors resulting from the subjective nature of continuum fitting,
we fit the spectra ``by-hand'' with the XIDL routine X\_CONTINUUM and
compared the change in rest equivalent width \EWr\ values.  The \EWr\
values measured from the spectra normalized by the automated program
are in good agreement with those measured from the spectra normalized
by hand. The root-mean-squared (RMS) fractional difference is $<\!5\%$
for \stis\ and $12\%$ for \fuse.  For column densities, the RMS
fractional difference is $<\!1\%$ for both instruments.

\section{Absorption-Line Systems}\label{sec.als}

\subsection{Identifying Systems}\label{subsec.idsys}

We search for IGM absorption systems---absorption lines physically
associated with one another--with allowance for variations in \eg
ionizing mechanism, density, temperature. The absorption features
detected in the \fuse\ and \stis\ spectra were sorted into Galactic
lines and intergalactic absorption lines.  The latter category was
further sorted into their respective systems by comparing their
redshifts, line profiles, and rest equivalent width \EWr\ ratios.
Lines of the same ionized species from a given absorption system
should have the same redshift, similar line profiles, and unsaturated
\EWr\ values that scale with oscillator strengths (see Figures
\ref{velplt.z00441} and \ref{cog.z00441}). Ions from the same phase of
the IGM tend to have roughly the same redshift and similar
profiles. For example, \ion{H}{1} and \ion{C}{3} of $\zabs=0.04222$
are well aligned in velocity space, and have similar asymmetric
profiles, whereas \ion{O}{6} is at a different redshift (velocity) and
has a dissimilar profile (see Figure \ref{velplt.z04226}); this
suggests the \ion{H}{1} and \ion{C}{3} absorption arises from one
phase of the IGM and \ion{O}{6} from another.

In order to thoroughly identify the absorption features in the spectra
of \pks, we developed an automated procedure to detect all features in
the spectra greater than a minimum significance and width.  We then
interactively identified the features not coincident with Galactic
lines.

\subsubsection{Automatic Line Detection and Doublet Search}\label{subsec.auto}

A purely interactive search (as described below) would be biased
towards systems with \Lya. \ion{O}{6} ions that are associated with
collisionally-ionized gas at $T\approx3\times10^{5}\K$ (\eg\ the
canonical WHIM) are likely to have associated \Lya\ profiles that are
broad and shallow \citep{richteretal04}. The \pks\ spectra does not
have sufficient S/N to reliably detect broad \Lya\ features. To avoid
biasing the search against warm-hot \ion{O}{6} gas, we searched for
doublets independently of \Lya.

In order to conduct a blind search for pairs, primarily doublets like
\ion{O}{6}, an automated feature-finding program was developed to
detect all possible absorption features with a minimum significance
$\sigma_{min}$ and width.  The spectrum was first convolved with a
Gaussian with FWHM $b_{min}$.  The convolved pixels were grouped into
potential features with significance $\sigma_{pix} \geq \sigma_{min}$
. The program does not attempt to separate blended lines into
component features. The final result is a list of central wavelengths,
observed equivalent width \EWo\ and error, and wavelength limits (used
to measure \EWo).

Next, a blind search for doublets (\ion{O}{6}, \ion{C}{4}, \ion{N}{5},
and \ion{Si}{4}) was performed. For example, the search assumed each
automatically detected feature between Galactic and quasar \ion{O}{6}
could be \ion{O}{6} 1031 and only identified a possible pair when
another feature was at the appropriate wavelength spacing within the
bounds of $\lambda1031$ translated to the appropriate, redshifted
wavelength of $\lambda1037$.  This procedure was repeated for the
\ion{C}{4}, \ion{N}{5}, and \ion{Si}{4} doublets and for \Lya, \Lyb\
and \Lya, \ion{C}{3} pairs.

The blind doublet search successfully identified Galactic \ion{O}{6},
\ion{C}{4}, \ion{N}{5}, and \ion{Si}{4} as well as the IGM \ion{O}{6}
systems at $\zabs=0.04222$, 0.06471, 0.09487, and 0.22555 verified by the
interactive search. The weak \ion{O}{6} absorbers at $\zabs=0.19161$
and 0.22752 are not $3\sigma$
features in both lines of the doublet and were not identified in the
automated search when $\sigma_{min}=3$.  The search also resulted in a
possible \ion{O}{6} doublet at $\zabs=0.01583$ with \Lya\ not detected
at $3\sigma$. However, this candidate \ion{O}{6} doublet is actually
the coincidence of Galactic \ion{Ar}{1} 1048 and \HH\ 1054.0 R(3).  In
this case, the misidentification was evident from the fact that the
\EWr\ ratio was inconsistent with an \ion{O}{6} doublet.

In general, the automated feature-finding program indicates features
in a spectrum that have a roughly Gaussian profile and a measured
significance greater than the minimum required. These features may or
may not be absorption lines. There is a balance between automatically
detecting weak absorption lines and including spurious features. We
calibrated the search parameters to maximize the detection of lines
with \EW\ $> 50 \mA$ while minimizing the inclusion of spurious
features (less than $10\%$).

\subsubsection{Interactive Search}\label{subsec.interac}

\tabletypesize{\scriptsize}
\begin{deluxetable*}{lccrcccccc}
\tablewidth{0pc}
\tablecaption{$W_{r}$ SUMMARY \label{tab:ewsumm}}
\tabletypesize{\scriptsize}
\tablehead{\colhead{Ion} & \colhead{$\lambda_{obs}$} &\colhead{$\lambda_{r}$} & \colhead{$z_{abs}$} & \colhead{$W_1$} 
& \colhead{$\sigma(W_1)$}& \colhead{$W_2$} & \colhead{$\sigma(W_2)$} &
\colhead{$W_f$} & \colhead{$\sigma(W_f)$} \\
 & (\AA) & (\AA) && (m\AA) & (m\AA) & (m\AA) & (m\AA) & (m\AA) & (m\AA)} 
\startdata
\cutinhead{\emph{FUSE}}
\ion{H}{1} 930& 934.798& 930.748& 0.00435&$ 168$&$ 47$&$  74$&$ 37$&$ 110$& 29\\
\ion{H}{1} 937& 941.898& 937.803& 0.00437&$ 293$&$ 60$&$ 247$&$ 40$&$ 262$& 33\\
\ion{H}{1} 949& 953.995& 949.743& 0.00448&$ 418$&$ 68$&$ 305$&$ 38$&$ 332$& 33\\
\ion{H}{1} 972& 976.903& 972.537& 0.00449&$ 367$&$ 86$&$ 350$&$ 44$&$ 353$& 39\\
\ion{H}{1} 937& 977.333& 937.803& 0.04215&$   7$&$ 74$&$  15$&$ 44$&$  13$& 38\\
\ion{C}{3} 977& 981.309& 977.020& 0.00439&$   7$&$ 63$&$ 156$&$ 41$&$ 111$& 35\\
\ion{H}{1} 949& 989.770& 949.743& 0.04214&$ 349$&$ 81$&$ 584$&$ 37$&$ 542$& 34\\
\enddata
\tablecomments{Note that the list is incomplete for wavelengths $< 1000$\AA\ where the data has poor S/N and significant line blending.
Columns 4,5 (6,7) refer to the SiC 1B (SiC 2A) channel for 905$ < \lambda < 1005\,$\AA; LiF1A for 990$ < \lambda < 1090\,$\AA; LiF 1B (LiF 2A) for 1088$ < \lambda < 1188$\AA; and STIS E140M for $\lambda > 1188\,$\AA.
[The complete version of this table is in the electronic edition of the Journal.  The printed edition contains only a sample.]}
\end{deluxetable*}

The automatic line-detection procedure described above generates a
list of unidentified features meeting a specified minimum set of
requirements. The features may be Galactic, intergalactic, or, in a
few cases, spurious. The automated pair searches (\eg \Lya, \Lyb)
supply a starting point for interactively identifying the features and
sorting them into IGM absorption systems. The potential feature list
gives the lines that should be identified and is used to determine the
completeness of the interactive search. The final, identified
absorption lines are listed in Table \ref{tab:ewsumm}.

Identifying the absorption lines first required disentangling Galactic
from intergalactic features.  The velocity plots of the \Lya, \Lyb\
pairs from the blind search were examined individually. If these pairs
were well aligned in velocity space, with similar line profiles, and
decreasing \EWr, we interactively searched for higher-order Lyman
lines (\eg \Lyg, \Lyd) and/or common metal lines, redshifted by the
assumed-\Lya\ redshift.  These lines were grouped as a possible
system. The rough priority of metal lines was: (a) \ion{C}{3},
\ion{O}{6}, \ion{C}{4} $\lambda\lambda1548,1550$; (b) \ion{N}{2}
$\lambda1083$, \ion{Si}{3} $\lambda1206$, \ion{C}{2} $\lambda1036$ or
$\lambda1334$; and (c) \ion{N}{3} $\lambda989$, \ion{Si}{4}
$\lambda\lambda1393,1402$ \citep[using atomic data listed
in][]{prochaskaetal04}. In this initial search, no knowledge of the
Galactic lines biased the identification of IGM absorption lines.

Second, all features from the automatic search corresponding to likely
Galactic lines were recorded as such, regardless (for now) whether the
same features were first identified as IGM lines. The likely Galactic
lines included various ionization states of iron, oxygen, nitrogen,
sulfur, carbon, silicon, phosphorus, argon, and aluminum. This set of
lines was defined in a stacked spectrum of normalized \stis\ E140M
spectra of 15 low-redshift quasars. In the \fuse\ channels with
$\lambda \lesssim 1000\Ang$, molecular hydrogen lines are abundant due
to the Lyman and Werner bands.  The \pks\ sight line has moderate to
low molecular hydrogen absorption with a line of sight $\log
N(\mbox{\HH}) = 16.3$ \citep{wakker06}.

All automatically detected features not already identified as Galactic
or associated with an intergalactic system was assumed to be \Lya\ if
between Galactic \Lya\ and \Lya\ at the redshift of \pks\
(\ie$1216\lesssim\lambda\lesssim1563\Ang$). For example, the strong
\Lya\ absorber at $\zabs=0.19243$ was detected automatically, not
paired with \Lyb, and not corresponding to a Galactic line.

For spectra of low-redshift quasars, line confusion (\eg blends) is
minimal. However, line coincidences do occur. As an example, consider
the Lyman series at $\zabs=0.09400$, which first appeared to be an
especially strong \ion{H}{1} absorber. The \Lyb, \Lyg, and \Lyd\
transitions were blended with Galactic lines, and higher-order Lyman
lines were confused with \HH\ absorption and with the higher-order
Lyman lines at $\zabs=0.09487$. Occasional blends also occur between
different IGM absorption systems. We disentangle these blends assuming
common line-strengths for the various transitions within a blend,
allowing for modest variations.  For instance, DSRS06 lists a \Lya\
absorber at $\zabs=0.08655$, whereas we identify it as \ion{Si}{3} at
$\zabs=0.09487$ because the would-be \Lyb\ at $\zabs=0.08655$ was less
than $3\sigma$ significance and the system at $\zabs=0.09487$ is a
strong \ion{H}{1} absorber expected to have associated metal-line
absorption.

Throughout this paper, only the statistical errors from photon
counting are quoted, but the true errors should account for the
combined statistical, continuum, and systematic errors. An estimate of
the combined error will change the detection limit with respect to the
statistical error. In \S\ \ref{subsec.conti}, we estimated the RMS
fractional difference due to continuum fitting to be $<\!5\%$ for
\stis\ and $12\%$ for \fuse.  The rest equivalent width is measured
with a simple boxcar summation.  The wavelength limits of the boxcar
window were defined interactively and are subjective.  The RMS
fractional difference due to changing the window by $15\%$ is $6\%$
for \stis\ and $13\%$ for \fuse. Column densities were affected by
less than $1\%$ by changing the window.

The errors from the continuum fitting and boxcar summation are
correlated, but as a first approximation, we will add them in
quadrature. Ultimately, for a feature to be detected at three times
the combined error, the feature must be detected at $3.1\sigma$ for
\stis\ and $3.6\sigma$ for \fuse, where $\sigma$ is the statistical
error only. For example, \ion{C}{3} at $\zabs=0.00438$ is a
$3.2\sigma$ detection, but folding in the $12\%$ continuum-fitting and
$13\%$ boxcar-summation errors for \fuse, the feature is 2.8 times the
combined error. \Lyb\ at $\zabs=0.12565$ is a similar case
in the \stis\ spectrum.  The majority of lines discussed in this paper
have significance greater than three times the combined error, and we
will use $3\sigma$, commonly quoted in the literature, as our
detection limit knowing a more rigorous examination of our errors does
not affect our results.

There are 28 \Lya\ features detected at $\ge\!3\sigma$ significance in
the spectra of \pks.  Of these, 15 are strong \Lya\ absorbers with
\logHI\ $\ge14$ (54\%) and eight with at least one metal line
(29\%). A ninth tentative metal-line system has \logHI\ $=13.1$. There
are five probable \ion{O}{6} systems. Line identification is complete
to 90\% in the region of \stis\ where intergalactic \Lya\ could be
detected with $\sigma_{min}=4\sigma$ and $b_{min} =
20\kms$,\footnote{The remaining 10\% of the automatically identified
features that could be \Lya\ appear to be spurious.} and
identification is complete to $>\!85\%$ in \fuse\ LiF 2A, 1B, and 1A
for $b_{min} = 40\kms$ features. Completeness was measured by
correlating the identified lines from the interactive search with the
automatically detected features discussed above.

\subsection{Column Densities, Doppler Parameters, and Metallicities}\label{subsec.meas}

For absorption systems exhibiting at least two members of the
\ion{H}{1} Lyman series, the \ion{H}{1} column density \logHI\ and
Doppler parameter \Dopb\ were measured with a curve-of-growth (COG)
analysis of the \EWr\ values that minimized $\chi^2$. For metal lines,
the apparent optical depth method (AODM) was used to measure the
column densities \citep{savageandsembach91}.

However, as discussed in \citet{foxsavageandwakker05}, the
AODM-measured column densities from low-S/N spectra 
systematically overestimate the true column densities due to spurious
high-AOD (low flux) pixels and the exponential nature of the AODM. 
From Monte Carlo tests, we measure this to be, at worst, a 0.2 and
0.25 dex effect for unsaturated features in \stis\ and \fuse,
respectively.

To measure metallicities, we used ionization corrections from
collisional ionization equilibrium (CIE) or photoionization models of
the metal-line systems, calculated with
CLOUDY\footnote{http://www.nublado.org} versions 94 and 06.02.09a,
respectively, as last described by \citet{ferlandetal98}.  The CIE
models are described in \citet{prochaskaetal04}.

To construct the photoionization models, the medium was assumed to be
a plane-parallel slab ionized by a \citet[][updated in
2005]{haardtandmadau96} quasar-only ultraviolet background. The number
density of hydrogen $n_{\rm H}$ was assumed to be 0.1 cm$^{-3}$,
though our models are insensitive to this parameter in the
optically-thin regime.  The ionization parameter \logU, metallicity
\Z,\footnote{\Z\ $=\log( \mathrm{N}(\mbox{M})/\mathrm{N}(\mbox{H}) ) -
\log( \mathrm{N}(\mbox{M})/\mathrm{N}(\mbox{H}) )_{\odot}$} neutral
column density \logHI, and redshift of the UV background were varied
to sample the parameter space. The ionization parameter is a
dimensionless ratio of the number of hydrogen-ionizing photons to the
total number of hydrogen atoms.

Abundance-independent ionic ratios of the same metal (\eg\
$\mathrm{N}(\mathrm{C}^{++})/\mathrm{N}(\mathrm{C}^{+3})$) and/or
abundance-dependent ratios of metals with similar ionization
potentials (\eg\
$\mathrm{N}(\mathrm{C}^{+3})/\mathrm{N}(\mathrm{O}^{+5})$) constrained
\logU\ or $T$ (see \S\ \ref{subsec.z09486}). The \logU\ or $T$ limits
define the ionization corrections for the measured metallicities (\eg\
[C/H]). Due to the simplifications and assumptions in the CLOUDY
models, the model abundances and metallicities are reliable to within
a factor of two.

Modeling of a multi-phase medium was generally not considered because
the systems presented in this paper simply have too few metal
lines. By the same token, multi-phase and collisional ionization
\emph{non}-equilibrium scenarios are not ruled out by the
observations. In select cases where the kinematics and abundances
suggest a multi-phase medium, we do use photoionization and CIE models
to describe the different components.

\section{Metal-line Systems}\label{sec.mls}

\tabletypesize{\scriptsize}

\begin{deluxetable}{lcccc}
\tablewidth{0pc}
\tablecaption{IONIC COLUMN DENSITIES\label{tab:ionsumm}}
\tabletypesize{\scriptsize}
\tablehead{\colhead{Ion} &\colhead{$\lambda_{r}$ (\AA)} & \colhead{$W_{r}$ (m\AA)} & 
\colhead{$\log N_{AODM}$} & \colhead{$\log N_{adopt}$}}
\startdata
\cutinhead{$z_{abs}=0.00438$, $\log \rm N_{\rm HI}=15.80 \pm 0.36$}
\ion{C}{       2}&1334.5323&$<  31 $&$ < 13.37$ &$< 13.37$ \\
\ion{C}{       3}& 977.0200&$ 111 \pm  35 $&$ > 13.48$ &$> 13.48$ \\
\ion{C}{       4}&1548.1950&$  44 \pm  13 $&$ 13.11 \pm 0.11 $&$ 13.11 \pm 0.11$ \\
\ion{C}{       4}&1550.7700&$<  28 $&$ < 13.34$ &$$ \nodata \\

\cutinhead{$z_{abs}=0.04222$, $\log \rm N_{\rm HI}=15.07 \pm 0.08$}
\ion{C}{       2}&1334.5323&$<  29 $&$ < 13.35$ &$< 13.35$ \\
\ion{C}{       3}& 977.0200&$ 168 \pm  12 $&$ 13.66 \pm 0.05 $&$ 13.66 \pm 0.05$ \\
\ion{C}{       4}&1548.1950&$  88 \pm  22 $&$ 13.47 \pm 0.10 $&$ 13.47 \pm 0.10$ \\
\ion{C}{       4}&1550.7700&$<  37 $&$ < 13.49$ &$$ \nodata \\
\ion{O}{       1}&1302.1685&$<  27 $&$ < 13.75$ &$< 13.75$ \\
\ion{O}{       6}&1031.9261&$ 190 \pm  18 $&$ 14.41 \pm 0.05 $&$ 14.46 \pm 0.04$ \\
\ion{O}{       6}&1037.6167&$ 173 \pm  15 $&$ 14.65 \pm 0.06 $&$ $ \nodata \\

\cutinhead{$z_{abs}=0.06471$, $\log \rm N_{\rm HI}=14.63 \pm 0.22$}
\ion{C}{       3}& 977.0200&$  56 \pm   8 $&$ < 13.09$ &$< 13.09$ \\
\ion{C}{       4}&1548.1950&$<  50 $&$ < 13.29$ &$< 13.29$ \\
\ion{C}{       4}&1550.7700&$<  52 $&$ < 13.64$ &$$ \nodata \\
\ion{O}{       1}&1302.1685&$<  26 $&$ < 13.72$ &$< 13.72$ \\
\ion{O}{       6}&1031.9261&$  47 \pm  11 $&$ 13.75 \pm 0.08 $&$ 13.81 \pm 0.06$ \\
\ion{O}{       6}&1037.6167&$  42 \pm  10 $&$ 13.96 \pm 0.09 $&$ $ \nodata \\

\cutinhead{$z_{abs}=0.09400$, $\log \rm N_{\rm HI}=15.06 \pm 0.06$}
\ion{C}{       2}&1334.5323&$<  29 $&$ < 13.33$ &$< 13.33$ \\
\ion{C}{       3}& 977.0200&$  80 \pm   9 $&$ > 13.26$ &$> 13.26$ \\
\ion{C}{       4}&1548.1950&$<  64 $&$ < 13.39$ &$< 13.39$ \\
\ion{C}{       4}&1550.7700&$<  57 $&$ < 13.60$ &$$ \nodata \\

\cutinhead{$z_{abs}=0.09487$, $\log \rm N_{\rm HI}=16.88 \pm 0.03$}
\ion{C}{       2}&1334.5323&$  73 \pm  18 $&$ < 13.67$ &$< 13.67$ \\
\ion{C}{       3}& 977.0200&$ 243 \pm  13 $&$ > 13.86$ &$> 13.86$ \\
\ion{C}{       4}&1548.1950&$<  75 $&$ < 13.50$ &$< 13.50$ \\
\ion{C}{       4}&1550.7700&$<  66 $&$ < 13.72$ &$$ \nodata \\
\ion{O}{       1}&1302.1685&$<  36 $&$ < 13.88$ &$< 13.88$ \\
\ion{O}{       6}&1031.9261&$  97 \pm  14 $&$ 14.00 \pm 0.06 $&$ 14.03 \pm 0.05$ \\
\ion{O}{       6}&1037.6167&$  42 \pm  13 $&$ 14.09 \pm 0.08 $&$ $ \nodata \\
\ion{Si}{       2}&1260.4221&$  31 \pm  11 $&$ < 12.48$ &$< 12.42$ \\
\ion{Si}{       3}&1206.5000&$ 172 \pm   8 $&$ 13.06 \pm 0.03 $&$ 13.06 \pm 0.03$ \\
\ion{Si}{       4}&1393.7550&$<  25 $&$ < 12.62$ &$< 12.62$ \\
\ion{Si}{       4}&1402.7700&$<  28 $&$ < 12.98$ &$$ \nodata \\

\cutinhead{$z_{abs}=0.14533$, $\log \rm N_{\rm HI}=15.39 \pm 0.03$}
\ion{C}{       2}&1036.3367&$<  30 $&$ < 13.60$ &$< 13.33$ \\
\ion{C}{       2}&1334.5323&$<  28 $&$ < 13.33$ &$$ \nodata \\
\ion{C}{       3}& 977.0200&$  66 \pm  10 $&$ 13.15 \pm 0.06 $&$ 13.15 \pm 0.06$ \\
\ion{O}{       1}&1302.1685&$<  33 $&$ < 13.81$ &$< 13.81$ \\
\ion{O}{       6}&1031.9261&$ 100 \pm  22 $&$ < 14.17$ &$< 14.17$ \\
\ion{O}{       6}&1037.6167&$ 111 \pm  30 $&$ < 14.46$ &$$ \nodata \\

\cutinhead{$z_{abs}=0.19161$, $\log \rm N_{\rm HI}=15.29 \pm 0.03$}
\ion{C}{       2}&1036.3367&$<  25 $&$ < 13.51$ &$< 13.30$ \\
\ion{C}{       2}&1334.5323&$<  28 $&$ < 13.30$ &$$ \nodata \\
\ion{C}{       3}& 977.0200&$  48 \pm   9 $&$ 13.07 \pm 0.08 $&$ 13.07 \pm 0.08$ \\
\ion{O}{       1}&1302.1685&$<  23 $&$ < 13.67$ &$< 13.67$ \\
\ion{O}{       6}&1031.9261&$  68 \pm  14 $&$ 13.85 \pm 0.09 $&$ 13.85 \pm 0.09$ \\
\ion{O}{       6}&1037.6167&$<  22 $&$ < 13.72$ &$$ \nodata \\
\ion{Si}{       3}&1206.5000&$  22 \pm   8 $&$ < 12.15$ &$< 12.11$ \\

\cutinhead{$z_{abs}=0.22555$, $\log \rm N_{\rm HI}=14.00 \pm 0.09$}
\ion{C}{       3}& 977.0200&$<  38 $&$ < 12.99$ &$< 12.99$ \\
\ion{O}{       1}&1302.1685&$<  37 $&$ < 13.91$ &$< 13.91$ \\
\ion{O}{       6}&1031.9261&$  52 \pm  12 $&$ 13.76 \pm 0.08 $&$ 13.91 \pm 0.05$ \\
\ion{O}{       6}&1037.6167&$  99 \pm  12 $&$ 14.28 \pm 0.05 $&$ $ \nodata \\

\cutinhead{$z_{abs}=0.22752$, $\log \rm N_{\rm HI}=13.11 \pm 0.11$}
\ion{O}{       6}&1031.9261&$  38 \pm   8 $&$ 13.56 \pm 0.09 $&$ 13.56 \pm 0.09$ \\
\ion{O}{       6}&1037.6167&$<  17 $&$ < 13.64$ &$$ \nodata \\

\enddata
\end{deluxetable}


This section summarizes the nine \Lya\ systems with at least one
metal line detected.  Velocity plots, COG analysis, and CLOUDY models
for each system are discussed.  Four of the nine systems have
\ion{C}{3} absorption only, one with \ion{C}{3} and the \ion{O}{6}
doublet, three with tentative \ion{O}{6} detections, and one with
\ion{C}{3}, \ion{Si}{3}, and a broad \ion{O}{6} doublet.  Because
\ion{C}{3} is the dominant line in photoionized gas, it may be more
readily detected in the moderate-S/N \pks\ spectra.  The metal-line
systems are summarized in Table \ref{tab:ionsumm}.

All but three of the metal-line systems have \logHI\ $\ge15$.  The two
systems with multiple metal lines are likely multi-phase based on
kinematic arguments (\eg velocity offsets, line profiles) and the poor
fit of single-phase models to the data. More specifically, a
single-phase, collisionally-ionized absorber does not have significant
\ion{C}{3} and \ion{O}{6} absorption without significant \ion{C}{4}
absorption.

The metallicities quoted are based on ionization corrections from the
best \logU\ or $T$ value from the CLOUDY models with \Z\ consistent
with the final, derived metallicity.  In cases where the \logU\
value is not well constrained by the observations, we adopt a central
value based on \logHI, as predicted by the empirical/theoretical
relation in \citet{prochaskaetal04}. The nature of the galaxy
environment of these systems will be discussed in \S\ \ref{sec.gal}.

\subsection{ $\zabs=0.00442$: \ion{C}{3}}\label{subsec.z00441}

\begin{figure} 
\epsscale{1.1}\plotone{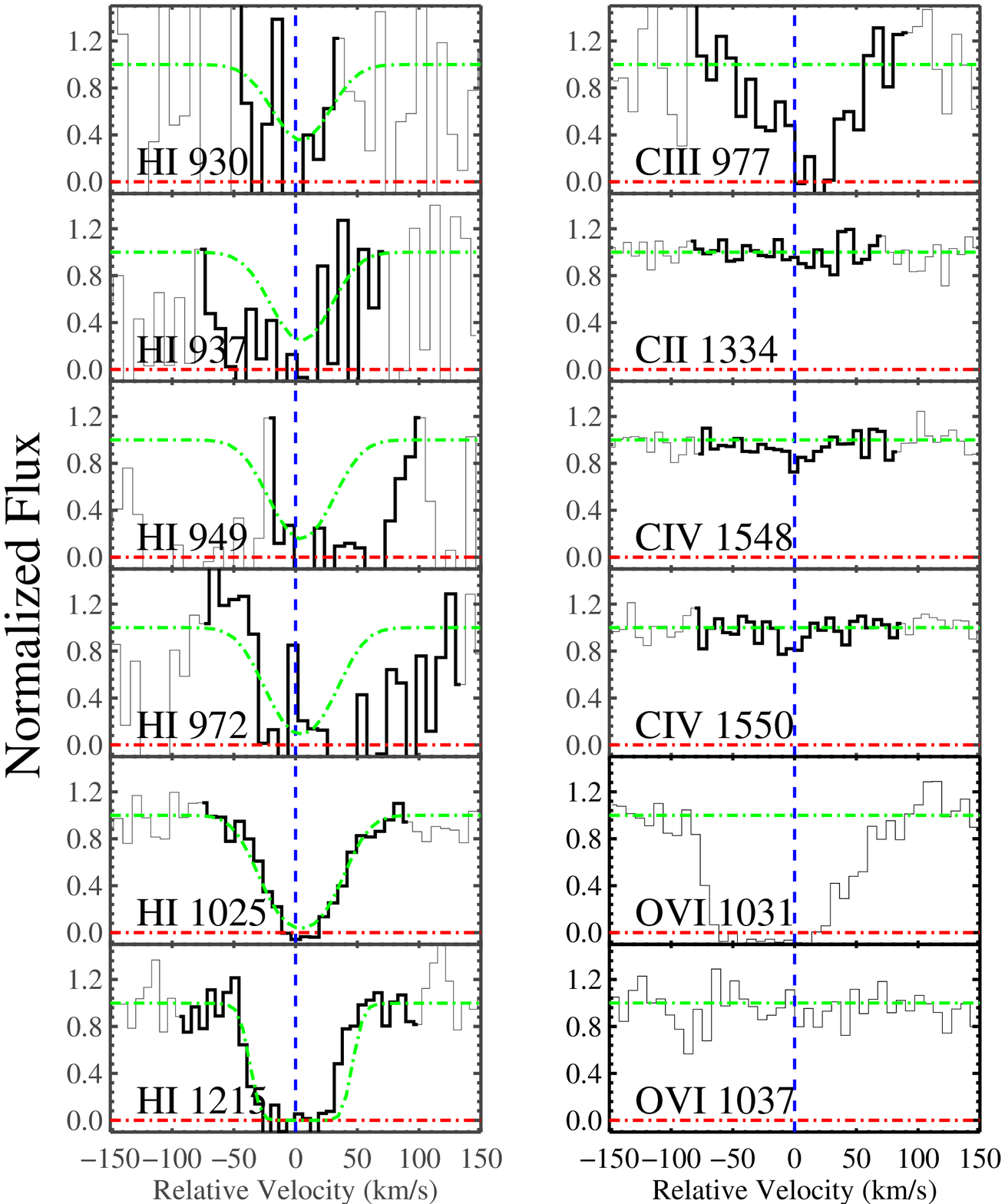}
\caption{Velocity plot for $\zabs=0.00438$.  Spectra, averaged over
  two pixels (thin black line), are stacked in velocity space with
  $v=0\kms$ at $\zabs=0.00438$, the optical-depth weighted centroid of
  \Lya\ (vertical blue dashed line). The region used to measure \EWr\ is
  highlighted (thick black line).  Also indicated is the flux at zero
  (red dash-dotted line).  The \ion{H}{1} Voigt profiles, based on the
  COG \logHI\ and \Dopb, are superimposed to show the predicted area
  under the curve (green dash-dotted line; for the metal lines, this line
  indicates the flux at unity). The Voigt profile centroid is fixed at
  the redshift of \Lyb\ $z_{\beta}=0.00439$. \Lya\ is detected in the
  wings of the damped Galactic \Lya\ feature, and the lines higher
  than \Lyb\ are detected in the SiC 2A and 1B channels, which have
  poor sensitivity. \ion{C}{2} and \ion{C}{4} 1550 are not detected at
  $3\sigma$. \Lyg, \Lyd, and \Lye\ are blended with Galactic
  \ion{C}{3}, \HH\ 954.0 R(4), and \HH\ 941.6 P(2), respectively. As
  shown, Galactic 
  \ion{C}{2} 1036 is coincident where \ion{O}{6} 1031 would be, and
  there is an absence of \ion{O}{6} 1037.
\label{velplt.z00441}
}
\end{figure}
\begin{figure} 
\includegraphics[clip,trim=0in 0in 0in 0in,width=0.37\textwidth,
  angle=90]{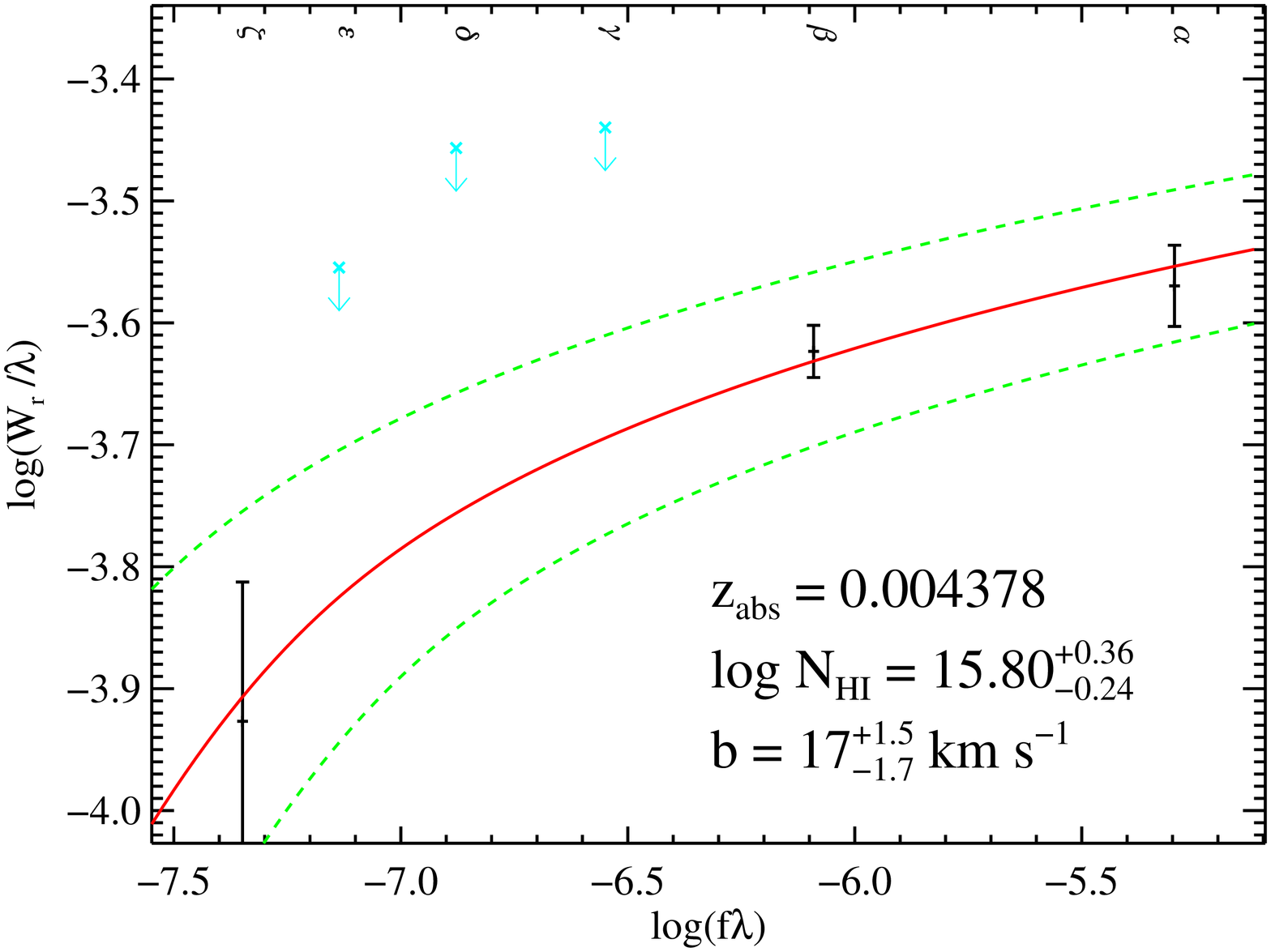}
\caption
{\ion{H}{1} COG for $\zabs=0.00438$.  The best-fit curve of growth
  (solid red line) with $1\sigma$ bounds (dashed green lines) is
  displayed over the measured \EWr\ values included in the fit (black
  hashes with $1\sigma$ error bars).  Blended lines are shown as upper
  limits (cyan cross-top arrows). For description of the upper limits,
  see Figure \ref{velplt.z00441}.  The \ion{H}{1} Lyman lines are
  labeled by letter or wavelength across the top.
\label{cog.z00441}
}
\end{figure}

This metal-line system was detected at $v \approx 1300\kms$, in the wings of
the damped Galactic \Lya\ profile (see Figure
\ref{velplt.z00441}). The \pks\ sight-line also passes through the
Virgo cluster at this redshift \citep{wakkeretal03}.  \Lya\ and
\ion{C}{3} are well aligned in velocity space with similar line
profiles, which imply the two absorbers are kinematically
similar. \Lya, \Lyb, and \Lyz\ were used to fit the COG: \logHI\
$=15.8^{+0.4}_{-0.2}$ and \Dopb\ $=17^{+1.5}_{-1.7}\kms$ (see Figure
\ref{cog.z00441}).

The \ion{H}{1} column density is not well constrained for this
system. \Lyb, which was detected in LiF 1A, has the highest detection
significance of the Lyman series. The higher-order lines fall in the
SiC 2A and 1B channels, which have poor sensitivity.  The AODM HI
column density for the saturated \Lyb\ line sets a lower limit of
\logHI\ $>14.9$.  DSRS06 measured \logHI\ $=14.872\pm0.286$ and
\Dopb\ $=18\pm2\kms$ from a Voigt profile fit to \Lya.  We examined
the velocity plot with Voigt profile outlines for [\logHI,\Dopb] =
[15.8,17.4] and [14.87,18]. The DSRS06 values clearly underestimate
the \Lyb\ absorption.

The ionic ratios
$\mathrm{N}(\mathrm{C}^{+})/\mathrm{N}(\mathrm{C}^{++})$ and
$\mathrm{N}(\mathrm{C}^{++})/\mathrm{N}(\mathrm{C}^{+3})$ constrained
the ionization parameter to $-3.4\le$ \logU\ $\le-2.1$ for the CLOUDY
models with \logHI\ $=15.75$. For \logU\ $=-2.1$, $-2\le$ [C/H]
$\le-0.9$.  From the kinematics of \Lya\ and \ion{C}{3} and CLOUDY
modeling, the $\zabs=0.00438$ system is well described as a
photoionized, metal-poor, single-phase medium.

\subsection{ $\zabs=0.04226$: \ion{O}{6}, \ion{C}{3}}\label{subsec.z04226}

\begin{figure} 
\epsscale{1.1}\plotone{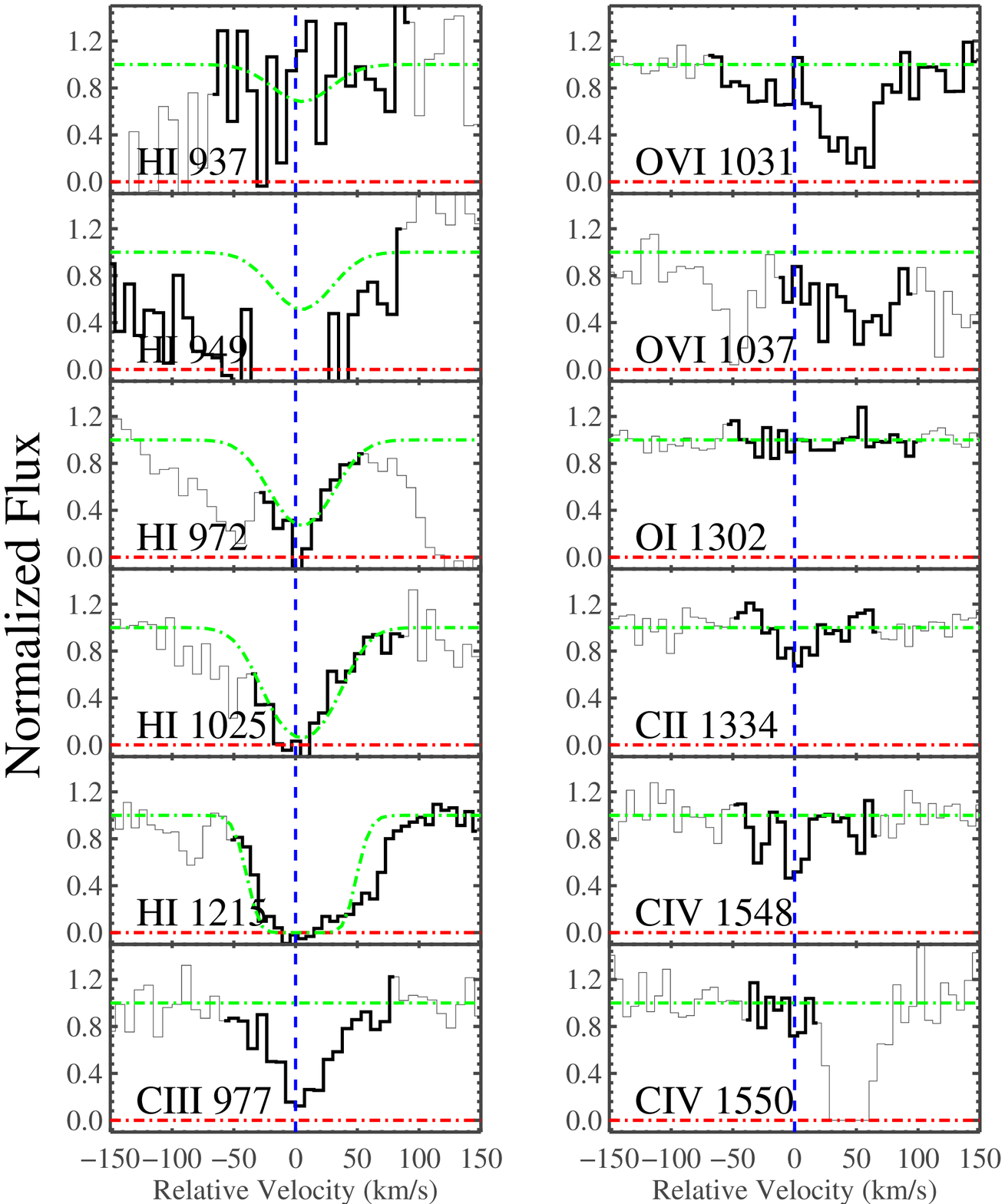}
\caption{Velocity plot for $\zabs=0.04222$ (see Figure
  \ref{velplt.z00441} description). The \Lya\ profile shows two
  components on either side of $v=0\kms$. \ion{C}{3} is more aligned
  with the stronger, blueward component of \Lya, while \ion{O}{6} is
  more aligned with the weaker, redward component. The Voigt profiles
  are based on a single-component COG model (see Figure
  \ref{cog.z04226}) with centroid fixed at
  $z_{\gamma}=0.04224$. \ion{O}{6} 1037 is blended with \HH\ 1081.7
  R(3) and is flanked by \HH\ 1081.2 P(2) to the blue and Galactic
  \ion{F}{2} 1081 to the red. \Lyb, \Lyg, and \Lyd\ are blended (see
  \S\ \ref{subsec.z04226}). \Lye, \ion{O}{1}, \ion{C}{2}, and
  \ion{C}{4} 1550 are not detected at $3\sigma$.  The location of
  \ion{C}{4} 1550 spans an echelle order gap in \stis.
\label{velplt.z04226}
}
\end{figure}
\begin{figure} 
\includegraphics[clip,trim=0in 0in 0in 0in,width=0.37\textwidth,
  angle=90]{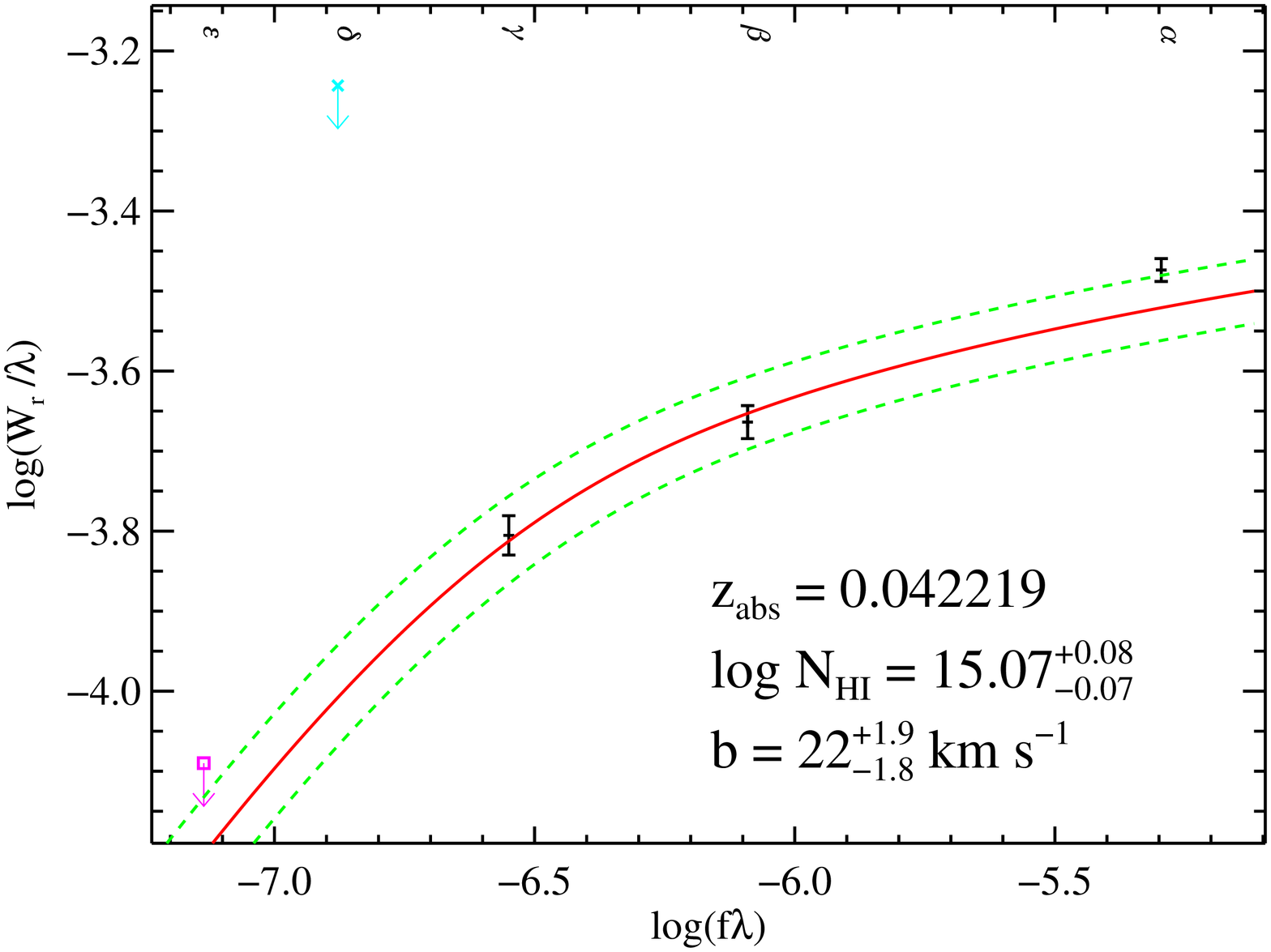}
\caption
{\ion{H}{1} COG for $\zabs=0.04222$ (see Figure \ref{cog.z00441}
  description). \Lye, which is not detected at $3\sigma$, is shown as
  a $2\sigma$ upper limit (magenta square-top arrow).  \Lyd\ is
  blended with Galactic \ion{N}{3}.
\label{cog.z04226}
}
\end{figure}

The \Lya\ profile shows two strong components. The blueward component
aligns well with \ion{C}{3} and the redward with \ion{O}{6} (see
Figure \ref{velplt.z04226}).  The \ion{H}{1} COG for this absorber
included \Lya, \Lyb, and \Lyg, and the values are well constrained:
\logHI\ $=15.07^{+0.08}_{-0.07}$ and \Dopb\ $=22^{+1.9}_{-1.8}\kms$ (see
Figure \ref{cog.z04226}).  \Lye, \ion{O}{1} $\lambda1302$, \ion{C}{2},
and \ion{C}{4} 1550 were not detected at $3\sigma$. \Lyd\ is blended
with Galactic \ion{N}{3}, \Lyg\ with \HH\ 1013.4 R(1), and \Lyb\ with
\ion{C}{3} at $\zabs=0.09400$.  \ion{O}{6} 1037 is blended with the
weak \HH\ 1081.7 R(3) line.

The \ion{H}{1} Lyman lines and \ion{C}{3} have similar line profiles
and appear well aligned in velocity space, while \ion{O}{6} is shifted
redward.  Since \Lya\ is saturated and multicomponent, its velocity is
not well constrained. Taking \Lyg\ as the reference line for the
stronger \ion{H}{1} component, \ion{O}{6} 1031 has $\dva\equiv
c(z_{abs}-z_{\gamma})/(1+z_{\gamma})=+54\kms$, while \ion{C}{3} is
perfectly aligned. The significant velocity offset between the
metal-line profiles suggests the metals reside in different phases of
gas, with overlapping \Lya\ absorption. The \ion{O}{6} absorption
appears associated with more tenuous narrow \Lya\ absorption, and
the doublet has a width similar to the redward \Lya\ component.

The \EWr\ ratio of \ion{O}{6} 1031 to 1037 ($1.1\pm0.14$) does not
agree with the expected $2:1$ ratio. The continuum fit around
\ion{O}{6} 1037 is poorly constrained because it is at the edge of LiF
1A and there are two absorption features close to 1037. This
potentially increases \EWr\ and \logOVI. \ion{O}{6} 1037 is also
blended with a weak \HH\ line. This system has the strongest
\ion{O}{6} absorption in the \pks\ sight line \logOVI\ $=14.5$ and
strong \ion{C}{3} absorption \logCIII\ $=13.7$.

In the CLOUDY models with \logHI\ $=15$, the ionic ratios
$\mathrm{N}(\mathrm{C}^{+})/\mathrm{N}(\mathrm{C}^{++})$ and
$\mathrm{N}(\mathrm{C}^{++})/\mathrm{N}(\mathrm{C}^{+3})$ constrain
\logU: $-3.2\le$ \logU\ $\le-1.9$.  At the central value of \logU\
$=-2.6$, one would require [O/C]~$\approx +3$ and a super-solar O
abundance to explain the column densities of C$^{++}$ and
O$^{+5}$. The abundance-dependent ionic ratio
$\mathrm{N}(\mathrm{C}^{+3})/\mathrm{N}(\mathrm{O}^{+5})$ sets \logU\
$>\mbox{[C/O]}-1.1$ and requires [C/O] $<-0.5$ for the oxygen and
carbon absorption to be from the same photoionized phase. In this
case, \logU\ $=-1.9$, [C/H] $=-1.1$, and [O/H] $=+0.3$ (see Table
\ref{tab:z0.0422x}). 

Since [O/H] $=+0.3$ is not likely, we consider a single-phase
photoionized model to be ruled out, as supported by the kinematics.
However, the oxygen absorption could arise in a photoionized phase
with \logU\ $=-1.1$, assuming [C/O] $=0$, and then [O/H] $=-1.2$.

We have considered collisional ionization models.  Under the
assumption of CIE, the carbon absorption is constrained to be in a
warm phase $5.3\times10^{4}<T<9.8\times10^{4}\K$.  Considering the
limit set by
$\mathrm{N}(\mathrm{C}^{+3})/\mathrm{N}(\mathrm{O}^{+5})$, the oxygen
would be from a warm-hot phase $T>2.4\times10^{5}\K$ with low
metallicity [O/H] $>-2$.  For this value, we have assumed the
\ion{H}{1} column density in the warm-hot phase is the same as
measured in the COG analysis for the warm phase.

In summary, we favor a two-phase photoionization model for this
system, as strongly supported by the kinematics. The strong, blueward
\ion{H}{1} component and the narrower, redward \Lya\ component have
strong, well-aligned \ion{C}{3} and \ion{O}{6} absorption, respectively.

\begin{deluxetable}{lccc}
\tablewidth{0pc}
\tablecaption{ELEMENTAL ABUNDANCES FOR ABSORBER AT $z$=0.04222\label{tab:z0.0422x}}
\tabletypesize{\scriptsize}
\tablehead{\colhead{Ion} &
\colhead{[X/H]} & \colhead{[X/C$^{++}$]}}
\startdata
C$^{+}$ & $<-0.09$ & $< 0.99$ \\
C$^{++}$ & $-1.07$ & $ 0.00$ \\
C$^{+3}$ & $-1.09$ & $-0.02$ \\
O$^{0}$ & $< 4.33$ & $< 5.40$ \\
O$^{+5}$ & $ 0.20$ & $ 1.28$
\enddata
\tablenotetext{a}{Assumes a photoionized gas with \logU\ $=-1.9$}
\end{deluxetable}

\subsection{ $\zabs=0.06468$: \ion{O}{6}}\label{subsec.z06468}

\begin{figure} 
\epsscale{1.1}\plotone{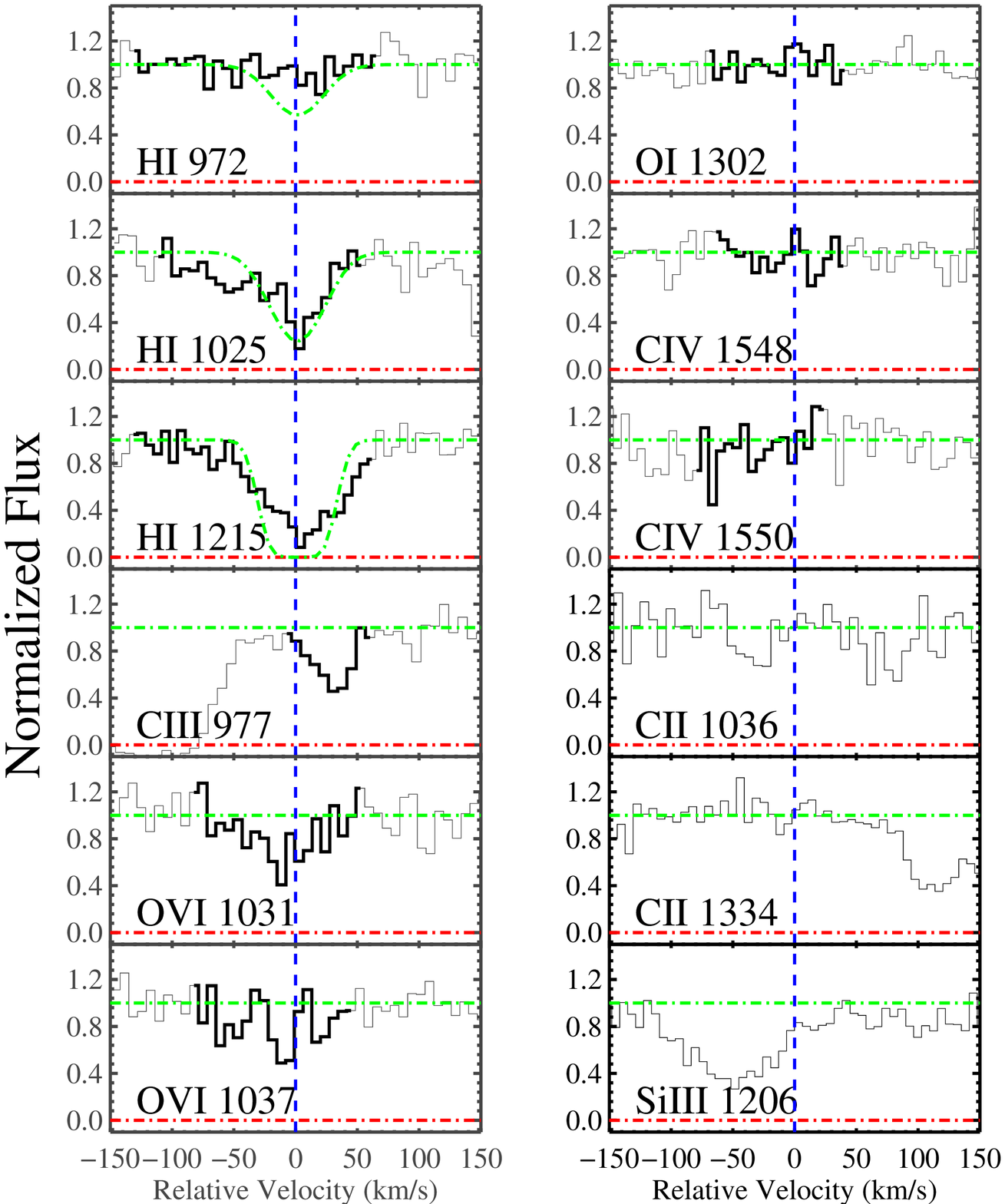}
\caption
{Velocity plot for $\zabs=0.06471$ (see Figure \ref{velplt.z00441}
  description). \Lya\ has a distinct asymmetric profile, probably due
  to unresolved components. The \ion{O}{6} doublet is detected at
  $\approx\!4.5\sigma$. The \EWr\ ratio of \ion{O}{6} 1031 to 1037 is
  $1.1\pm0.4$.  \Lyg, \ion{O}{1} and \ion{C}{4} are not detected at
  $3\sigma$. \ion{C}{3} is treated as an upper limit since it is
  coincident with \HH\ 1040.4 P(2). The Voigt profile centroid is
  fixed at $z_{\beta}=0.06472$.
\label{velplt.z06468}
}
\end{figure}
\begin{figure} 
\includegraphics[clip,trim=0in 0in 0in 0in,width=0.37\textwidth,
  angle=90]{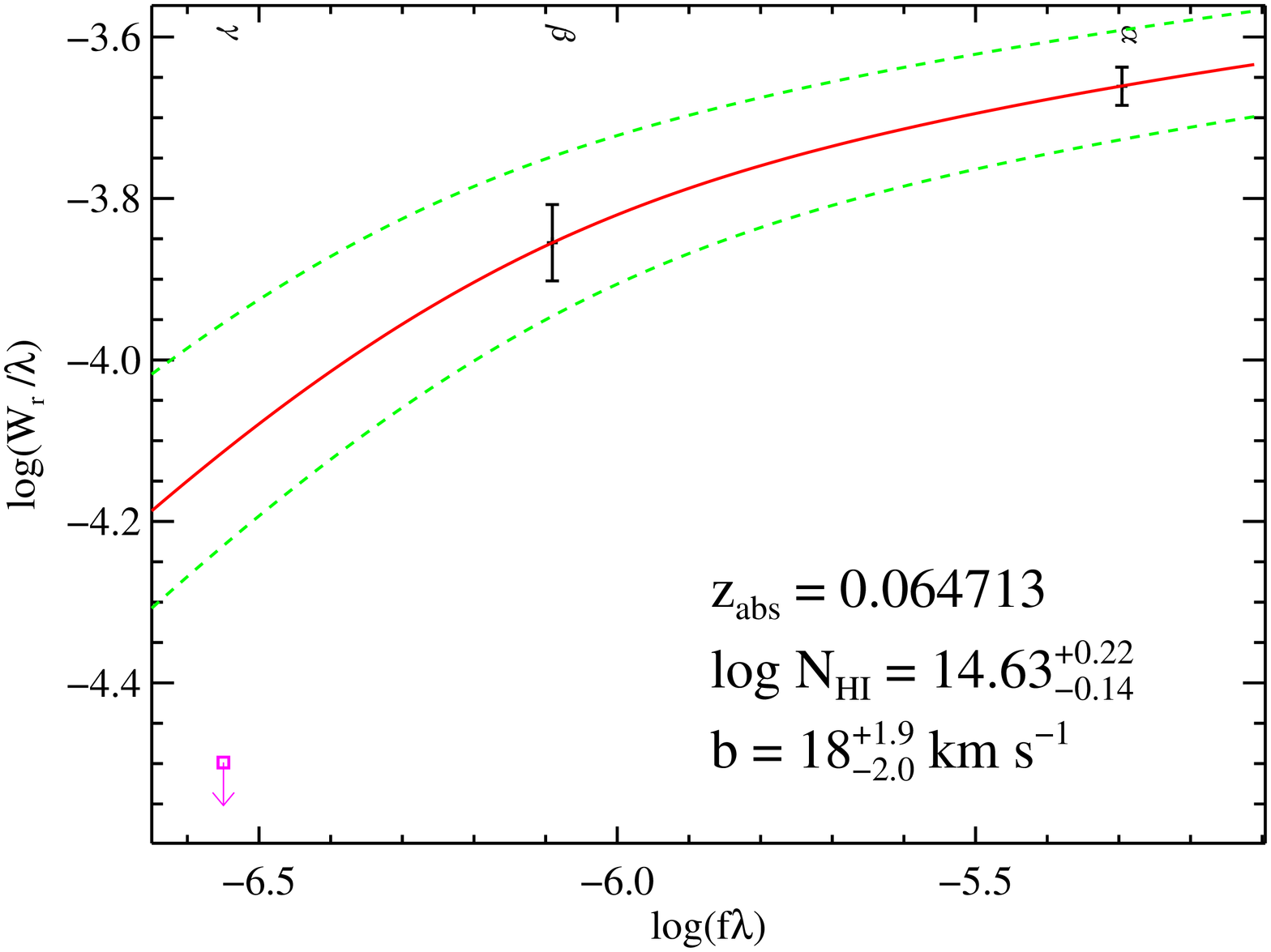}
\caption
{\ion{H}{1} COG for $\zabs=0.06471$ (see Figure \ref{cog.z00441}
  description).  \Lyg\ is not detected at $3\sigma$ but \emph{is}
  consistent within $1\sigma$ with the single-component COG fit. This
  \ion{H}{1} system 
  is possibly multi-component, as evidenced by the asymmetric \Lya\
  and \Lyb\ profiles in Figure \ref{velplt.z06468}.
\label{cog.z06468}
}
\end{figure}

Though a tentative detection, \ion{O}{6} is well aligned with \Lya\
(see Figure \ref{velplt.z06468}). The \EWr\ ratio of \ion{O}{6} 1031
to 1037 is $1.1\pm0.4$, and \logOVI\ $=13.8$ (see Table
\ref{tab:ionsumm}). At the redshift of \Lya, there is no \ion{C}{3}
detected. However, there are two features in the vicinity: $\zabs=
0.09487$ \Lyd\ at $\dva > 100\kms$ and \HH\ 1040.4 P(2) at $\dva
\approx0\kms$, with respect to the centroid of the unsaturated
\Lyb. The \HH\ P(2) profile may be blended, and we treat the whole
feature as an upper limit for \ion{C}{3} at $\zabs=0.06471$: \logCIII\
$<13.1$. \ion{O}{1} and \ion{C}{4} 1550 are not detected at
$3\sigma$.

The $\zabs=0.06471$ system is one of two tentative metal-line systems
with \logHI\ $<15$.  The \ion{H}{1} COG includes only \Lya\
and \Lyb, but the column density is well constrained because \Lyb\ is
unsaturated: \logHI\ $=14.6^{+0.2}_{-0.14}$ and \Dopb\
$=18^{+1.9}_{-2}\kms$ (see Figure \ref{cog.z06468}). The upper limit to
the equivalent width of \Lyg\ \emph{is} consistent within $1\sigma$
of the value predicted by the COG.  The \ion{H}{1} absorption features
are asymmetric and should probably be fit by a two-component COG, but
the total \nhi\ value is well constrained by our COG analysis.

Assuming a photoionized gas, the ionic ratio
$\mathrm{N}(\mathrm{C}^{+3})/\mathrm{N}(\mathrm{O}^{+5})$ constrain
\logU\ $>\mbox{[C/O]}-1.5$ in the CLOUDY models with \logHI\ $=14.75$.
For \logU\ $=-1.5$ and assuming [C/O] $=0$, [O/H] $>-0.9$ and [C/H]
$<-1$ (see Table \ref{tab:z0.0647x}).  The $\zabs=0.06471$ could be
a single-phase photoionized medium

A single-phase CIE model is \emph{not} ruled out by the
observations. $\mathrm{N}(\mathrm{C}^{+3})/\mathrm{N}(\mathrm{O}^{+5})$
constrains $T \ge 2.2\times10^{5}\K$, for which [O/H] $=-1.8$ and
[C/H] $=-1.8$. This system may represent a detection of the WHIM
because of its temperature and the non-detection of \ion{C}{4}
absorption.

\begin{deluxetable}{lccc}
\tablewidth{0pc}
\tablecaption{ELEMENTAL ABUNDANCES FOR ABSORBER AT $z$=0.06471\label{tab:z0.0647x}}
\tabletypesize{\scriptsize}
\tablehead{\colhead{Ion} &
\colhead{[X/H]} & \colhead{[X/O$^{+5}$]}}
\startdata
C$^{++}$ & $<-1.04$ & $<-0.13$ \\
C$^{+3}$ & $<-1.00$ & $<-0.09$ \\
O$^{0}$ & $< 5.40$ & $< 6.32$ \\
O$^{+5}$ & $-0.92$ & $ 0.00$
\enddata
\tablenotetext{a}{Assumes a photoionized gas with \logU\ $=-1.5$}
\end{deluxetable}

\subsection{ $\zabs=0.09400$: \ion{C}{3}}\label{subsec.z09399}

\begin{figure} 
\epsscale{1.1}\plotone{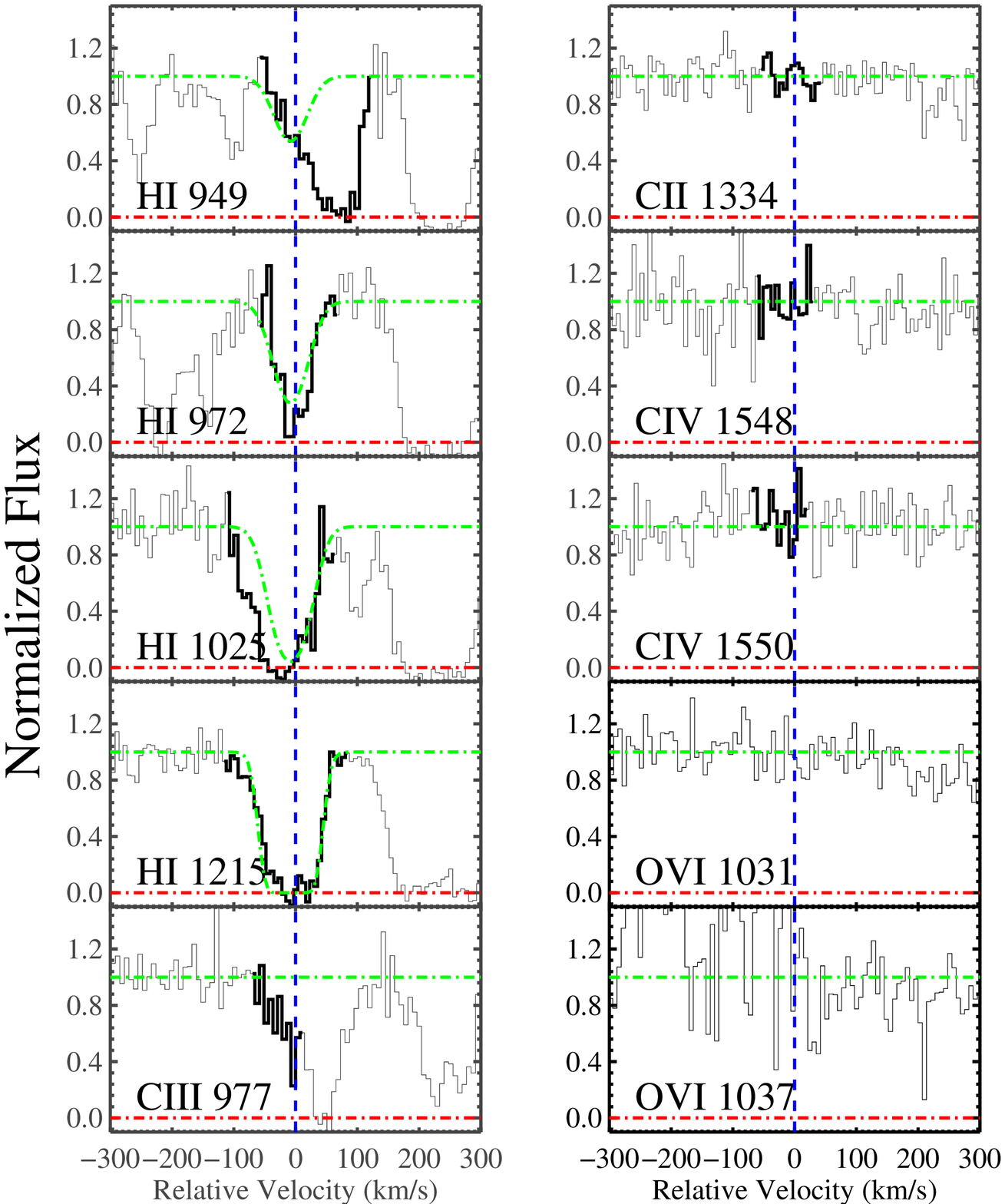}
\caption
{Velocity plot for $\zabs=0.09400$ (see Figure \ref{velplt.z00441}
  description).  \Lyb, \Lyg, and \Lyd\ are blended with Galactic
  \ion{Fe}{2} 1121, Galactic \ion{Fe}{2} 1064, and Galactic \ion{O}{1}
  1039, respectively. \ion{C}{3} is blended with \Lyb\ at
  $\zabs=0.04222$.  \ion{C}{2} and the \ion{C}{4} doublet are not
  detected at $3\sigma$. The region where the \ion{O}{6} doublet would
  be is shown. The Voigt profile centroid is fixed at
  $z_{\gamma}=0.09397$.
\label{velplt.z09399}
}
\end{figure}
\begin{figure} 
\includegraphics[clip,trim=0in 0in 0in 0in,width=0.37\textwidth,
  angle=90]{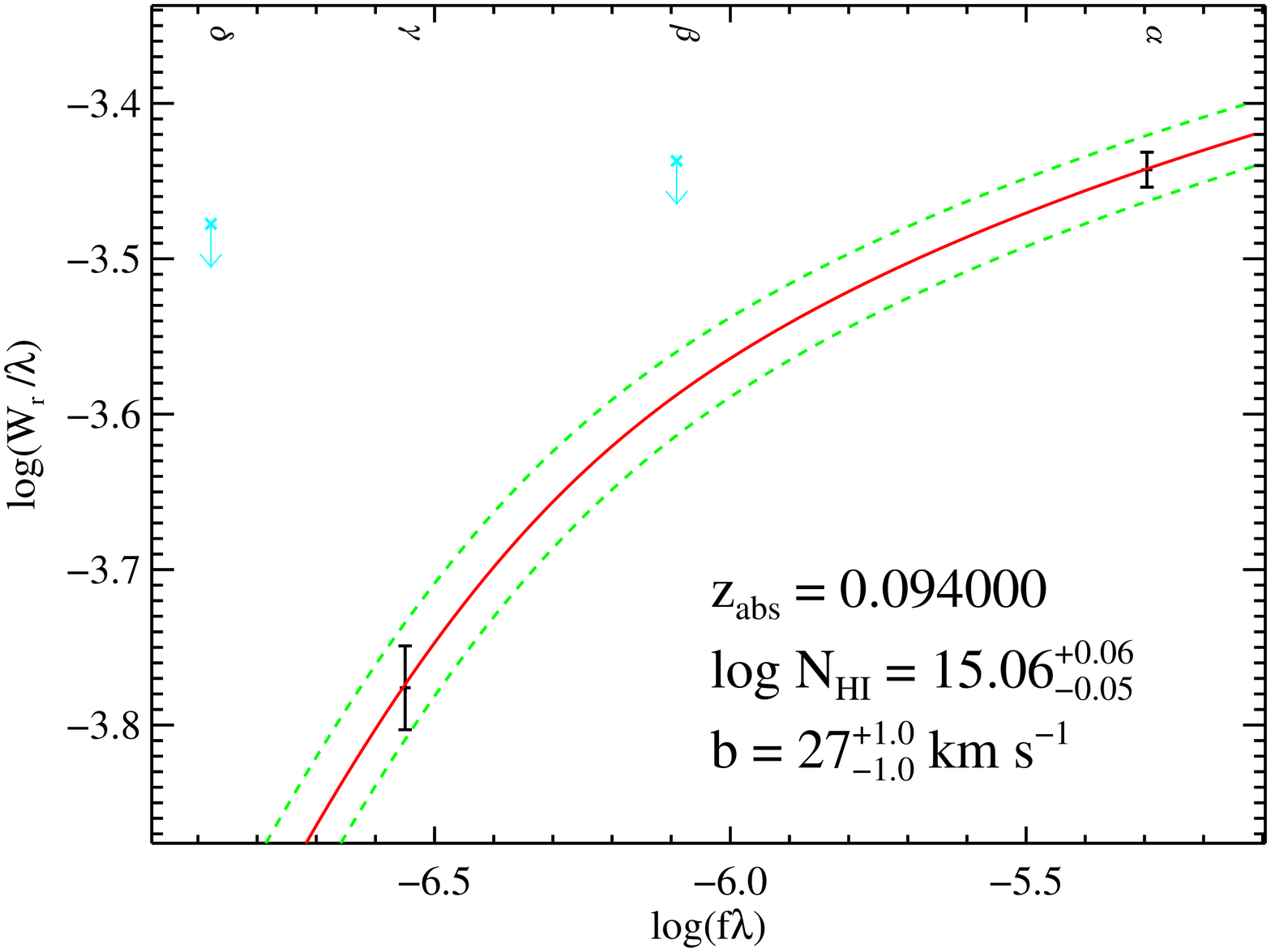}
\caption
{\ion{H}{1} COG for $\zabs=0.09400$ (see Figure \ref{cog.z00441}
  description). The COG analysis only includes \Lya\ and \Lyg\ because
  \Lyb\ and \Lyd\ were blended (see Figure \ref{velplt.z09399}). \Lyg\
  is blended with a weak Galactic line.  Therefore, we conservatively
  consider the best fit to be an upper limit.
\label{cog.z09399}
}
\end{figure}

\Lyb\ and \Lyd\ are significantly blended with Galactic \ion{Fe}{2}
1121 and Galactic \ion{O}{1} 1039, respectively (see Figure
\ref{velplt.z09399}). \Lyg\ is somewhat blended with Galactic
\ion{Fe}{2} 1064.  The degree of blending is apparent from the line
profiles in Figure \ref{velplt.z09399} as well as a velocity plot of
the Galactic lines (not shown).  The COG includes only \Lya\ and the
blended \Lyg\ and yields an upper limit: \logHI\ $\le15.06$ for \Dopb\
$=27^{+1}_{-1}\kms$ (see Figure \ref{cog.z09399}).

\ion{C}{3} is blended with \Lyb\ at $\zabs=0.04222$, though the part
included in Figure \ref{velplt.z09399} is well aligned. Since
\ion{C}{3} is partially deblended, there is a lower limit on the
column density \logCIII\ $>13.3$ from the AODM (see Table
\ref{tab:ionsumm}). \ion{C}{2} and the \ion{C}{4} doublet are not
detected at $3\sigma$.

In the CLOUDY models with \logHI\ $=15$, the ionic ratios
$\mathrm{N}(\mathrm{C}^{+})/\mathrm{N}(\mathrm{C}^{++})$ and
$\mathrm{N}(\mathrm{C}^{++})/\mathrm{N}(\mathrm{C}^{+3})$ constrain
the ionization parameter: $-3.6\le$ \logU\ $\le-1.5$. For \logU\
$=-1.5$, $-1.3\le$ [C/H] $\le+0.3$. This system is well modeled by a
single-phase, photoionized medium.

\subsection{ $\zabs=0.09487$: Partial Lyman Limit System}\label{subsec.z09486}

\begin{figure*} 
\epsscale{1.1}\plottwo{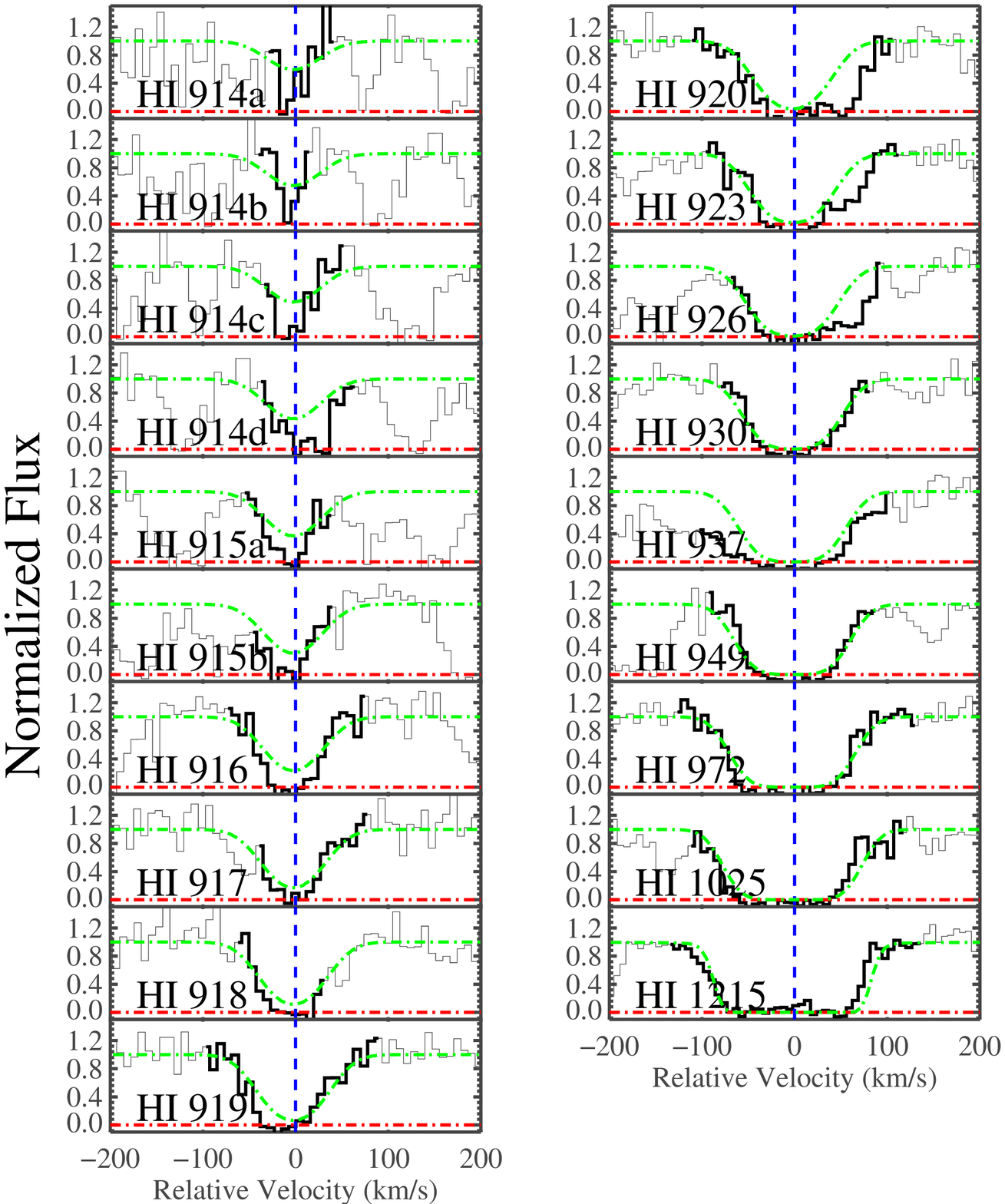}{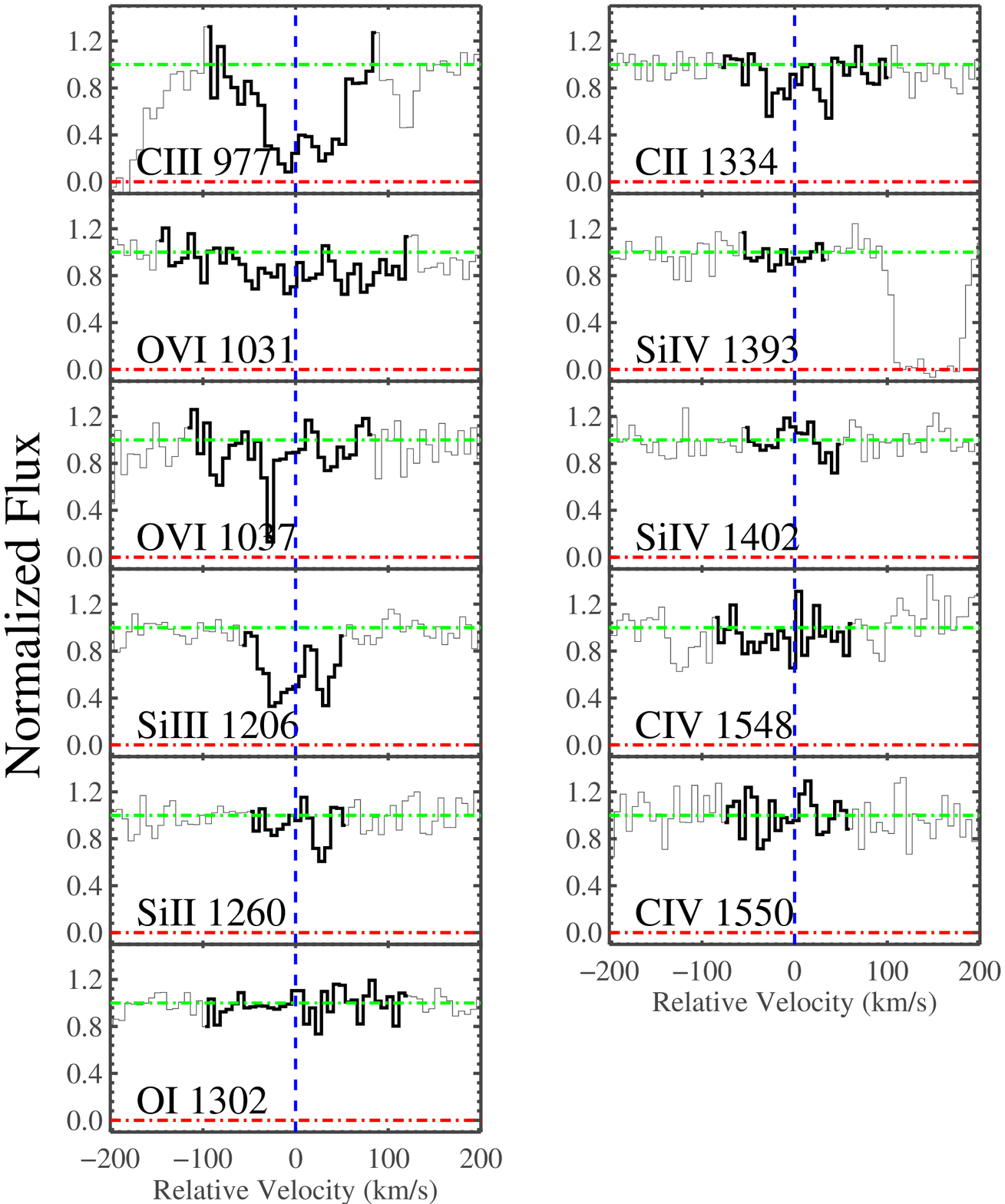}
\caption
{Velocity plot for the partial Lyman limit system at $\zabs=0.09487$
 (see Figure \ref{velplt.z00441} description). \Lye\ lies in the wings
 the damped Galactic \Lyb, and the higher-order Lyman lines \ion{H}{1}
 914d, 916, 920, 923, and 926 are blended or coincident with \HH. For
 \ion{H}{1} 917 and 918, the \HH\ blends, 1004.0 R(2) and 1005.4
 P(2), respectively, are excluded from the \EWr\ measure. This system
 was used to shift the \fuse\ channels onto the \stis\ wavelength
 solution. The Voigt profile centroid is fixed at
 $z_{915\mathrm{a}}=0.09486$.  \Lya, \ion{C}{3}, \ion{C}{2}, and \ion{Si}{3} are
 well aligned, and they have similar, multi-component profiles. A
 broad \ion{O}{6} doublet is detected at $>\!3\sigma$.  The \EWr\
 ratio of \ion{O}{6} 1031 to 1037 is $2.3\pm0.8$, in agreement with
 the predicted value for the unsaturated regime. \ion{Si}{2},
 \ion{O}{1}, and the \ion{Si}{4} and \ion{C}{4} doublets
 are not detected at $3\sigma$.  (Note: the horizontal limits are
 from -200 to $+200\kms$.)  
\label{velplt.z09486a}
}
\end{figure*}
\begin{figure} 
\includegraphics[clip,trim=0in 0in 0in 0in,width=0.37\textwidth,
  angle=90]{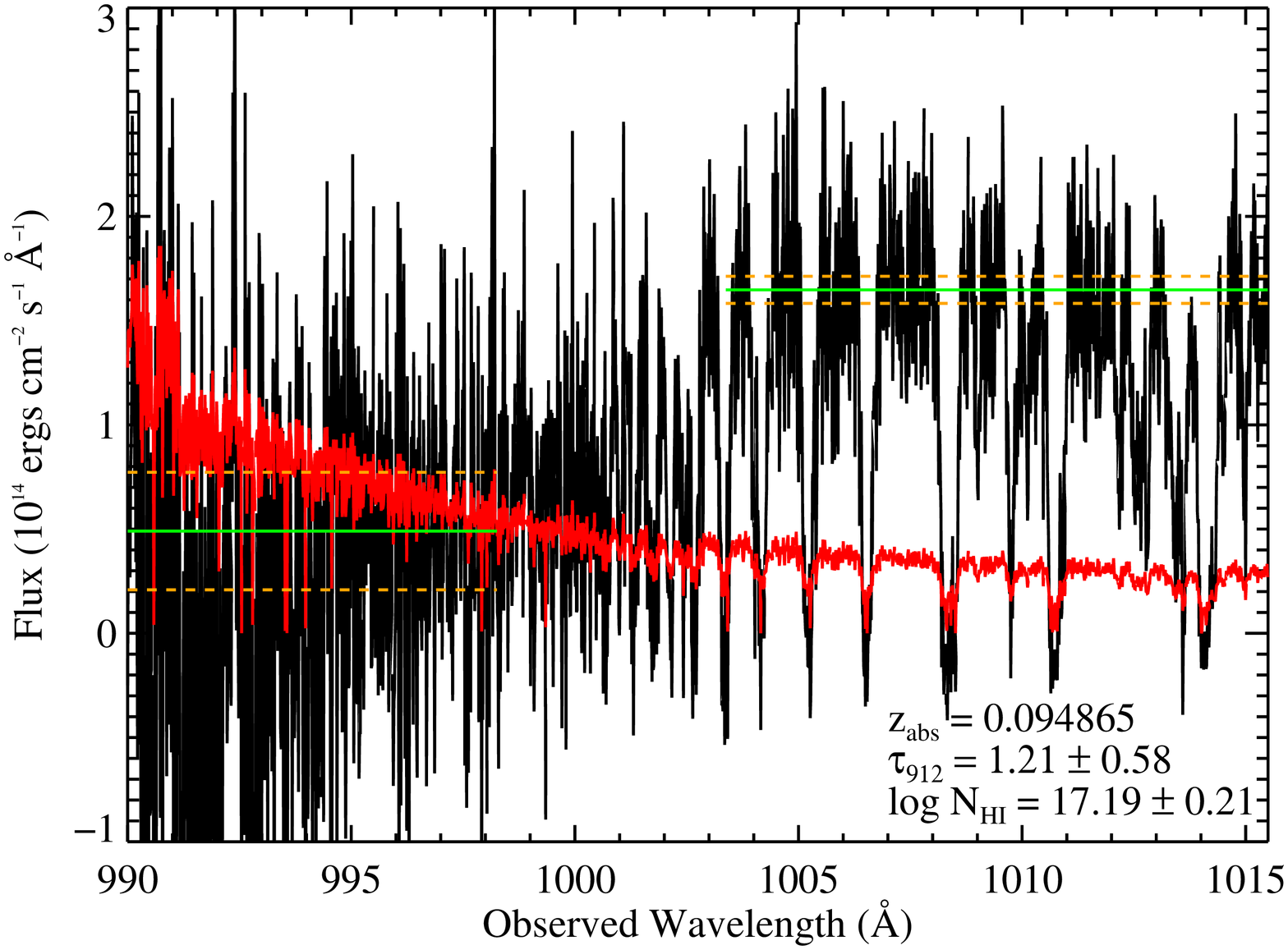}
\caption
{Lyman limit at $\zabs=0.09487$. We show our fits and conservative
  error estimates (solid green and dashed orange lines, respectively)
  for the quasar continuum redward of the $\lambda912$ break and the
  flux blueward. From the flux decrement, we measure the optical depth
  at the limit and the \ion{H}{1} column density.  The red line is the
  error array of the spectrum.  This portion of the spectrum is at the
  blue edge of LiF 1A, where the sensitivity decreases.
\label{lls.z09486}
}
\end{figure}
\begin{figure} 
\includegraphics[clip,trim=0in 0in 0in 0in,width=0.37\textwidth,
  angle=90]{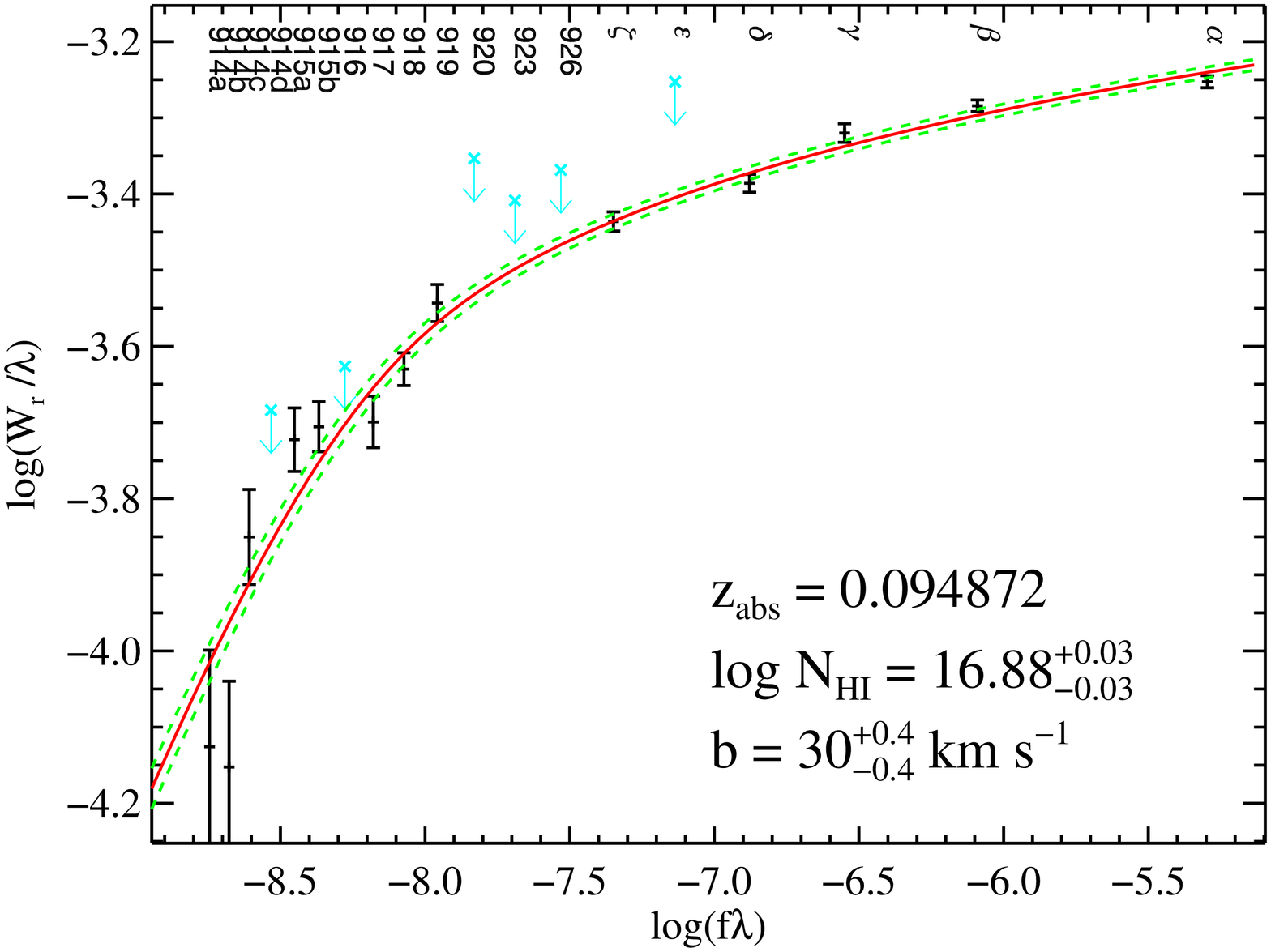}
\caption
{\ion{H}{1} COG for $\zabs=0.09487$ (see Figure \ref{cog.z00441}
  description). This system is fit well by a single-component COG
  model despite \Lya\ having a multi-component line profile (see
  Figure \ref{velplt.z09486a}).
\label{cog.z09486}
}
\end{figure}
\begin{figure} 
\includegraphics[clip,trim=0in 0in 0in 0in,width=0.37\textwidth,
  angle=90]{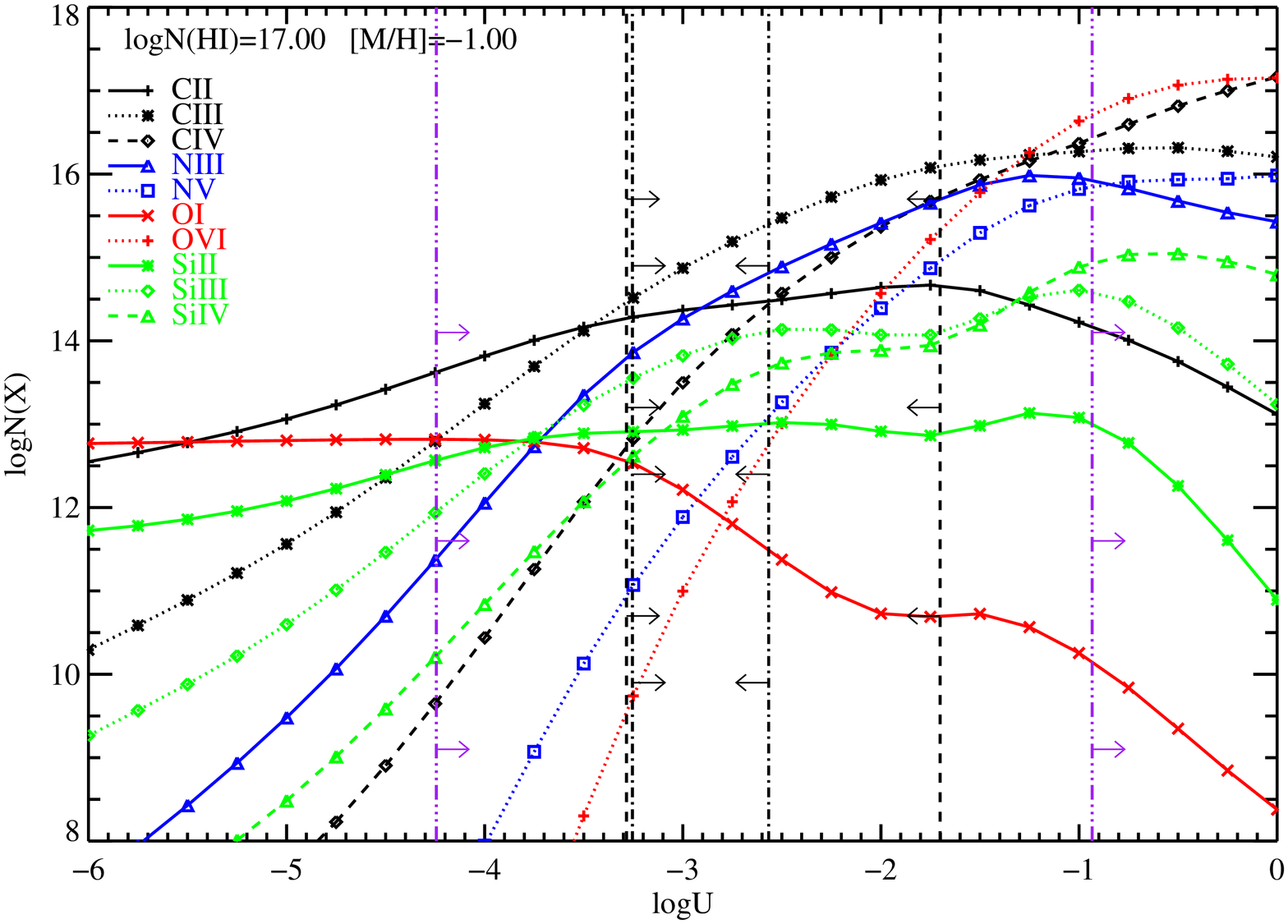}
\caption
{Column densities from CLOUDY photoionization model for
  $\zabs=0.09487$. The model is parameterized by \logHI\ $=16.75$ and
  metallicity scaled to one-tenth solar abundance \Z\ $=-1$.  The
  ionic ratios of the detected metal lines for $\zabs=0.09487$
  constrain \logU:
  $\mathrm{N}(\mathrm{C}^{+})/\mathrm{N}(\mathrm{C}^{++})$ and
  $\mathrm{N}(\mathrm{C}^{++})/\mathrm{N}(\mathrm{C}^{+3})$ (black
  dashed lines, left and right, respectively);
  $\mathrm{N}(\mathrm{Si}^{+})/\mathrm{N}(\mathrm{Si}^{++})$ and
  $\mathrm{N}(\mathrm{Si}^{++})/\mathrm{N}(\mathrm{Si}^{+3})$ (black
  dash-dotted lines, left and right, respectively); and
  $\mathrm{N}(\mathrm{O}^{0})/\mathrm{N}(\mathrm{C}^{++})$ and
  $\mathrm{N}(\mathrm{C}^{+3})/\mathrm{N}(\mathrm{O}^{+5})$ (purple
  dash-dot-dot-dotted lines, from left to right).
\label{cldy.z09486}
}
\end{figure}
\begin{figure} 
\includegraphics[clip,trim=0in 0in 0in 0in,width=0.37\textwidth,
  angle=90]{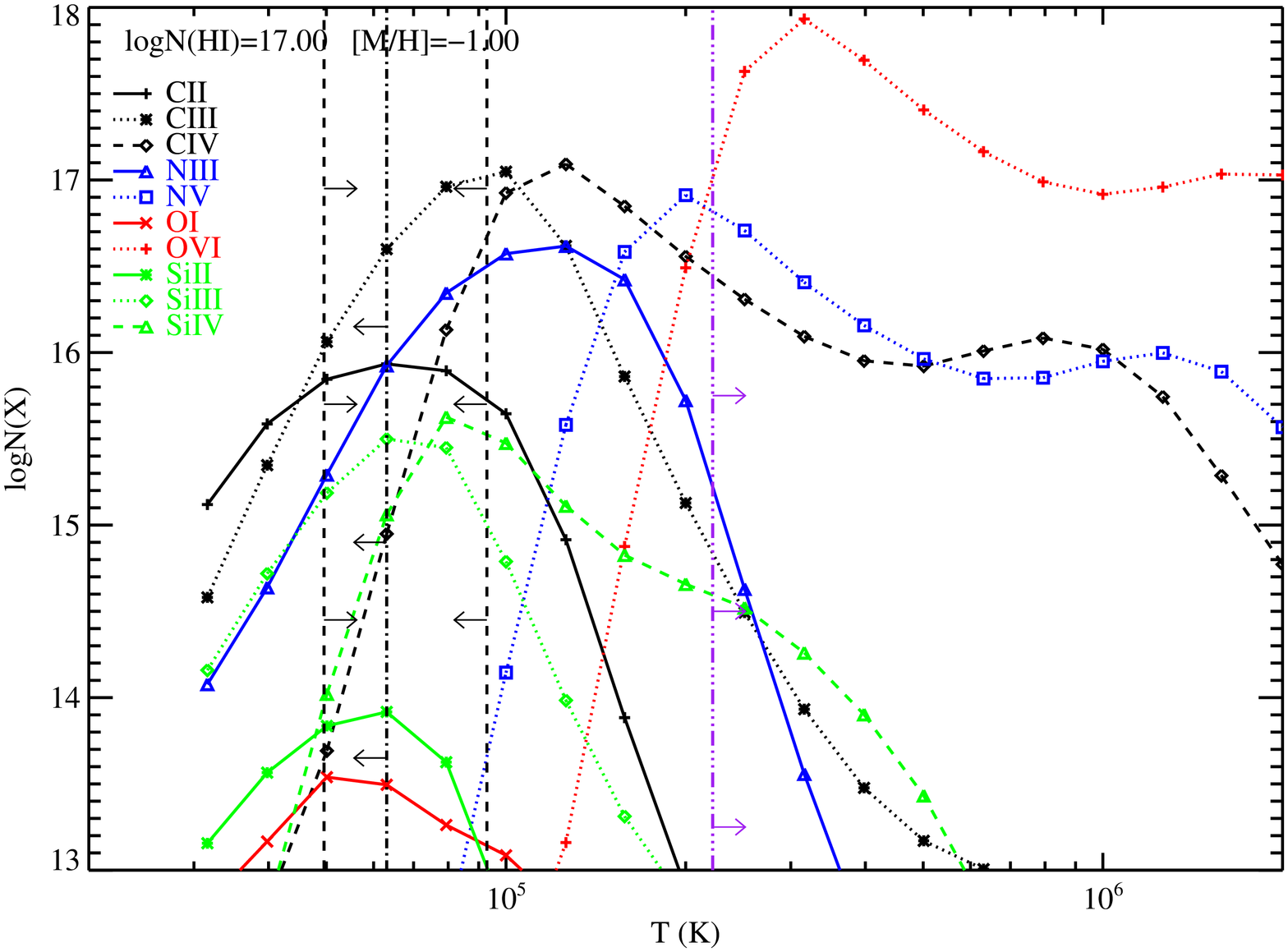}
\caption
{Column densities from CLOUDY collisional ionization equilibrium model
  for $\zabs=0.09487$. The model is parameterized by \logHI\ $=16.75$
  and \Z\ $=-1$ (same as in Figure \ref{cldy.z09486}).  The ionic
  ratios of detected metal lines for $\zabs=0.09487$ constrain the
  temperature:
  $\mathrm{N}(\mathrm{C}^{+})/\mathrm{N}(\mathrm{C}^{++})$ and
  $\mathrm{N}(\mathrm{C}^{++})/\mathrm{N}(\mathrm{C}^{+3})$ (dashed
  lines, left and right, respectively);
  $\mathrm{N}(\mathrm{Si}^{+})/\mathrm{N}(\mathrm{Si}^{++})$ (dash-dotted
  line); and $\mathrm{N}(\mathrm{C}^{+3})/\mathrm{N}(\mathrm{O}^{+5})$
  (purple dash-dot-dot-dotted line).
\label{cldy.z09486coll}
}
\end{figure}

This system shows strong \ion{H}{1} Lyman absorption from \Lya\ to
\ion{H}{1} 914a (see Figure \ref{velplt.z09486a}). \Lye\ is in the
wings of Galactic \Lyb\ absorption, and the higher-order Lyman lines
\ion{H}{1} 926 up to \ion{H}{1} 914a fall in a region riddled with
\HH\ lines.  Though the Lyman break $\lambda912$ is in a region of LiF
1A with low sensitivity, \logHI\ can be measured from the flux
decrement at the limit. The optical depth
$\tau_{912}=\ln(f_{QSO}/f_{912})=\sigma_{912} \mnhi$, where $f_{QSO}$
($f_{912}$) is the flux redward (blueward) of the break and the cross
section at the limit $\sigma_{912}=7.9\times10^{-18}\cm{-2}$ (see
Figure \ref{lls.z09486}). We measure \logHI\ $=17.2\pm0.2$, consistent
with the COG value discussed below.

The majority of the Lyman series from \Lya\ to \ion{H}{1} 914a were
used in the \ion{H}{1} COG analysis: \logHI\ $=16.88^{+0.03}_{-0.03}$
and \Dopb\ $=30.3^{+0.4}_{-0.4}\kms$ (see Figure \ref{cog.z09486}).  A
single-component COG fits the data well, though this system has
multiple components, as seen in the \Lya, \ion{C}{3}, \ion{Si}{2}, and
\ion{Si}{3} line profiles.

\ion{C}{3} and \ion{Si}{3} are well-aligned with \Lya.  This system is
the strongest \ion{C}{3} absorber in the \pks\ sight line \logCIII\
$>13.9$.  A broad, well-aligned \ion{O}{6} doublet is detected with an
\EWr\ ratio of $2.3\pm0.8$ and \logOVI\ $=14$.  \ion{Si}{2}
$\lambda1260$, \ion{O}{1}, and the \ion{Si}{4} and
\ion{C}{4} doublets are not detected at $3\sigma$.

For the CLOUDY models with \logHI\ $=17$, the ionic ratios
$\mathrm{N}(\mathrm{C}^{+})/\mathrm{N}(\mathrm{C}^{++})$ and
$\mathrm{N}(\mathrm{Si}^{++})/\mathrm{N}(\mathrm{Si}^{+3})$ set
$-3.3\le$ \logU\ $\le-2.6$ (see Figure \ref{cldy.z09486}).  For \logU\
$=-2.9$, we derive $-2<$ [C/H] $<-1.6$ and [Si/H] $=-1.7$ (see Table
\ref{tab:z0.0949x}).  \ion{O}{6} is very broad, and this indicates at
least a \emph{kinematically} different phase from the \ion{C}{3} and
\ion{Si}{3} absorption. Likely, \ion{O}{6} is thermally broadened, and
we should consider the CLOUDY CIE models.  The system could not be
reasonably described by a single-phase CIE model since there would not
be significant absorption of \ion{C}{3} and \ion{O}{6} at one
temperature without significant \ion{C}{4} absorption (see Figure
\ref{cldy.z09486coll}). For CIE, the temperature limit
$T>2.2\times10^{5}$, set by
$\mathrm{N}(\mathrm{C}^{+3})/\mathrm{N}(\mathrm{O}^{+5})$, yields
[O/H] $>-3.8$ assuming the total \nhi\ value of this absorber, which
is most likely dominated by the photoionized phase.

The $\zabs=0.09487$ partial Lyman limit system is a metal-poor
($-3.8\lesssim$ \Z\ $\lesssim-1.6$) and two-phased medium. \ion{C}{3} and
\ion{Si}{3} are from one phase; they are narrow, multi-component
features from a photoionized medium. The broad \ion{O}{6} indicates
another phase that is likely collisionally ionized but is also
reasonably described by a photoionization model.


\begin{deluxetable}{lccc}
\tablewidth{0pc}
\tablecaption{ELEMENTAL ABUNDANCES FOR ABSORBER AT $z$=0.09487\label{tab:z0.0949x}}
\tabletypesize{\scriptsize}
\tablehead{\colhead{Ion} &
\colhead{[X/H]} & \colhead{[X/C$^{++}$]}}
\startdata
C$^{+}$ & $<-1.60$ & $< 0.41$ \\
C$^{++}$ & $>-2.02$ & $> 0.00$ \\
C$^{+3}$ & $<-1.12$ & $< 0.90$ \\
O$^{0}$ & $< 0.95$ & $< 2.96$ \\
O$^{+5}$ & $ 1.72$ & $ 3.74$ \\
Si$^{+}$ & $<-1.41$ & $< 0.61$ \\
Si$^{++}$ & $-1.72$ & $ 0.29$ \\
Si$^{+3}$ & $<-1.51$ & $< 0.51$
\enddata
\tablenotetext{a}{Assumes a photoionized gas with \logU\ $=-2.9$}
\end{deluxetable}

\subsection { $\zabs=0.14529$: \ion{C}{3}}\label{subsec.z14529}

\begin{figure} 
\epsscale{1.1}\plotone{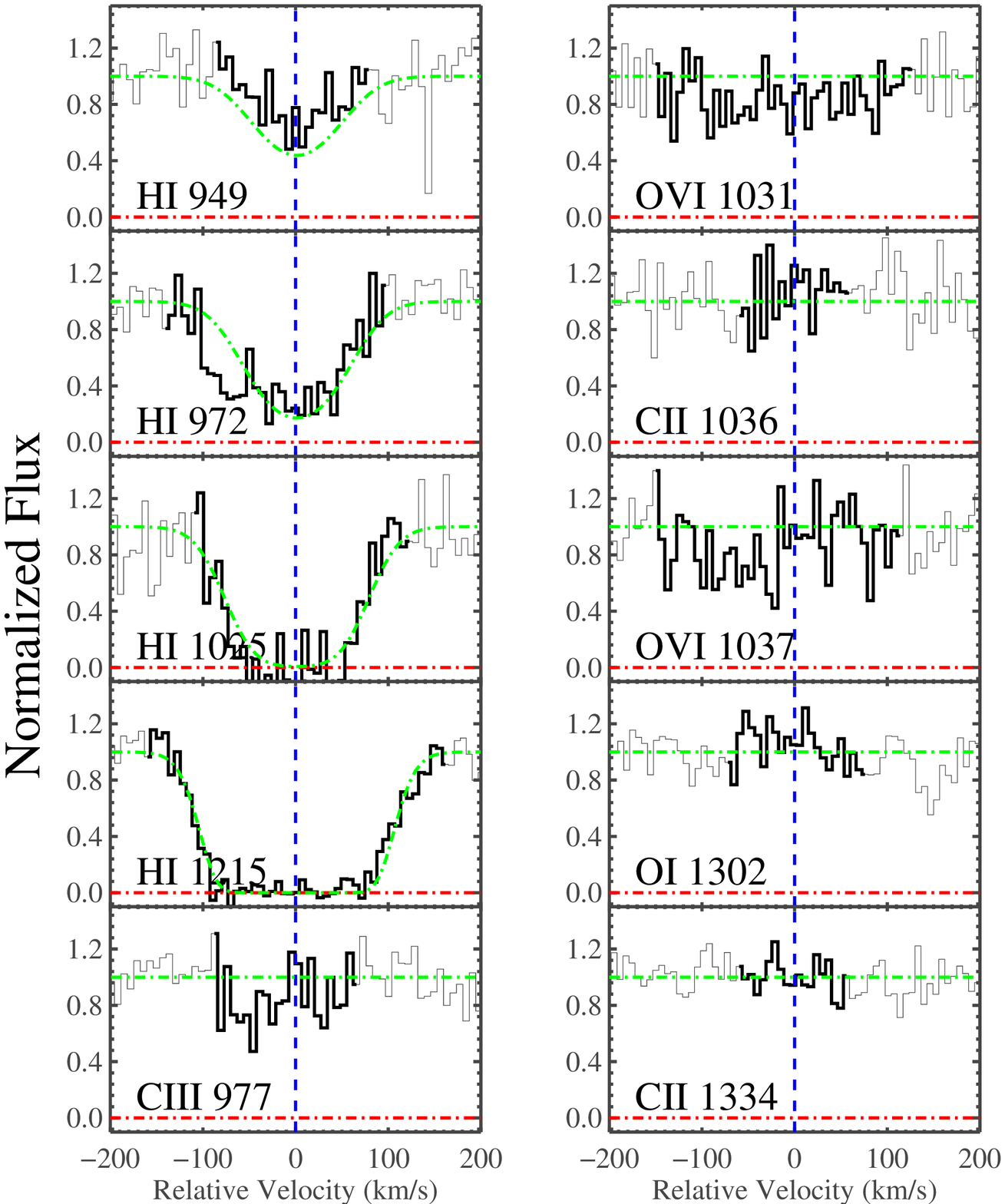}
\caption
{Velocity plot for $\zabs=0.14533$ (see Figure \ref{velplt.z00441}
  description). \Lyd\ is near the edge of the LiF 2A
  spectrum. \ion{C}{3} is a $>\!6\sigma$ detection (see Table
  \ref{tab:ionsumm}). Both \ion{C}{2} lines and \ion{O}{1} are not
  detected at $>\!3\sigma$. The \ion{O}{6} doublet is not a confirmed
  detection; \ion{O}{6} 1037 is coincident with \Lyb\ at
  $\zabs=0.15835$ and is at the edge of LiF 1B and in a noisy
  region of \stis. The Voigt profile centroid is fixed at
  $z_{\beta}=0.14534$.  (Note: the horizontal limits are
 from -200 to $+200\kms$.)  
\label{velplt.z14529}
}
\end{figure}
\begin{figure} 
\includegraphics[clip,trim=0in 0in 0in 0in,width=0.37\textwidth,
  angle=90]{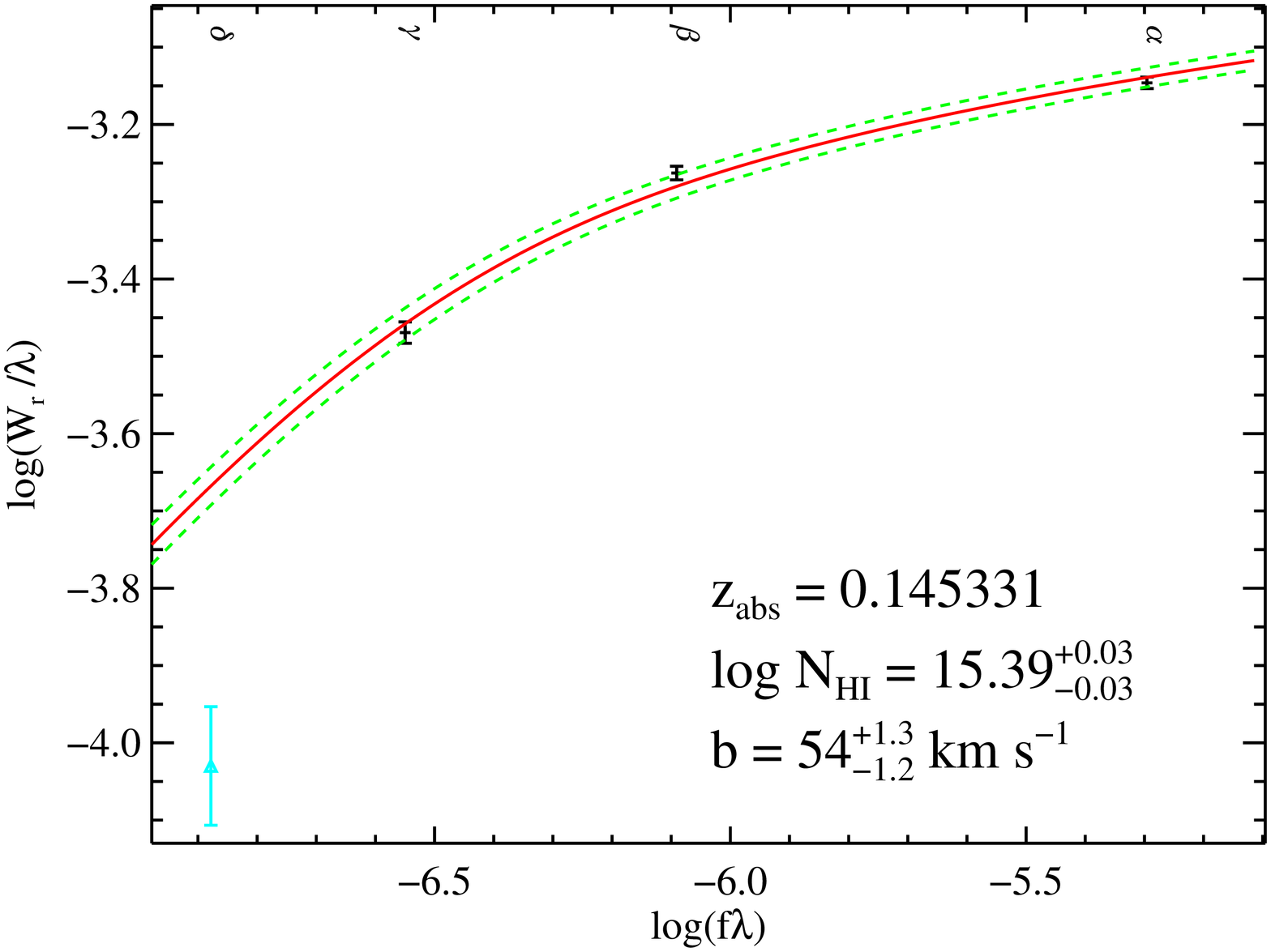}
\caption
{\ion{H}{1} COG for $\zabs=0.14533$ (see Figure \ref{cog.z00441}
  description). The difference between the predicted and measured
  \EWr\ of \Lyd\ is significant and suggests that \ion{H}{1} is
  multi-component and poorly fit by this single-component COG.
\label{cog.z14529}
}
\end{figure}

Although \Lya\ is as broad as that of the partial Lyman limit system
discussed previously, the system at $\zabs=0.14533$ has a
significantly lower \nhi\ value (see Figure
\ref{velplt.z14529}). \Lya, \Lyb, and \Lyg\ were used to fit the
\ion{H}{1} COG: \logHI\ $=15.39^{+0.03}_{-0.03}$ and \Dopb\
$=54^{+1.3}_{-1.2}\kms$ (see Figure \ref{cog.z14529}). \Lyd\ was excluded
because it lies near the edge of LiF 2A; it deviates from the value
predicted by the COG by $>\!3\sigma$. This discrepancy may also
indicate that the system is multicomponent and poorly modeled by a
single-component COG.

\ion{C}{3} is well aligned with the broad \Lya, and \logCIII\ $=13.2$
(see Table \ref{tab:ionsumm}). A detection of the \ion{O}{6} doublet
is not confirmed because \ion{O}{6} 1037 is at the edge of LiF 1B and
in the low-sensitivity region of \stis. The equivalent width of
\ion{O}{6} 1037 is greater than that of 1031 because 1037 is 
coincident with \Lyb\ at $\zabs=0.15835$. An upper limit is given by
\ion{O}{6} 1031: \logOVI\ $<14.2$. \ion{O}{1} and \ion{C}{2} are not
detected at $3\sigma$ significance.

In the CLOUDY model with \logHI\ $=15.5$, the ionic ratios
$\mathrm{N}(\mathrm{C}^{+})/\mathrm{N}(\mathrm{C}^{++})$ and
$\mathrm{N}(\mathrm{C}^{++})/\mathrm{N}(\mathrm{O}^{+5})$ set $-3.7\le$
\logU\ $\le \mbox{[C/O]}-1.3$.  Assuming \logU\ $=-1.9$, the value
predicted by \citet{prochaskaetal04}, [C/H] $=-1.9$ and [O/H] $<-0.4$
for [C/O] $=0$.  In the CLOUDY CIE model, the same ratios set
$4.1\times10^{4}\le T \le1.9\times10^{5}\K$. Assuming the width of
\ion{H}{1} is due purely to thermal broadening, then $T=b^{2}m/(2
k)<1.8\times10^{5}\K$, where $m$ is the mass of hydrogen and $k$ is
the Boltzmann constant. For the latter $T$ bound, [C/H] $=-1.8$ and
[O/H] $<-0.8$.  This system can be modeled by a single-phase
photoionized or collisionally-ionized medium.

\subsection{ $\zabs=0.19156$: \ion{C}{3}}\label{subsec.z19156}

\begin{figure}
\epsscale{1.1}\plotone{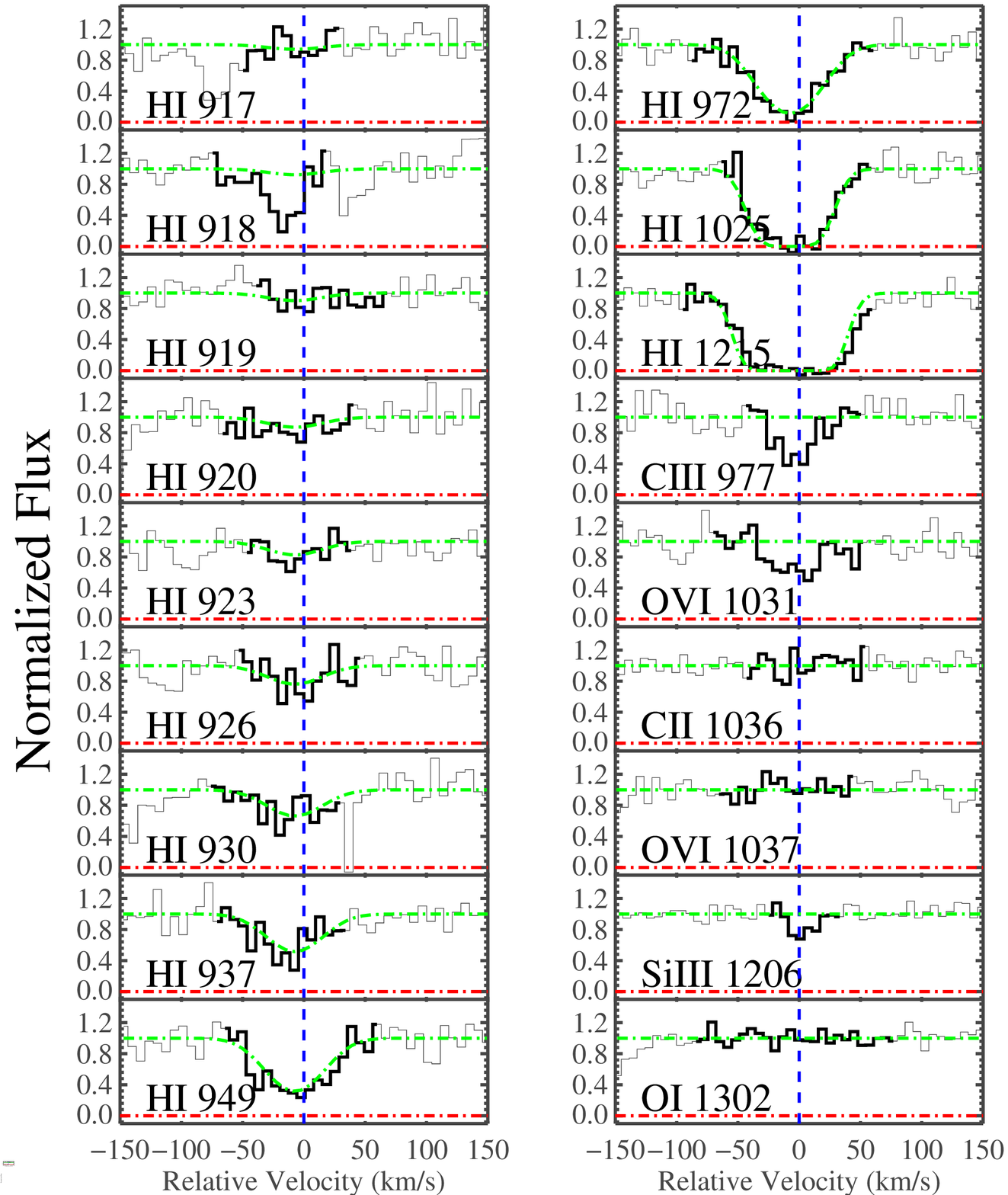}
\caption
{Velocity plot for $\zabs=0.19161$ (see Figure \ref{velplt.z00441}
  description). \ion{H}{1} 917, both \ion{C}{2} lines, \ion{O}{1},
  \ion{Si}{3}, and \ion{O}{6} 1037 are not detected at
  $3\sigma$. \ion{H}{1} 918 is blended with \HH\ 1094.0 P(1). The
  Voigt profile centroid is fixed at $z_{\delta}=0.19158$.
\label{velplt.z19156}
}
\end{figure}
\begin{figure}
\includegraphics[clip,trim=0in 0in 0in 0in,width=0.37\textwidth,
  angle=90]{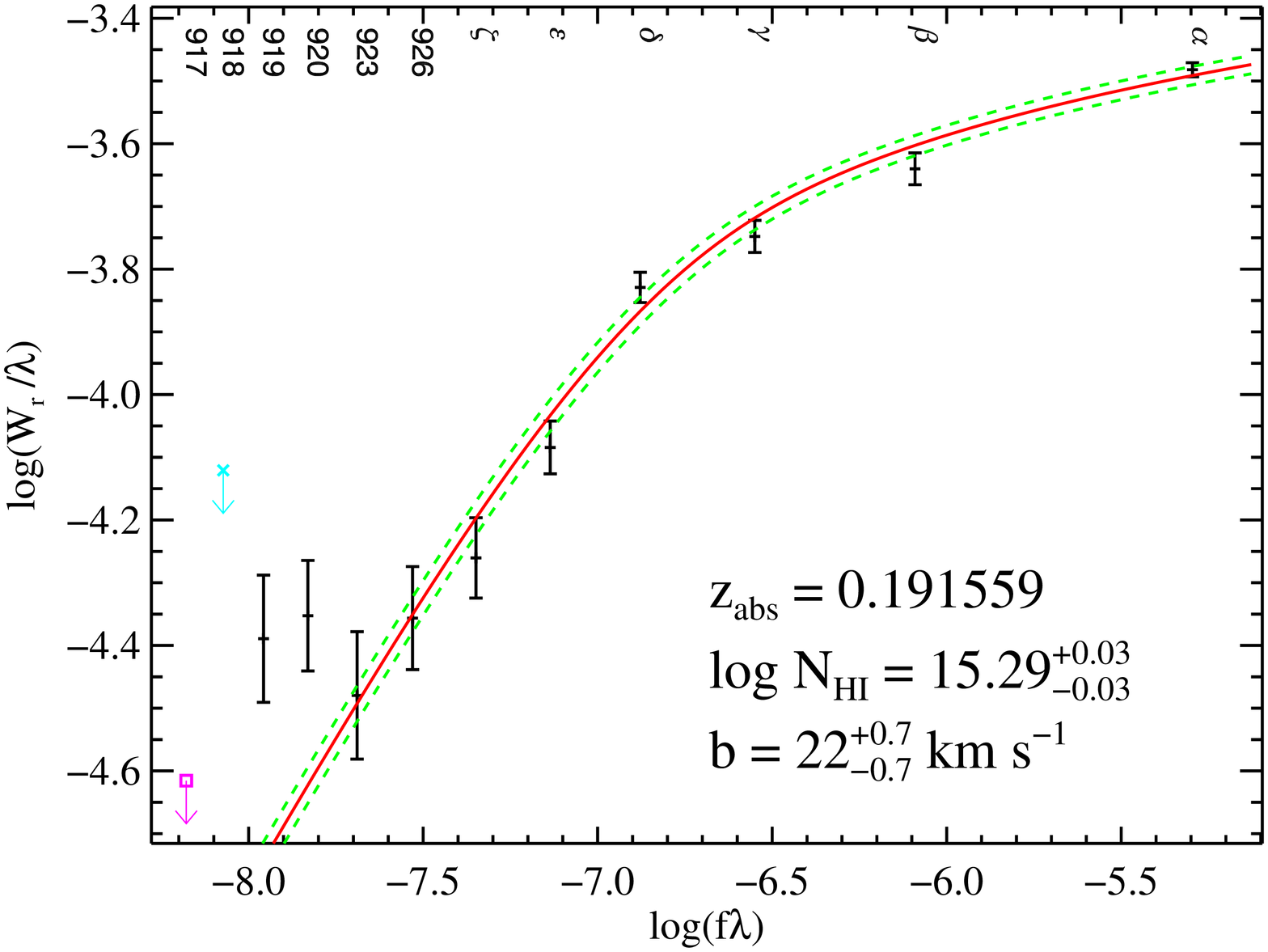}
\caption
{\ion{H}{1} COG for $\zabs=0.19161$ (see Figure \ref{cog.z00441}
  description).  
\label{cog.z19156}
}
\end{figure}

This system is another strong Lyman absorber with \ion{C}{3} well
aligned with \Lya\ (see Figure \ref{velplt.z19156}). The \ion{H}{1}
COG is consistent from \Lya\ to \ion{H}{1} 917: \logHI\
$=15.29^{+0.03}_{-0.03}$ and \Dopb\ $=22.4^{+0.7}_{-0.7}\kms$.
\ion{H}{1} 917, \ion{O}{6} 1037, \ion{O}{1}, both
\ion{C}{2} lines, and \ion{Si}{3} are not detected at $3\sigma$.  \ion{H}{1} 918 is
blended with \HH\ 1094.0 P(1) (see Figure \ref{cog.z19156}).

This system does not have strong \ion{C}{3} absorption: \logCIII\
$=13.1$ (see Table \ref{tab:ionsumm}). In the CLOUDY models for
\logHI\ $=15.25$, the ionic ratios
$\mathrm{N}(\mathrm{C}^{+})/\mathrm{N}(\mathrm{C}^{++})$ and
$\mathrm{N}(\mathrm{C}^{++})/\mathrm{N}(\mathrm{O}^{+5})$ constrain
$-3.7\ge$ \logU\ $\le \mbox{[C/O]}-1.4$.  For \logU\ $=-1.7$, [C/H]
$=-1.9$, [Si/H] $<-1.5$, and [O/H] $\le-1.1$ for [C/O] $=0$. This
system can be described as a single-phase photoionized medium.

\subsection{ $\zabs=0.22555$: \ion{O}{6}}\label{subsec.z22552}

\begin{figure}
\plotone{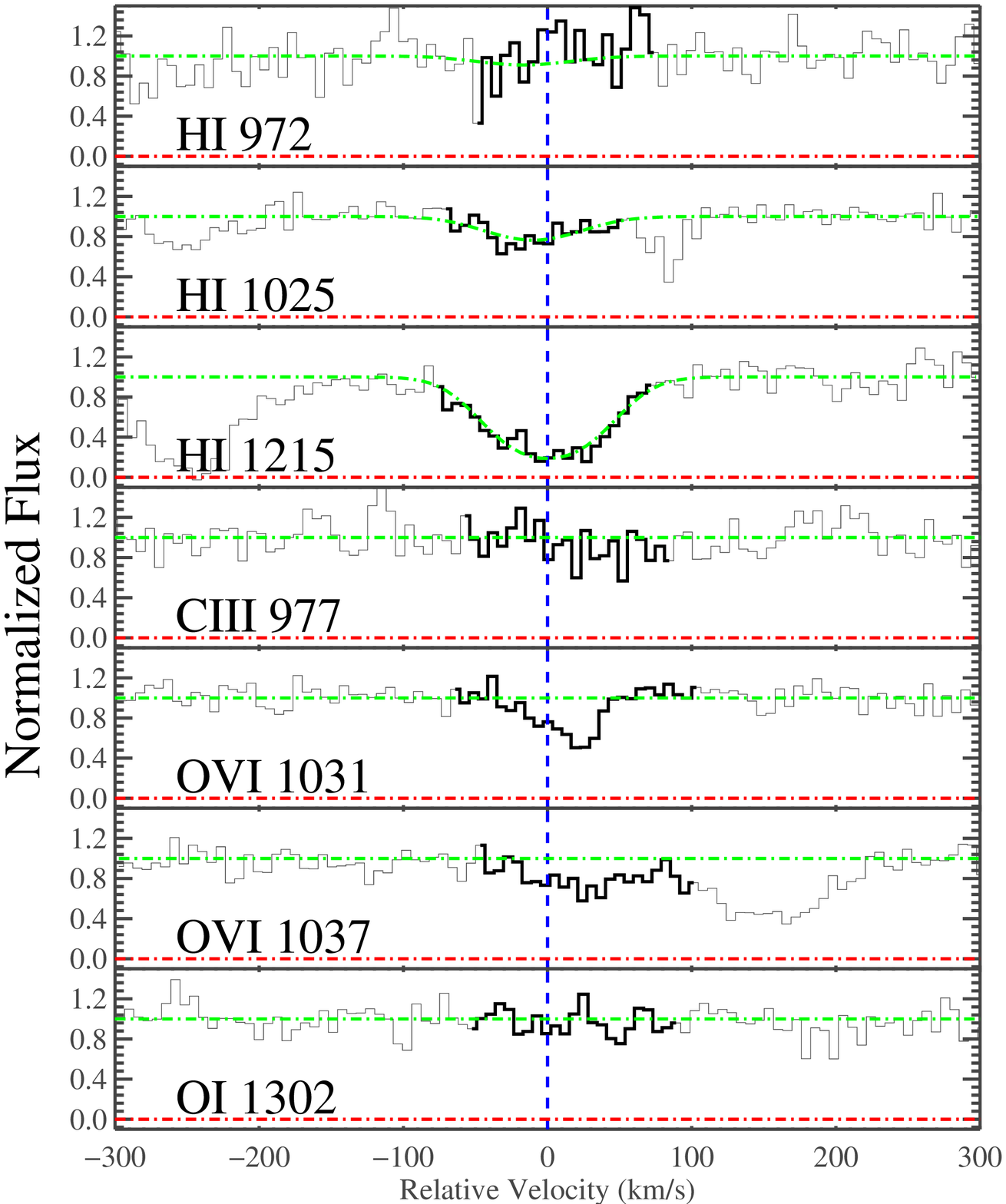}
\caption
{Velocity plot for $\zabs=0.22555$ (see Figure \ref{velplt.z00441}
  description). \ion{O}{6} 1037 is
  slightly blended with \Lya\ at $\zabs=0.04222$ on the red side.
  \Lyg, \ion{C}{3}, and \ion{O}{1} are not detected at $3\sigma$. The
  Voigt profile centroid is fixed at $z_{\alpha}=0.22555$.
\label{velplt.z22552}
}
\end{figure}
\begin{figure}
\includegraphics[clip,trim=0in 0in 0in 0in,width=0.37\textwidth,
  angle=90]{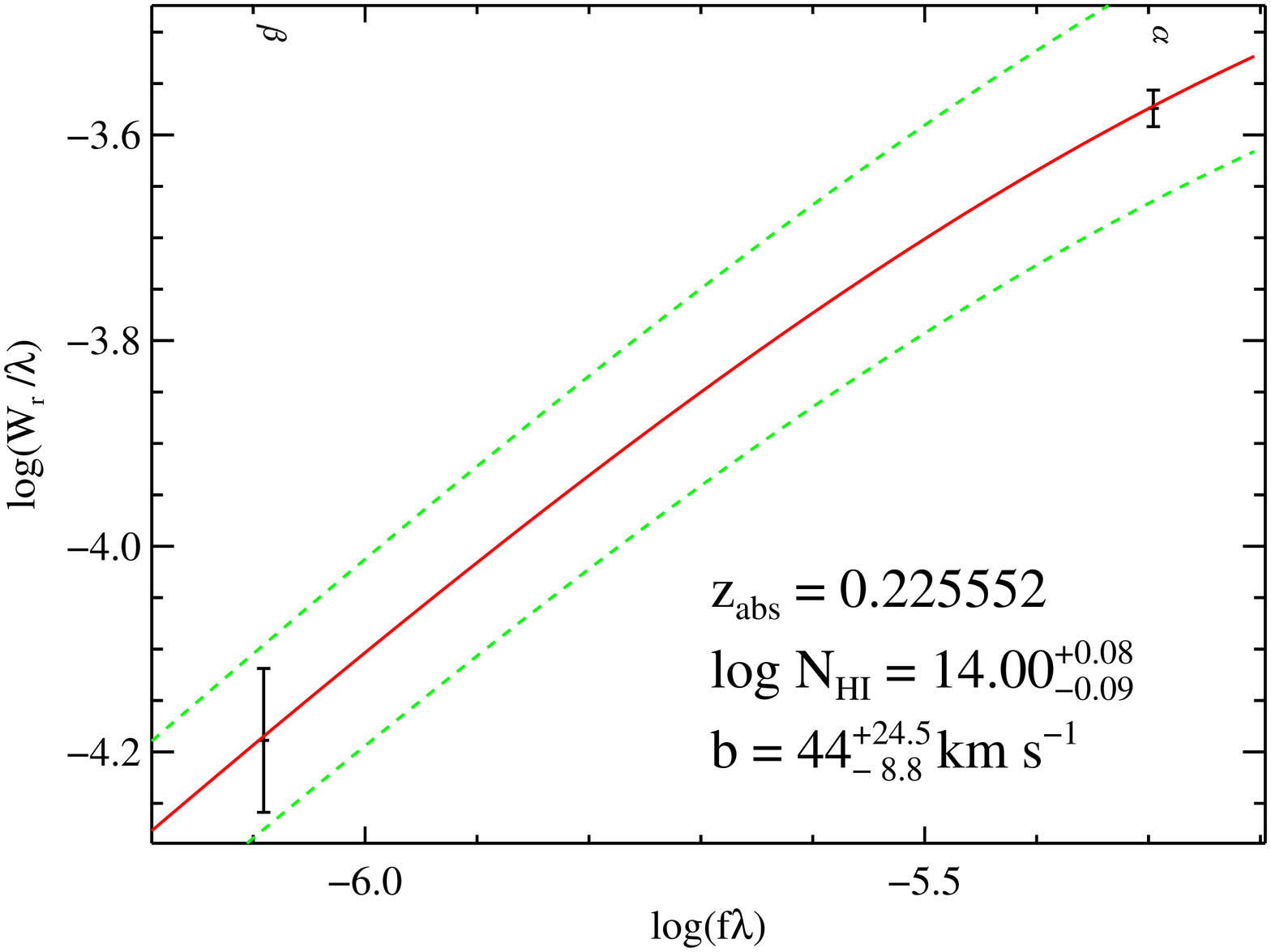}
\caption
{\ion{H}{1} COG for $\zabs=0.22555$ (see Figure \ref{cog.z00441}
  description).
\label{cog.z22552}
}
\end{figure}

This system has the second lowest \logHI\ of the potential metal-line
systems: \logHI\ $=14^{+0.08}_{-0.09}$ and \Dopb\
$=44^{+25}_{-9}\kms$. The COG analysis is based on \Lya\ and \Lyb. 
\logOVI\ $=13.9$.  The equivalent width of \ion{O}{6} 1037 is greater
than that of 1031; \ion{O}{6} 1037 is partially blended with \Lya\ at
$\zabs=0.04658$, and the continuum fit is poor. \ion{C}{3} is not detected,
and \logCIII\ $<13$. The velocity plot and the COG are presented in Figures
\ref{velplt.z22552} and \ref{cog.z22552}, respectively. 

The ionic ratio
$\mathrm{N}(\mathrm{C^{++}})/\mathrm{N}(\mathrm{O^{+5}})$ constrains
\logU\ $>\mbox{[C/O]}-1.4$ or $T \ge 1.9\times10^{5}\K$. For these
lower bounds and assuming [C/O] $=0$, [O/H] $=-0.3$ and [C/H]
$\lesssim-0.3$.  The ionization parameter predicted from
\citet{prochaskaetal04} is \logU\ $=-0.8$ for which the oxygen
abundance would be unreasonably small. Tentatively, we consider this
system to be collisionally ionized. Observations that covered the
\ion{C}{4} doublet would better constrain the ionization mechanism.

\subsection{ $\zabs=0.22752$: \ion{O}{6}}\label{subsec.z22752}

The weak \Lya\ absorber at $\zabs=0.22752$ may potentially have
\ion{O}{6} absorption associated with it. \Lya\ and \ion{O}{6} 1031
are detected at $>\!3\sigma$. From the AODM, \logHI\ $=13.1\pm0.1$ and
\logOVI\ $=13.56\pm0.09$. \ion{O}{6} 1031 is likely blended with \Lya\
at $\zabs=0.04222$. No other common absorption line is evident or
observable: \ion{C}{2} 1036 is coincident with \ion{O}{6} at
$\zabs=0.22555$ and \Lya\ at $\zabs=0.04658$; \ion{C}{3} is coincident
with Galactic \ion{N}{1}; and \ion{C}{4} is shifted out of the \stis\
wavelength coverage. No CLOUDY models were examined for this system.

\section{Strong \Lya\ Absorbers without Metals}\label{sec.lya}

\tabletypesize{\scriptsize}
 
\begin{deluxetable}{clcll}
\tablewidth{0pc}
\tablecaption{Ly$\alpha$ ABSORBERS SUMMARY \label{tab.lyasumm}}
\tabletypesize{\scriptsize}
\tablehead{\colhead{$\lambda_{obs}$} & \colhead{$z_{abs}$} &
\colhead{$W_{r}$} & 
\colhead{$\log\mathrm{N}_{\rm HI}$\tablenotemark{a}} & \colhead{$b$\tablenotemark{a}} \\
(\AA) & & (m\AA) & & (km s$^{-1}$) }
\startdata
\cutinhead{$\log \rm N_{\rm HI}\ge14.0$\tablenotemark{b}}
1272.299& 0.04658& $ 228\pm 12$ &$14.00^{+0.13}_{-0.16}$ & $23^{+ 9.1}_{- 3.5}$ \\
1335.915& 0.09891& $ 362\pm  7$ &$14.22^{+0.04}_{-0.05}$\tablenotemark{c} & $35^{+ 2.6}_{- 2.0}$ \\
1421.538& 0.16935& $ 212\pm 13$ &$14.04^{+0.18}_{-0.28}$\tablenotemark{c} & $20^{+17.0}_{- 3.2}$ \\
1449.602& 0.19243& $ 222\pm  8$ &$14.02^{+0.17}_{-0.27}$\tablenotemark{c} & $21^{+22.2}_{- 3.3}$ \\
1488.674& 0.22457& $ 298\pm 13$ &$13.95^{+0.09}_{-0.11}$ & $42^{+32.0}_{- 9.0}$ \\
1518.071& 0.24875& $ 263\pm 14$ &$14.10^{+0.06}_{-0.06}$ & $25^{+ 3.3}_{- 2.7}$ \\
1522.250& 0.25219& $ 447\pm 12$ &$14.74^{+0.06}_{-0.06}$ & $32^{+ 1.5}_{- 1.5}$ \\
\cutinhead{$\log \rm N_{\rm HI}<14.0$\tablenotemark{b}}
1269.721& 0.04446& $ 149\pm  9$ &$13.56\pm0.03$& \nodata \\
1287.019& 0.05869& $  86\pm  9$ &$13.35\pm0.05$& \nodata \\
1288.219& 0.05968& $  71\pm  7$ &$13.21\pm0.06$& \nodata \\
1365.581& 0.12332& $ 135\pm 13$ &$13.51\pm0.04$& \nodata \\
1368.425& 0.12565& $  66\pm 10$ &$13.17\pm0.07$& \nodata \\
1408.170& 0.15835& $  71\pm 11$ &$13.18\pm0.07$& \nodata \\
1450.246& 0.19296& $ 220\pm  8$ &$13.82^{+0.12}_{-0.15}$ & $31^{+49.1}_{- 7.6}$ \\
1503.392& 0.23668& $  59\pm 11$ &$13.15\pm0.09$& \nodata \\
1507.475& 0.24004& $  81\pm  9$ &$13.28\pm0.06$& \nodata \\
1520.922& 0.25110& $ 201\pm 16$ &$13.67\pm0.04$& \nodata \\
1523.069& 0.25286& $ 280\pm 11$ &$13.89^{+0.10}_{-0.12}$ & $45^{+78.8}_{-12.4}$ \\
1524.594& 0.25412& $ 227\pm 16$ &$13.84\pm0.03$& \nodata \\
\enddata
\tablenotetext{a}{Where $b$ not given, $\log \mathrm{N}_{\rm HI}$ from the AODM and is typically a lower limit. Otherwise, $\log \mathrm{N}_{\rm HI}$ and $b$ calculated from COG analysis where at least one other Lyman line also detected.}
\tablenotetext{b}{Strong Ly$\alpha$ features have $\log \rm N_{\rm HI}
  \ge14.0$ from either the AODM or the COG analysis.}
\tablenotetext{c}{COG analysis notes: for $z_{abs}=0.09891$ Ly$\alpha$
  blended with G \ion{C}{2}* 1335; $z_{abs}=0.16935$ Ly$\beta$ blended with G
  \ion{N}{1} 1199; for $z_{abs}=0.19243$ Ly$\beta < 3\sigma$,
  Ly$\alpha$, Ly$\gamma$ COG} 
\end{deluxetable}


In addition to the nine metal-line systems described above, we
identified 15 \Lya\ features detected at $>\!3\sigma$ significance
(see Table \ref{tab.lyasumm}). There are seven \Lya\ lines with
\logHI\ $\ge14$ that we identify as strong.  Given the absence of
metal-line absorption, the identification of these lines as \Lya\
should be considered less secure. However all of the strong \Lya\
lines show corresponding \Lyb\ absorption at $>\!3\sigma$
significance, which lends credibility to our identification.  In Table
\ref{tab.lyasumm}, we quote either the AODM column density or \logHI\
and \Dopb\ from the COG analysis when there is at least one other
Lyman line detected.  The \logHI\ values from the AODM are lower
limits when \Lya\ is saturated.

The \Lya\ absorbers at $\zabs=0.19243$ and 0.19296 are within
$\dva<350\kms$ of the metal-line system at $\zabs=0.19161$. \Lya\ at
$\zabs=0.22457$ is within $\dva<500\kms$ of the tentative \ion{O}{6}
absorbers at $\zabs=0.22555$ and 0.22752.  The implications of these
close systems will be discussed in the following section. 

The \Lya\ absorber at $\zabs=0.25219$ has larger \logHI\ than the the
tentative \ion{O}{6} systems at $\zabs=0.06471$, 0.22555, and 0.22752. As
mentioned previously, metal-line absorption roughly scales with
\logHI; the $\zabs=0.25219$ system should show some metal-line
absorption since the other three systems do. These systems appear
to be at the edge of our ability to detect metal-line absorbers.

\section{Comparisons with Previous Analyses}\label{sec.comp}

\begin{figure}
\includegraphics[clip,trim=0in 0in 0in 0in,width=0.36\textwidth,
  angle=90]{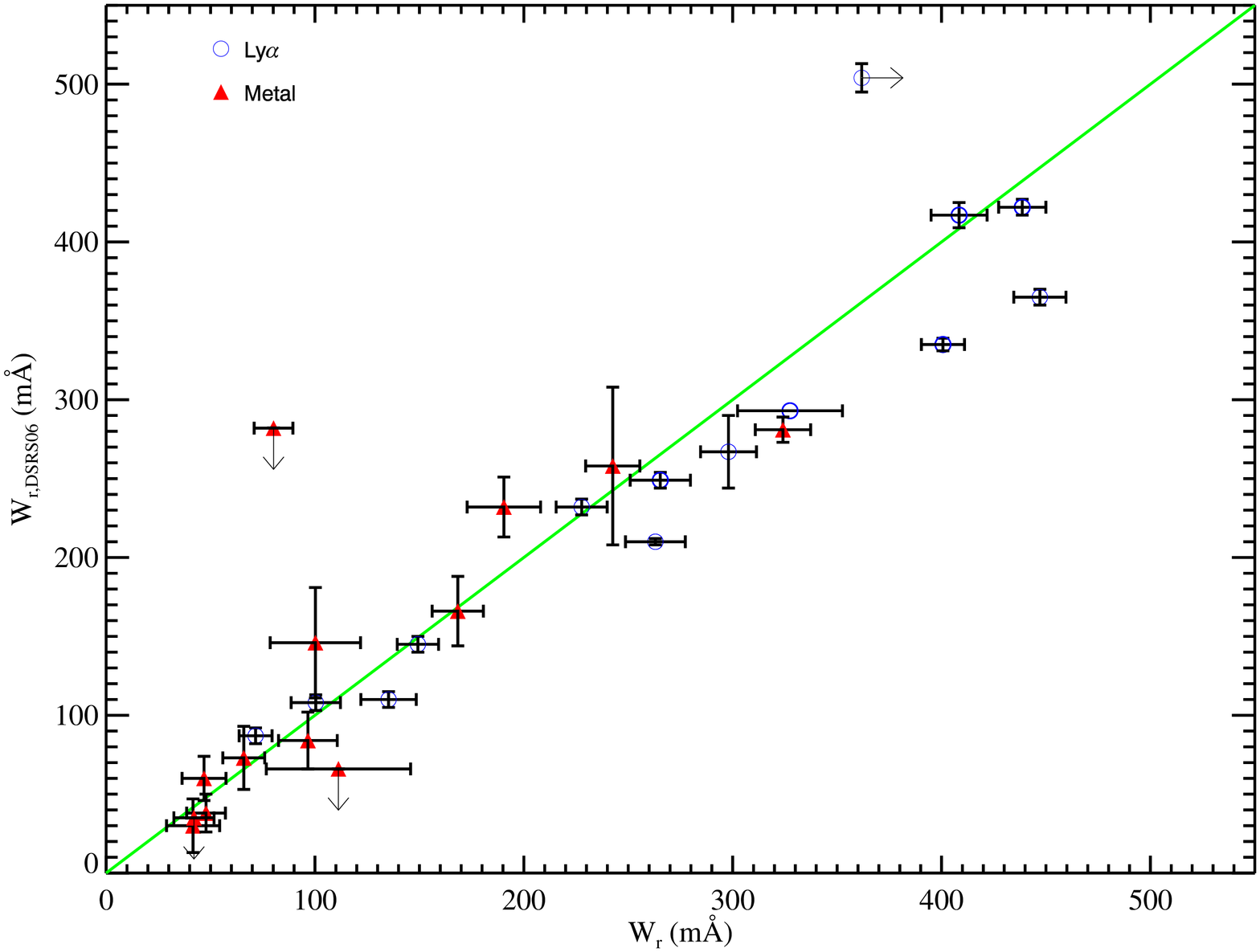}
\caption
{Comparison of rest equivalent widths from our analysis with
  \citet{danforthetal06}. The DSRS06 values are plotted over the
  values measured in the current paper, and the one-to-one relation is
  shown (solid green line). The (blue) circles are for \Lya\ \EWr, and
  the (red) triangles are for metal lines (\ie \ion{C}{3} and
  \ion{O}{6}).  DSRS06 use COG concordance plots, Voigt profile fits,
  and/or the AODM to measure $\log \rm N_{DSRS06}$ and $b_{DSRS06}$,
  typically assuming a single component. From $\log \rm N_{DSRS06}$
  and $b_{DSRS06}$, they measure $W_{r,DSRS06}$. They tend to
  underestimate $W_{r,DSRS06}$ compared to our values, which are a
  simple sum of the absorbed flux and include unresolved components.
\label{dsrs06comp_EW}
}
\end{figure}

\begin{figure}
\epsscale{1.1}
\plotone{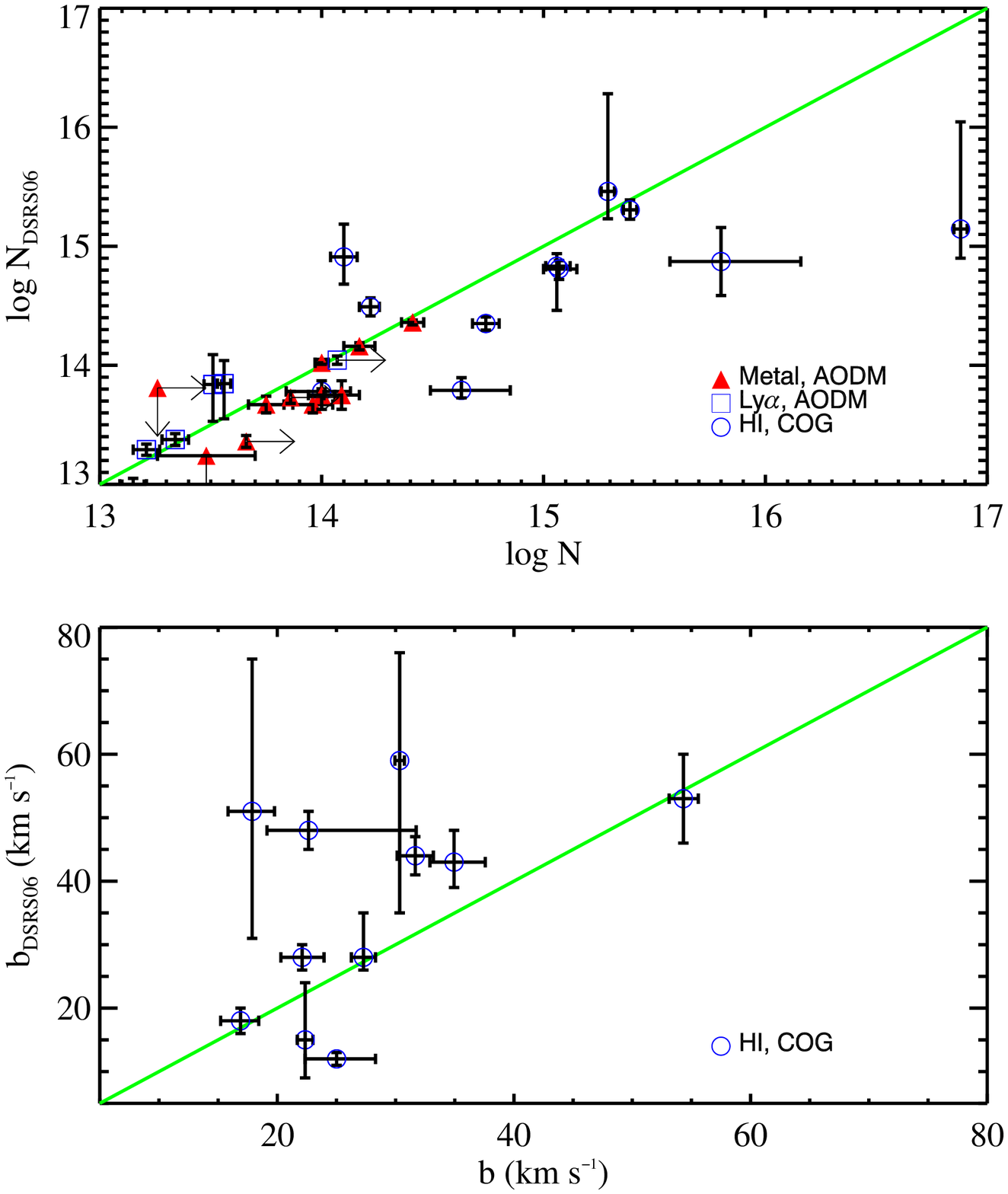}
\caption{ Comparison of column densities and Doppler parameters
  reported by DSRS06 against those from our analysis. The categories
  denoted by symbol (and color) refer to the current work. The
  \ion{H}{1} COG analysis $\log \rm N$ and \Dopb\ are (blue) circles,
  the AODM $\log \rm N$ are (red) triangles for metal lines (\ie
  \ion{C}{3} and \ion{O}{6},) and (blue) squares for \Lya. We
  generally agree with $\log \rm N_{DSRS06}$ because the sum of
  potential components does not greatly affect the total column
  density (top panel). On the other hand, unresolved components tend
  to increase $b_{DSRS06}$ compared our values, as discussed in DSRS06
  (bottom panel).
\label{dsrs06comp_Nb}
}
\end{figure}

The STIS dataset (PI: M. Lemoine) was acquired to measure an
intergalactic D/H value from the $z=0.095$ partial Lyman limit system.
Unexpected line-blending has apparently precluded such an analysis
(Lemoine, priv. comm.) and the data was not studied for this purpose.
\pks\ was included, however, in the compilation of
\citet{danforthetal06} who studied \Lya, \Lyb, \ion{O}{6}, and
\ion{C}{3} lines along 31 AGN sightlines at $z<0.3$.  We have compared
our results against DSRS06 to search for systematic effects related to
different procedures of data reduction and analysis.  In particular,
we have derived equivalent widths differently than DSRS06; our
analysis adopts values from a simple boxcar summation whereas DSRS06
implemented line-profile fits using the VPFIT software package.

Figure~\ref{dsrs06comp_EW} presents a comparison of the rest
equivalent width (\EWr) measurements of DSRS06 against our values for
\Lya\ and metal-line transitions.  We find that the two sets of
measurements are in good agreement for \EWr\ values of metal-line
transitions.  Similarly, there is relatively good agreement between
the two studies for \Lya\ lines at low rest equivalent widths (\EWr\
$< 300\mA$).  The only notable difference is that the DSRS06 rest
equivalent width errors $\sigma(\mbox{\EWr})$ for the \Lya\ lines are
systematically lower than our values; DSRS06 report
$\sigma(\mbox{\EWr}) \le 5\mA$ for the majority of their lines.  While
line-profile fitting techniques can recover more precise measurements
of the equivalent width than a boxcar summation, we contend that a
$5\mA$ error cannot be achieved from this dataset (S/N~$\approx 3$ to 6 per
pixel).  Even for strong lines where one might be justified in
assuming the core has zero flux with zero uncertainty, the wings of
the line-profile have equivalent width errors of greater than $10\mA$.
We can only speculate on the implications of adopting very small
errors on \EWr\ for \Lya\ transitions.  DSRS06 performed concordance
COG analysis of \nhi\ and \Dopb\ values in a similar manner as the
analysis presented here.  Because the \Lya\ line has the largest \EWr\
value in the Lyman series, adopting a very small uncertainty will
drive the COG analysis to best model the \Lya\ transition.  In
particular, this will imply $b_{\rm COG} = b_{\rm Ly\alpha}$ which
DSRS06 emphasize generally overestimates the true Doppler parameter of
the `cloud' dominating the optical depth.  We will return to this
point below.

There are more serious discrepancies between our results and DSRS06
for stronger \Lya\ lines (\EWr\ $> 300\mA$).  First, we identify five
\Lya\ lines with \logHI\ $>13.8$ that DSRS06 have not, at
$\zabs=0.16935$, 0.19243, 0.19296, 0.25286, and 0.25412.  The
$\zabs=0.16935$ \Lya\ is a multi-component system with \Lyb\ lost in
Galactic \ion{N}{1}, and $\zabs=0.19243$ \Lya\ has a similar
problem. The $\zabs=0.19296$ and 0.25286 \Lya\ have \Lyb\ detected at
$>\!3\sigma$. These features
are not Galactic lines nor misidentified metal lines from other
intervening systems. Although DSRS06 detect other strong \Lya\ lines
without \Lyb\ absorption, these lines were not reported in their
survey.  Second, we have derived systematically larger \EWr\ values
for strong \Lya\ lines. Most notable are the five \Lya\ lines in
Figure \ref{dsrs06comp_EW} that deviate by more than $3\sigma$ from the
\EWr\ values reported by DSRS06. The majority of the discrepancy is
probably due to these features being multi-component; we quote the
total \EWr\ of the feature whereas DSRS06 generally only report the
strongest single component.  We find similar differences when
comparing the DSRS06 results for PKS0405--123 against the results
reported in \cite{prochaskaetal04}.

We also compare the column densities and Doppler parameter values for
absorption lines analyzed by DSRS06 (see Figure~\ref{dsrs06comp_Nb}).
The metal-line column densities are considered first.  In contrast to
the equivalent width measurements for these transitions, we find that
our values are systematically larger than those reported by DSRS06.
Most worrisome is that we report several lower limits to the column
density of \ion{C}{3} because the line is clearly saturated in the
FUSE observations whereas DSRS06 report not a single lower limit.  For
example, we report \logCIII\ $> 13.9$ for the \ion{C}{3} transition in
the $z=0.09487$ absorber whereas DSRS06 report \logCIII\ $= 13.73 \pm
0.05$. The differences in \ion{O}{6} column densities are $<\!0.3$
dex, and probably due to continuum placement. The broad, shallow
\ion{O}{6} detection at $\zabs=0.09487$ differs by 0.3 dex whereas the
stronger $\zabs=0.04222$ feature differs only by 0.1 dex.

Regarding the \ion{H}{1} column densities, we note trends similar to
those for the \EWr\ values: at low column densities there is good
agreement between the two analyses, but at larger \nhi\ our values are
systematically larger.  The difference is most acute for the two
systems at \logHI\ $> 15.5$: $\zabs=0.00438$ and 0.09487. As mentioned
in \S\ \ref{subsec.z00441}, DSRS06 use a profile fit to
$\zabs=0.00438$ \Lya, which falls in the Galactic damped \Lya\
profile, to determine \nhi. Our COG analysis includes \Lyb, which is
the only feature of the system in a good region of the spectra.  The
difference for $\zabs=0.09487$ is due to DSRS06 only including \Lya\
through \ion{H}{1} 926 in their concordance COG analysis. For this
system, the higher-order \ion{H}{1} Lyman lines are most important for
measuring the \nhi\ value.

Finally, we have compared the Doppler parameter values from the two
analyses (Figure~\ref{dsrs06comp_Nb}; lower panel).  At low \Dopb\
values, we find reasonable agreement, but at moderate values our
results are systematically lower than the values reported by DSRS06.
We suspect the discrepancy is related to the very small errors adopted
for their \Lya\ equivalent widths (Figure~\ref{dsrs06comp_EW}).  In
this case, a COG analysis will yield a Doppler parameter which better
describes \Lya\ and, as DSRS06 emphasize, $b_{\rm Ly\alpha}$ is
systematically larger than $b_{\rm COG}$.  Because DSRS06 generally
adopt equivalent values from the literature
\citep[e.g.][]{pentonstockeandshull04}, it is possible that this
systematic effect is only present in the few sightlines analyzed by
DSRS06 (\eg \pks, PKS0405--123).  We also note that the larger \Dopb\
values likely lead to a systematic underestimate of \nhi\ which
explains at least part of the offset of their values from our results
for \logHI\ $> 14$.

\citet{trippetal07ph} searched for \ion{H}{1} and \ion{O}{6}
absorption in archival \stis\ spectra of 16 low-redshift quasars. They
fit Voigt profiles and applied the AODM to measure equivalent widths and
column densities of absorption lines, including individual
components. For \pks, they report three \ion{O}{6} systems at
$z_{abs,T07}=0.19159$, 0.22563, and 0.22744 that 
correspond to the systems at $\zabs=0.19161$, 0.22555, and 0.22752,
respectively, from \S\ \ref{sec.mls}. We briefly summarize the
\citet{trippetal07ph} results for the \pks\ sightline. The \ion{H}{1}
system at $z_{abs,T07}=0.19159$ has a strong, narrow component
coincident with a tentative weak, shallow component. \ion{C}{3},
\ion{Si}{3}, and only \ion{O}{6} 1031 are detected at $>\!3\sigma$ and
well-aligned with the \ion{H}{1} lines. The $z_{abs,T07}=0.22563$
\ion{H}{1} absorption is single component but is offset
($\dva=-18\kms$) from the \ion{O}{6} doublet, which is detected at
$>\!3\sigma$ in both lines. Weak \Lya\ (\logHI\ $=13$) and the
\ion{O}{6} doublet, also detected at $>\!3\sigma$ in both lines, are
well-aligned at $z_{abs,T07}=0.22744$.

We differ from \citet{trippetal07ph} most with respect to the measured
integrated equivalent widths. The discrepancy is strongest for the
\ion{O}{6} doublet measurements, but the difference are typically less
than $2\sigma$. The quoted Doppler parameters (for \ion{H}{1}) and column
densities (\ion{H}{1}, \ion{O}{6} doublet) are in excellent
agreement. 

In conclusion, we qualitatively agree with one main result of the
\citet{danforthetal06} and \citet{trippetal07ph} surveys: typically
\ion{O}{6} absorption is found in multi-phase systems.  Except for one
line, we do not disagree with the identification of lines from the two
surveys. The exception is for a line at $\lambda_{obs}\approx1321\Ang$
that DSRS06 identify as \Lya\ at $\zabs=0.0865$ and we list as
\ion{Si}{3} 1206 at $\zabs=0.09487$. Discrepancies in measured
quantities are due to differences in reduction (\eg spectra
extraction, continuum fitting) and analysis (\eg measuring \EWr,
accommodating blending), which greatly affect error estimates.

\section{Galaxy Survey}\label{sec.gal}

\begin{figure} 
\includegraphics*[clip,trim=0.5in 0in 0in 1in,width=0.37\textwidth,
  angle=90]{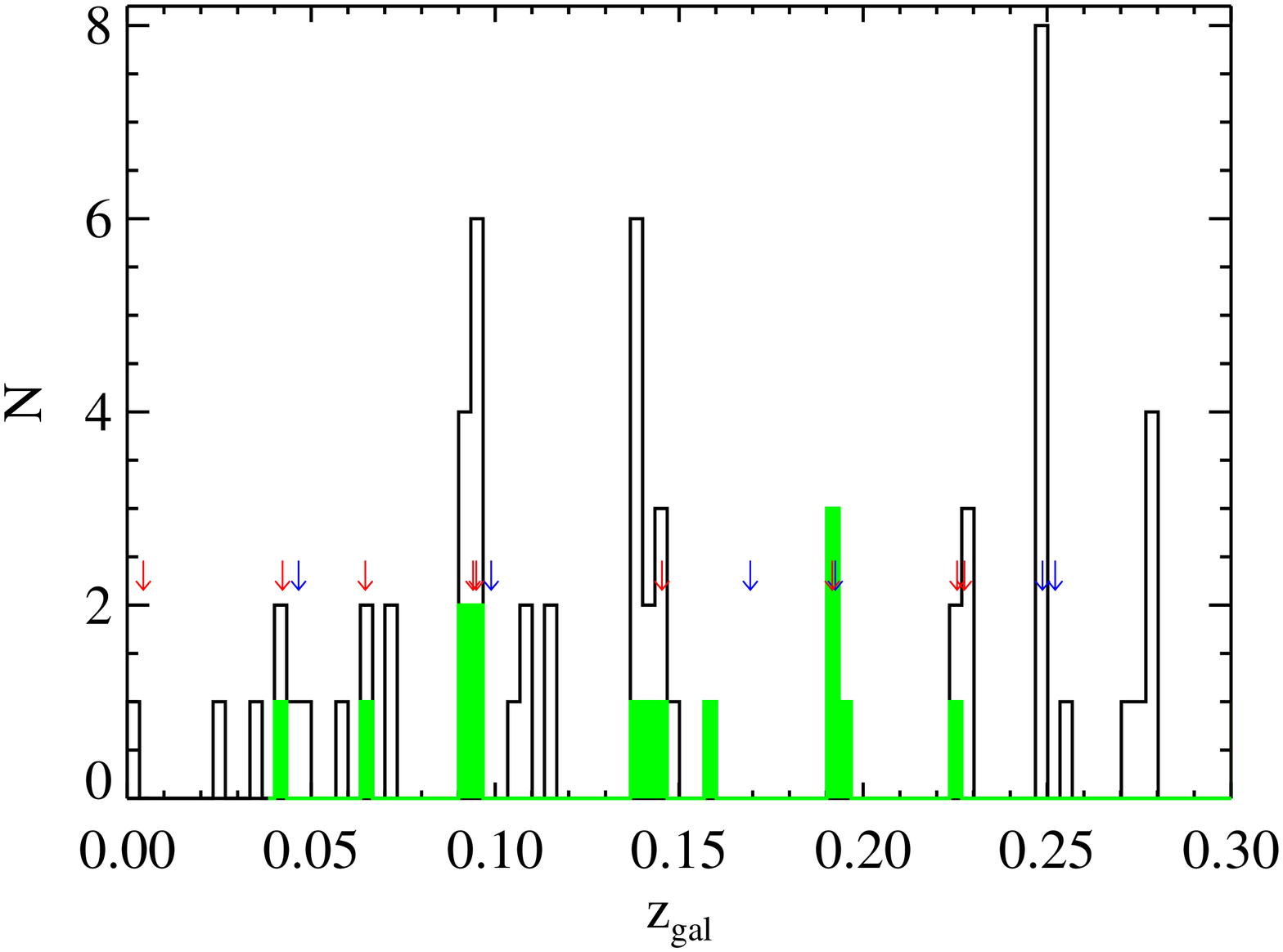}
\caption
{Histogram of 64 galaxies with $z_{gal} \le 0.3$ in \pks\ field,
  binned to $1000\kms$ (open).  The solid (green) histogram is
  galaxies within $5'$, 95\% complete to $R\approx19.5$.  The arrows
  indicate the redshifts of the nine metal-line systems (red) and the
  seven strong \Lya\ absorbers (blue; see Table
  \ref{tab.lyasumm}). There are groups of absorbers with
  $\vert\dva\vert<500\kms$ and with nearby galaxies that may be large-scale
  filaments at $\zabs\approx0.094$, 0.192, and 0.225.
\label{gal.hist}
}
\end{figure}

\begin{figure*} 
\epsscale{1.}\plotone{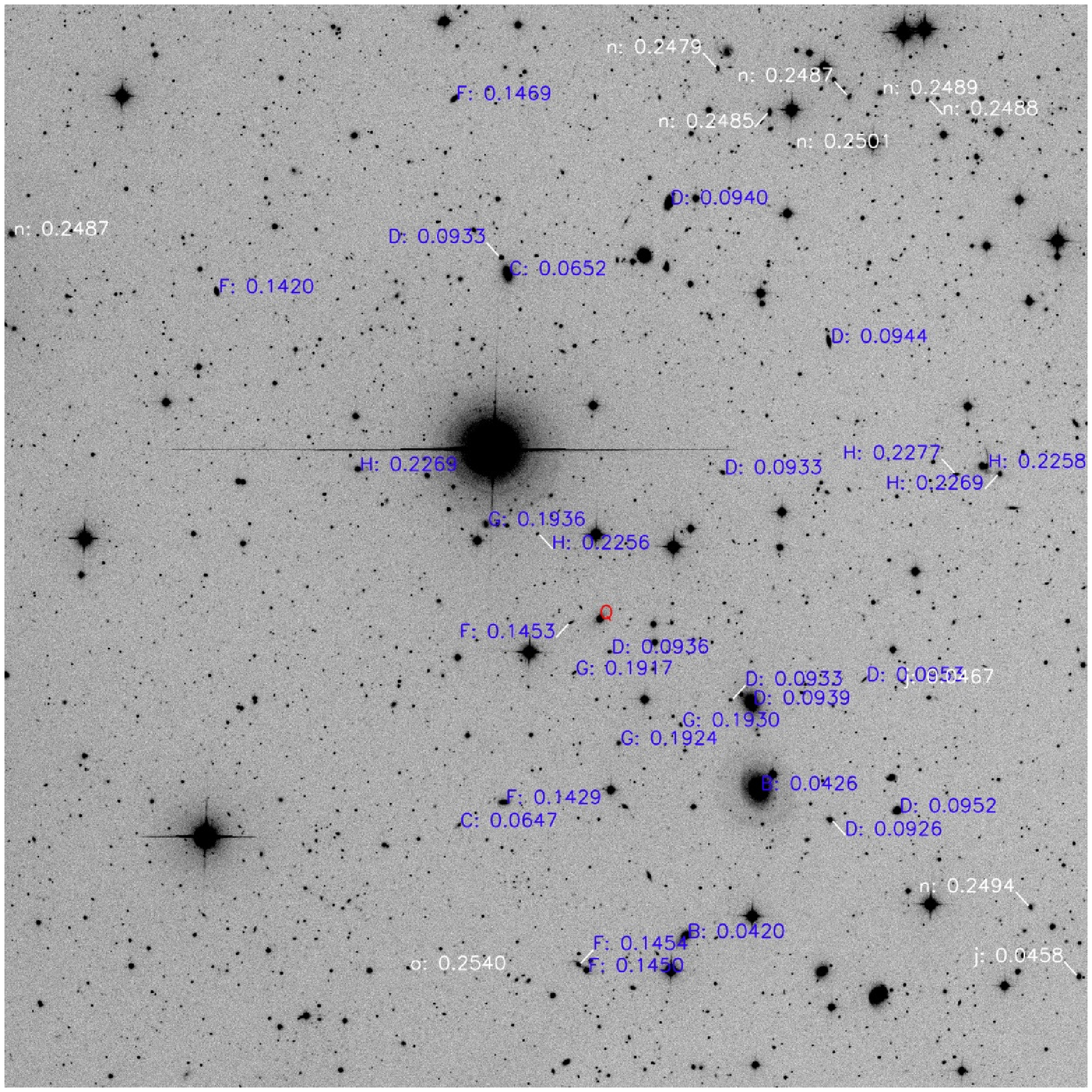}
\caption
{Galaxies with $\vert\dvg\vert \le1000\kms$ from metal-line systems
  and strong \Lya\ absorbers. The former have (blue) capital letters
  indicating the system they neighbor, while the latter have (white)
  lower case letters. The systems at $\zabs=0.09400$ and $0.09487$
  have most all neighboring galaxies in common; they are not labeled
  twice. Similarly for the systems at $\zabs=0.22555$ and $0.22752$.
  The \ion{O}{6} systems are at $\zabs=0.04222$, $0.06471$, 
  $0.09487$, and $0.22555$ (B, C, D, and H, respectively).  Aside from
  $\zabs=0.14533$ (F), the metal-line systems are more likely probing
  the intra-group medium. (North is up, and east is left. \pks\ is
  indicated by a (red) capital Q. There are about a dozen galaxies in
  the southeast corner that are either at higher redshift than \pks\
  or not within $1000\kms$ of an intervening system.) The image is
  about $20'$ on a side.
\label{gal.impact}
}
\end{figure*}

\tabletypesize{\scriptsize}
\begin{deluxetable*}{rrrccccc}
\tablewidth{0pc}
\tablecaption{OBJECT SUMMARY \label{tab:obj}}
\tabletypesize{\scriptsize}
\tablehead{\colhead{ID} & \colhead{RA} &\colhead{DEC} & \colhead{$R$} & \colhead{S/G$^a$} 
& \colhead{Area}& \colhead{flg$^b$} & \colhead{$z$} \\
 & & & (mag) &  & ($\square''$) }
\startdata
4&13:04:56.0&--10:29:18&$18.11 \pm  0.02$&0.09&   6.8& 7&$0.27252$\\
5&13:04:54.6&--10:39:58&$17.89 \pm  0.01$&0.13&   4.6& 7&$0.04576$\\
107&13:06:21.4&--10:30:37&$17.29 \pm  0.01$&0.25&   7.3& 7&$0.10819$\\
152&13:06:20.7&--10:27:53&$17.98 \pm  0.01$&0.18&   4.3& 7&$0.11608$\\
171&13:06:19.5&--10:34:42&$17.34 \pm  0.01$&0.22&   4.2& 7&$0.13839$\\
179&13:06:20.0&--10:26:11&$17.96 \pm  0.02$&0.01&   6.1& 7&$0.24866$\\
234&13:06:17.0&--10:37:51&$18.94 \pm  0.03$&0.89&   4.2& 7&$0.35805$\\
306&13:06:13.4&--10:44:51&$18.57 \pm  0.02$&0.98&   4.5& 7&$0.00000$\\
329&13:06:15.4&--10:27:22&$18.70 \pm  0.02$&0.13&   4.9& 7&$0.36601$\\
768&13:06:03.8&--10:27:12&$17.58 \pm  0.01$&0.12&   6.0& 7&$0.14202$\\
\enddata
\tablenotetext{a}{Star/galaxy classifier calculated by SExtractor.
Values near unity indicate a stellar-like point-spread function.}
\tablenotetext{b}{This binary flag has the following code:
1: Photometry; 2: Spectrum taken; 4= Redshift determined.}
\tablecomments{[The complete version of this table is in the electronic edition of the Journal.  The printed edition contains only a sample.]}
\end{deluxetable*}

A number of studies have examined the relationship between galaxies
and absorption-line systems at $z \lesssim 0.1$.  Regarding metal-line
systems, the majority of recent analyses can be characterized as a
detailed study of a single or few absorbers
\citep[e.g.][]{skm+04,jbt05,trippetal06}, an analysis of a complete
sightline and its surrounding galaxies \citep{sts+04,prochaskaetal06},
or a survey comprising multiple fields and absorbers
\citep{stockeetal06}.  These studies have examined metal-line systems
associated with a diverse set of ions (Si$^+$, C$^+$, O$^0$,
O$^{+5}$), metallicities, and \ion{H}{1} column densities.
Furthermore, the galaxy surveys have a wide range of magnitude limits
and field-of-view areas.  Not surprisingly, a range of conclusions
have been drawn regarding the association of galaxies and absorbers
including: (i) a physical association of the gas with individual
galactic halos \citep{chenandprochaska00}, (ii) outflows from dwarf
galaxies \citep{skm+04}, and large-scale (e.g.\ filamentary)
structures \citep{stockeetal06,trippetal06}.  Indeed, all of these may
contribute to metal-line systems, presumably with a dependence on the
metallicity, ionization state, and column density of the gas.  Other
analyses begin with a well-defined galaxy survey and search for
absorption associated with galaxies at small impact parameters to the
sightline \citep{lbt+95,chenlanzettaandwebb01,chenetal01}.  These
authors conclude that the presence of a galaxy within $\approx
200\kpc$ of a quasar sightline results in a high probability of
showing coincident \Lya\ and \ion{C}{4} absorption.

Our analysis of \pks\ has identified nine metal-line systems showing
a diverse set of characteristics, and we might expect, therefore, them
to arise in a range of galactic environments.  We have obtained
spectra of objects in the field surrounding \pks\ using the WFCCD
camera on the 100$''$ Dupont telescope at Las Campanas Observatory
during UT 2001-04-16 to 2001-04-20 (see Table \ref{tab:obj}). We refer
the reader to \citet{prochaskaetal06} for details of the imaging and
spectral data reduction and analysis procedures. The survey of the
\pks\ field is 95\% complete within $5'$ and 70\% within $10'$ to the
limiting magnitude $R\approx19.5$.  We have redshift information for
82 galaxies in the \pks\ field, 64 of which are at $z_{gal}< \zqso$.
At the highest redshifts $\zabs\approx0.25$, the survey covers a
physical radius of $\approx3\,\mathrm{Mpc}$ but not to faint intrinsic
magnitudes ($L \approx L_*$).\footnote{In this paper, we assume a
Hubble constant $H_{0}=75\,h_{75}\kms\,\mathrm{Mpc}^{-1}$ and the
absolute magnitude for $L_*$ at $z=0$ is $M_{R}=-21.04$
\citep{blantonetal03}. This value is one magnitude fainter than the
value used in \citet{prochaskaetal06}.}  For the lowest redshift
absorbers ($z<0.02$), we do not have the coverage to comment on
large-scale structures (\ImPar\ $\approx1$--$3\hinv$Mpc) as in
\eg\citet{pentonetal02} and \citet{prochaskaetal06}.  For example, the
field-of-view covers only \ImPar\ $\lesssim 25\kpc$ around the
\ion{C}{3} system $\zabs=0.00438$, an absorber that is likely
affiliated with the Virgo cluster.

\begin{figure} 
\includegraphics[clip,width=0.36\textwidth,angle=90]{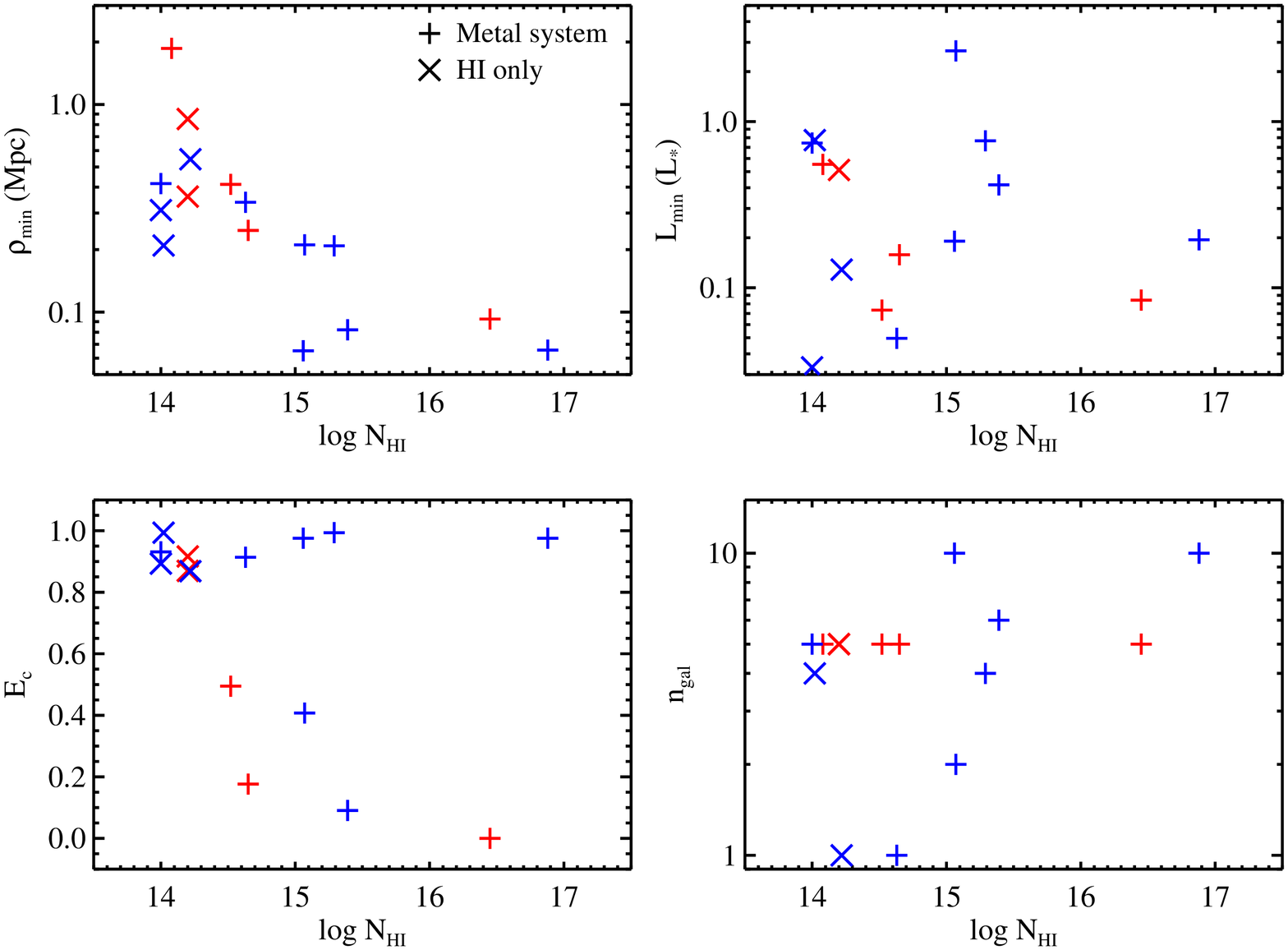}
\caption{Impact parameter ($\rho_{min}$), luminosity $(L_{min})$, and spectral
coefficient ($E_C$) of the closest galaxy with $L>0.1\,L_*$, \ImPar\ $<
5$\,Mpc, and $\vert\dvg\vert < 1000 \kms$ for the absorbers with
$\mnhi > 10^{14} \cm{-2}$ at $z_{abs} < 0.2$ along the sightlines to
PKS0405--123 (red) and \pks\ (blue).  The point types distinguish
between metal-line systems (plus signs) and absorbers that only show
\Lya\ absorption (crosses).  There appears to be a trend toward
lower $\rho_{min}$ values for higher \nhi\ value of the
absorbers. (Note: in the current work, the absolute magnitude for
$L_{\ast}$ at $z=0$ is one magnitude fainter than that used in
\citet{prochaskaetal06}.)
\label{gal.param}
}
\end{figure}

Table~\ref{tab.galsumm} lists\footnote{See also
http://www.ucolick.org/$\sim$xavier/WFCCDOVI/} the galaxies are
associated with the IGM systems by $\dvg\equiv
c(z_{abs}-z_{gal})/(1+z_{abs}) \leq 1000\kms$. The velocity constraint
comfortably covers the peculiar velocities expected for large-scale
structures.  In Figure~\ref{gal.hist}, we show a histogram of the
galaxy redshifts for the field surrounding \pks\ and, in
Figure~\ref{gal.param}, the impact distribution of galaxies with
$|\dvg| < 1000 \kms$ from a metal-line system.  Although an exact
comparison of galaxy-absorber correlations cannot be performed between
systems because the survey varies in field-of-view and depth with
redshift, it is evident that the metal-line systems arise in a diverse
set of galactic environments.  For example, the partial LLS at
$z=0.0949$ is associated with a group of galaxies and quite likely is
found within the halo of a $L \approx 0.2\,L_*$ galaxy at \ImPar\
$\approx 65 \hinv\kpc$.  In contrast, the \ion{O}{6} absorber at
$z=0.0646$ is at least 300~$\hinv\kpc$ away from any galaxy with $L >
0.01\,L_*$ and one identifies no obvious large scale-structure at this
redshift.  Let us now turn to discuss a few of these absorbers in
greater detail.

There are three groups of metal-line absorbers with $\vert\dva\vert <
500\kms$ at 
$\zabs\approx 0.094$, 0.192, and 0.225. These groups may represent
filamentary structures where the \ion{H}{1} and metal-line
absorption arise in
the gas between the galaxies populating this large-scale structure
\citep{bowenpettiniandblades02}.  The $\zabs\approx0.094$ group has
ten detected galaxies with $65<$ \ImPar\ $<800\hinv\kpc$ and $0.1 <
L/L_* < 6$.  The brightest galaxy is at $\dvg=-259\kms$ from the
partial Lyman-limit system $\zabs=0.09487$, with \ImPar\
$=331\hinv\kpc$. Both absorption systems near $\zabs\approx0.094$ have
\ion{C}{3} absorption, and the partial LLS has \ion{Si}{3} and a broad
\ion{O}{6} doublet.  \citet{chenlanzettaandwebb01} have found that
\ion{C}{4} absorption is strongly correlated with galaxies with
$\dvg<250\kms$ and \ImPar\ $<100\hinv\kpc$.  We have searched for such
absorption associated with the galaxy at $z=0.09358$ with \ImPar\ $=65
\hinv\kpc$ but unfortunately this places the doublet in the
high-wavelength, low-sensitivity end of the \stis\ E140M spectrum. We
can place an upper limit on the absorption: $\log \rm N(\rm
C^{+3})<13.5$.  Given the small impact parameter of this galaxy to the
\pks\ sightline, we tentatively associate this galaxy with the partial
LLS at $z=0.09487$.  This association is challenged by the observed
velocity offset $\dvg = -354\kms$; an association would require the
gas to have a large inflow/outflow.

The group at $\zabs\approx 0.192$ has a strong \ion{H}{1} Lyman
absorber with \ion{C}{3}, \ion{O}{6} 1031, and marginal \ion{Si}{3}
absorption and two \Lya\ systems with \logHI\ $>13.8$.  There are four
detected galaxies in this group, with $200<$ \ImPar\ $\le520\hinv\kpc$
and $0.4< L/L_*<2.5$.  In the $\zabs\approx0.225$ group, there are
five bright ($L>0.7\,L_*$) galaxies detected in the field with all but
one at \ImPar\ $>1\hinv\,$Mpc. The $0.7L_{*}$ galaxy in this group
is \ImPar\ $=416\hinv\kpc$ from the tentative \ion{O}{6} systems at
$\zabs=0.22555$ ($\dvg=4\kms$) and 0.22752 ($\dvg=-477\kms$).

There are very few galaxies in the survey around the remaining
metal-line systems at $\zabs= 0.04222$, 0.06471, and 0.14533  
even though we probe similar impact parameters as the group at
$z=0.2$.
The \ion{O}{6} systems at $\zabs= 0.04222$ and 0.06471 each have two
galaxies with $\vert\dvg\vert<160\kms$ and $200<$ \ImPar\
$<500\hinv\kpc$. At these redshifts, the survey is 100\% within
$200\kpc$ ($2'$) to $L = 0.01\,L_{*}$ and $0.03\,L_{*}$, respectively
($R=20$\,mag). For the $\zabs= 0.06471$ system, the fainter galaxy ($L
\approx 0.05\,L_*$) is closer in $\dvg$ and impact parameter, while
the opposite is true for the other \ion{O}{6} absorber.  In addition,
the two galaxies around $\zabs=0.06471$ have larger \ImPar\ than the
galaxies near $\zabs=0.04222$. Since the $\zabs=0.06471$ \ion{O}{6}
system has lower \logHI\ and \logOVI, it may probe the extended halos
of the nearby galaxies. The $\zabs=0.04222$ system, with \ion{O}{6}
and \ion{C}{3}, probably probes a multi-phase medium closer to the
galaxies.  \citet{stockeetal06} report that \ion{O}{6} systems with no
\ion{C}{3} absorption have larger nearest-galaxy distances than
systems with both lines detected.

There are four galaxies surrounding the $\zabs=0.14533$ absorber with
$\dvg<500\kms$ and $0.4<L/L_*<2.5$, suggesting the gas arises in an
intra-group medium \citep{mulchaey96}.  However, one of these galaxies
has a very small velocity offset and impact parameter ($\dvg=-7\kms$
and \ImPar\ $=82\hinv\kpc$) and may host the broad \ion{H}{1} and
\ion{C}{3} absorption.  As for the other strong \Lya\ absorbers, there
are surrounding galaxies with $\dvg<1000\kms$, except for
$\zabs=0.16935$ and 0.22457. However, no one bright and close galaxy
appears as the source of the gas.  The $\zabs=0.16935$ 
\Lya\ absorber is obviously multi-component with no
galaxies with $\dvg<1000\kms$ and brighter than $R=19.5$\,mag. The
$\zabs=0.22457$ \Lya\ absorber is $\dvg<750\kms$ from the metal-line
systems at $\zabs\approx0.225$ and is probably associated.

It is illustrative to compare the galaxy-absorber connection by
examining the properties of the `nearest' galaxy to each absorber and
a characteristic of the large-scale structure.  Before proceeding,
however, we wish to caution that the nearest galaxy in this context
corresponds to the galaxy with smallest impact parameter that (i) has
$\vert\dvg\vert < 1000\kms$ and (ii) is brighter than the magnitude
limit.  In many cases, there may be no direct physical association
between the galaxy and the absorber.  Figure~\ref{gal.param} presents
the impact parameter $\rho_{min}$, luminosity $L_{min}$, and spectral
coefficient $E_C$ of the galaxy at closest impact parameter to all of
the absorbers with $\mnhi > 10^{14} \cm{-2}$ along the sightlines to
PKS0405--123 \citep{prochaskaetal04} and \pks\ (this paper).  In
addition, the lower-right panel shows the number of galaxies with
$L>0.1\,L_*$, \ImPar\ $< 5$\,Mpc, and $\vert\dvg\vert < 1000 \kms$ with
respect to the absorber.  In terms of the impact parameter, one notes
a qualitative trend of decreasing $\rho_{min}$ with increasing \nhi\
that suggests a physical association between individual galaxies and
absorbers for $\mnhi \gtrsim 10^{14.5} \cm{-2}$ \citep[see
also][]{cpw+05}.  At lower column densities ($\mnhi \lesssim 10^{14}
\cm{-2}$), there is no discernible trend (for \Lya-only or metal-line
systems) which suggests these absorbers are predominantly associated
with large-scale structures (\eg intra-group material, filamentary
structures).

If this qualitative picture is correct, one may comment on the
luminosities of the galaxies hosting absorbers.  Based on the systems
with $\rho_{min} < 100\hinv\kpc$, all of the galaxies are sub-$L_*$
although we note that the partial LLS at $z=0.16$ toward PKS0405--123
also shows an approximately 2$L_*$ galaxy at \ImPar\ $< 100 \hinv\kpc$
\citep{sfy+93}.  Of particular interest to examining the enrichment
history of the IGM is to study the luminosity function of galaxies
dominating such absorbers.  We will address these issues in greater
depth in a future paper summarizing our full set of galaxy surveys.
Lastly, we comment that the average $n_{gal}$ value may rise with
\nhi\ but that there is apparently significant scatter in this crude
measure of galactic environment.

\tabletypesize{\scriptsize}
 
 
\begin{deluxetable*}{rlllrrrr}
\tablewidth{0pc}
\tablecaption{SUMMARY OF GALAXIES NEIGHBORING ABSORPTION SYSTEMS\label{tab.galsumm}}
\tabletypesize{\scriptsize}
\tablehead{\colhead{ID} & \colhead{$z_{gal}$} & \colhead{$R$} & 
\colhead{$L$} & \colhead{$\delta v$} &\colhead{$\rho$} & 
\colhead{$E_{C}$} & \colhead{$L_{C}$} \\
 & & & ($L_{*}$) & (km s$^{-1}$) & 
($\mathrm{h}_{75}^{-1}\,\mathrm{kpc}$) & & }
\startdata
\cutinhead{$z_{abs}=0.04222$, $\log\mathrm{N}_{\rm HI}=15.1$, $\log \rm N (O^{+5})=14.5$}
 2447 & 0.04256 & $ 14.1$ &  2.71 & $   99$ &   212 &  0.41 &  0.57 \\ 
 2226 & 0.04196 & $ 16.3$ &  0.35 & $  -76$ &   294 & -0.17 &  0.42 \\ 
\cutinhead{$z_{abs}=0.04658$, $\log\mathrm{N}_{\rm HI}=14.0$}
 2886 & 0.04674 & $ 19.1$ &  0.03 & $   44$ &   311 &  0.89 & -0.05 \\ 
    5 & 0.04576 & $ 17.9$ &  0.09 & $ -235$ &   588 &  0.79 & -0.16 \\ 
\cutinhead{$z_{abs}=0.06471$, $\log\mathrm{N}_{\rm HI}=14.6$, $\log \rm N (O^{+5})=13.8$}
 1576 & 0.06468 & $ 19.4$ &  0.05 & $   -9$ &   338 &  0.91 &  0.10 \\ 
 1712 & 0.06523 & $ 15.4$ &  1.95 & $  146$ &   484 &  0.97 & -0.04 \\ 
\cutinhead{$z_{abs}=0.09487$, $\log\mathrm{N}_{\rm HI}=16.9$, $\log \rm N (O^{+5})=14.0$\tablenotemark{a}}
 2033 & 0.09358 & $ 18.8$ &  0.19 & $ -354\phn(-115)$ &    65 &  0.98 & -0.03 \\ 
 2415 & 0.09328 & $ 19.2$ &  0.12 & $ -435\phn(-196)$ &   292 &  0.99 & -0.04 \\ 
 2435 & 0.09393 & $ 15.2$ &  5.21 & $ -259\phn\phm{1}(-20)$ &   331 &  0.97 & -0.19 \\ 
 2391 & 0.09331 & $ 18.1$ &  0.33 & $ -427\phn(-189)$ &   360 &  0.61 &  0.44 \\ 
 2790 & 0.09531 & $ 19.3$ &  0.12 & $  120\phn\phm{-}(359)$ &   527 &  0.87 &  0.03 \\ 
 2685 & 0.09258 & $ 17.3$ &  0.73 & $ -626\phn(-388)$ &   575 &  0.81 & -0.03 \\ 
 2867 & 0.09523 & $ 16.9$ &  1.10 & $   98\phn\phm{-}(337)$ &   684 &  0.85 &  0.16 \\ 
 2682 & 0.09442 & $ 16.9$ &  1.10 & $ -125\phn\phm{-}(114)$ &   686 &  0.91 &  0.22 \\ 
 1711 & 0.09332 & $ 18.5$ &  0.23 & $ -424\phn(-185)$ &   702 &  0.13 &  0.93 \\ 
 2193 & 0.09402 & $ 16.2$ &  1.97 & $ -234\phn\phm{-11}(4)$ &   795 &  0.85 &  0.29 \\ 
\cutinhead{$z_{abs}=0.09891$, $\log\mathrm{N}_{\rm HI}=14.2$}
 2790 & 0.09531 & $ 19.3$ &  0.12 & $ -983$ &   527 &  0.87 &  0.03 \\ 
\cutinhead{$z_{abs}=0.14533$, $\log\mathrm{N}_{\rm HI}=15.4$, $\log \rm N (O^{+5})=14.2$}
 1909 & 0.14530 & $ 18.9$ &  0.42 & $   -7$ &    82 &  0.09 &  0.82 \\ 
 1708 & 0.14292 & $ 17.3$ &  1.87 & $ -630$ &   564 &  0.97 &  0.03 \\ 
 1920 & 0.14536 & $ 17.9$ &  1.13 & $    7$ &   953 & -0.08 &  0.66 \\ 
 1921 & 0.14500 & $ 17.1$ &  2.26 & $  -87$ &   966 &  0.42 &  0.57 \\ 
  768 & 0.14202 & $ 17.6$ &  1.37 & $ -867$ &  1374 &  0.86 &  0.31 \\ 
 1564 & 0.14685 & $ 17.2$ &  2.06 & $  399$ &  1501 &  0.78 &  0.36 \\ 
\cutinhead{$z_{abs}=0.19161$, $\log\mathrm{N}_{\rm HI}=15.3$, $\log \rm N (O^{+5})=13.9$\tablenotemark{b}}
 1926 & 0.19171 & $ 18.9$ &  0.77 & $   26\phn(-182)$ &   209 &  0.99 & -0.01 \\ 
 2051 & 0.19239 & $ 18.3$ &  1.32 & $  197\phn\phm{1}( -11)$ &   435 &  0.99 & -0.08 \\ 
 2232 & 0.19296 & $ 19.0$ &  0.73 & $  342\phn\phm{-}( 134)$ &   465 & -0.16 &  0.70 \\ 
 1664 & 0.19361 & $ 18.0$ &  1.91 & $  503\phn\phm{-}( 295)$ &   520 &  0.99 & -0.03 \\ 
\cutinhead{$z_{abs}=0.22555$, $\log\mathrm{N}_{\rm HI}=14.0$, $\log \rm N (O^{+5})=13.9$\tablenotemark{c}}
 1821 & 0.22557 & $ 19.4$ &  0.74 & $    4\phn(-477)$ &   416 &  0.93 &  0.25 \\ 
 1253 & 0.22690 & $ 17.7$ &  3.63 & $  331\phn(-151)$ &  1133 &  1.00 & -0.07 \\ 
 3074 & 0.22772 & $ 18.9$ &  1.15 & $  531\phn\phm{-1}(  50)$ &  1539 &  0.87 &  0.29 \\ 
 3152 & 0.22584 & $ 18.0$ &  2.51 & $   69\phn(-412)$ &  1650 &  0.92 &  0.23 \\ 
 3208 & 0.22687 & $ 17.8$ &  3.08 & $  323\phn(-159)$ &  1696 &  0.99 & -0.05 \\ 
\cutinhead{$z_{abs}=0.24875$, $\log\mathrm{N}_{\rm HI}=14.1$}
 2573 & 0.25012 & $ 19.0$ &  1.30 & $  328$ &  2162 &  0.99 & -0.11 \\ 
 3314 & 0.24942 & $ 17.8$ &  3.99 & $  160$ &  2213 &  0.68 &  0.31 \\ 
 2510 & 0.24852 & $ 18.0$ &  3.23 & $  -57$ &  2250 &  1.00 & -0.04 \\ 
 2378 & 0.24786 & $ 18.7$ &  1.65 & $ -213$ &  2364 &  0.67 &  0.44 \\ 
 2726 & 0.24873 & $ 18.3$ &  2.38 & $   -5$ &  2442 &  0.99 & -0.11 \\ 
 2832 & 0.24892 & $ 18.1$ &  2.96 & $   40$ &  2518 &  0.98 & -0.12 \\ 
 2973 & 0.24882 & $ 19.1$ &  1.19 & $   17$ &  2595 &  0.98 & -0.12 \\ 
  179 & 0.24866 & $ 18.0$ &  3.38 & $  -22$ &  2997 &  0.94 &  0.15 \\ 
\cutinhead{$z_{abs}=0.25219$, $\log\mathrm{N}_{\rm HI}=14.7$}
 1413 & 0.25398 & $ 19.0$ &  1.30 & $  430$ &  1710 &  0.72 &  0.45 \\ 
 2573 & 0.25012 & $ 19.0$ &  1.30 & $ -495$ &  2162 &  0.99 & -0.11 \\ 
 3314 & 0.24942 & $ 17.8$ &  3.99 & $ -663$ &  2213 &  0.68 &  0.31 \\ 
 2510 & 0.24852 & $ 18.0$ &  3.23 & $ -880$ &  2250 &  1.00 & -0.04 \\ 
 2726 & 0.24873 & $ 18.3$ &  2.38 & $ -828$ &  2442 &  0.99 & -0.11 \\ 
 2832 & 0.24892 & $ 18.1$ &  2.96 & $ -783$ &  2518 &  0.98 & -0.12 \\ 
 2973 & 0.24882 & $ 19.1$ &  1.19 & $ -806$ &  2595 &  0.98 & -0.12 \\ 
  179 & 0.24866 & $ 18.0$ &  3.38 & $ -844$ &  2997 &  0.94 &  0.15 \\ 
\enddata
\tablenotetext{a}{($z_{abs}=0.09400$,
 $\log\mathrm{N}_{\rm HI}=15.1$, $\log \rm N (O^{+5})<13.8$)}
\tablenotetext{b}{($z_{abs}=0.19243$, $\log\mathrm{N}_{\rm HI}=14.0$)}
\tablenotetext{c}{($z_{abs}=0.22752$,
 $\log\mathrm{N}_{\rm HI}=13.1$, $\log \rm N (O^{+5})=13.6$)} 
\tablecomments{The galaxy summary is restricted to those galaxies
 within $1000\,\mathrm{km\,s}^{-1}$ of the absorption system.  The
impact parameter refers to physical separation, not comoving. Galaxy
redshifts were determined from fitting the four SDSS star and galaxy
eigenfunctions to the spectra \citep[see][]{prochaskaetal06}. The coefficient
of the first eigenfunction $E_C$ and a composite of the last three
eigenfunctions $L_C$ are used to define galaxy type. Early-type
galaxies have $E_C > 0.8$ and $L_C < 0.4$, while late-type galaxies
have $E_C < 0.8$ and $L_C > 0.4$.}
\end{deluxetable*}


\section{Discussion}\label{sec.disc}

\tabletypesize{\scriptsize}
\begin{deluxetable*}{llllllllll}
\tablewidth{0pt} 
\tabletypesize{\scriptsize}
\tablecaption{METAL ABSORBERS SUMMARY\label{tab.abssum} } 
\tablehead{ 
\colhead{$z_{abs}$} &
\colhead{$\log \rm N_{\rm HI}$} & \colhead{$b_{\rm HI}$} &
\colhead{$\log \rm N(\rm C^{++})$} &  
\colhead{$\log \rm N(\rm O^{+5})$} & \colhead{Ion.} &
\colhead{$\log U$} & \colhead{[M/H]$_{\rm phot}$} & \colhead{$\log T_{\rm
    coll}$} & \colhead{[M/H]$_{\rm coll}$} 
}
\startdata

0.00438 & 15.8  & 17 & $>13.5$ & $<14.0$ & Photo & $-2.1$ & [-2, -0.9] & \nodata & \nodata     \\
0.04222 & 15.07 & 22 &  $13.7$ & $14.5$  & Multi & $-1.9$; $-1.1$ & $\approx-1.1$ & $>5.4$  & $-2$    \\
0.06471 & 14.6 & 18 & $<13.1$ & $13.8$  & ???   & $>\rm{[C/O]}-1.5$ & $\approx1$ & $>5.3$  & $-1.8$ \\
0.09400 & 15.06 & 27 & $>13.3$ & $<14.0$ & Photo & 1.5 & [-1.3, +0.3] & \nodata & \nodata     \\
0.09487 & 16.88 & 30 & $>13.9$ & $14.0$  & Multi & $-2.9$ & [-2, -1.6] & $>5.3$  & $-3.8$      \\
0.14533 & 15.39 & 54 &  $13.2$ & $<14.2$ & Singl & -1.9 & [-1.9,-0.4] & $<5.3$   & [-1.8,-0.8]     \\
0.19161 & 15.29 & 22 &  $13.1$ & $13.9$ & Photo & -1.7 & [-1.9, -1.1] & \nodata & \nodata  \\
0.22555 & 14.00 & 44 &  $<13.0$ & $13.9$ & Coll? & $>\rm{[C/O]}-1.4$ &
 $\lesssim-0.3$ & $>5.3$ & $\lesssim-0.3$ \\
0.22752 & 13.1 & \nodata & \nodata & $13.6$ & \nodata & \nodata & \nodata & \nodata & \nodata 
\enddata

\tablecomments{Ionization mechanism from relative abundances of measured species. In several cases, the mechanism is multi-phase or ambiguous. The bracketed values indicate the range of acceptable values. }

\end{deluxetable*}

We have presented the reduction and analysis of archival \hst/\stis\
and \fuse\ UV spectra of the low-redshift quasar \pks\ ($\zqso =
0.2784$). We have identified $90\%$ of the potential \Lya\ features in
\stis\ and $>\!85\%$ of the features in \fuse\ with $>\!4\sigma$
significance and FWHM $=20\kms$ and $40\kms$, respectively . We also
performed a blind search for doublets without \Lya\ absorption; there
were no such systems in the \pks\ spectra. There are 28 \Lya\ systems;
15 are strong absorbers with \logHI\ $>14$.  Of those strong systems,
eight are metal-line systems: four with \ion{C}{3} only, two with
\ion{C}{3} and \ion{O}{6} absorption, and two tentative
\ion{O}{6}-only systems (see Table \ref{tab.abssum}). There is also a
tentative \ion{O}{6} absorber with \logHI\ $=13.1$ at $\zabs=0.22752$.

The unblocked redshift path length \Dz\ for detecting \Lya,
\ion{C}{3}, or the \ion{O}{6} doublet was measured for regions where
the \EWr\ $\ge50\mA$ absorption line(s) could be detected to
$>\!3\sigma$ significance, excluding regions blocked by Galactic or
IGM lines and within $1500\kms$ of \pks\ ($\zqso = 0.2784$). We quote
the 68\% confidence limits assuming Poisson statistics. With 28 \Lya\
absorbers and \Dz\ $=0.236$, \dNLyadz\ $=118^{+14}_{-12}$, which is
consistent with other comparable published values, \dNLyadz\
$\gtrsim\!100$ for \EWr\ $\ge50\mA$
\citep[\eg][]{trippetal98,pentonshullandstocke00}. Similar to other
published values
\citep[\eg][]{richteretal04,prochaskaetal04,danforthandshull05}, we
derive \dNOVIdz\ $=7^{+9}_{-4}$ for the one doublet with both lines
detected at $3\sigma$ and with \EWr\ $>50\mA$ for \Dz\ $=0.152$.  On
the other hand, we measure \dNCIIIdz\ $=36^{+13}_{-9}$ from the five
detections with \EWr\ $>50\mA$ over \Dz\ $=0.138$. For their entire
sample, DSRS06 measure \dNCIIIdz\ $= 12^{+3}_{-2}$.  We agree with
DSRS06 on the number of \ion{O}{6} and \ion{C}{3} absorbers in the
\pks\ sightline; the difference in redshift density for these species
is likely due to fluctuations between sightlines.

The four systems with only one metal line are modeled well by a
single-phase absorber with \Z\ $\approx-1$, the currently favored
value for $\mnhi > 10^{14} \cm{-2}$ absorbers in the low-$z$ IGM
\citep{prochaskaetal04,danforthetal06}.  The $\zabs=0.00438$,
$0.09400$, and $0.19161$ systems are likely photoionized media. The
$\zabs=0.14533$ system may also be photoionized or collisionally
ionized, assuming the \Lya\ width is due to all thermal broadening.

With only an upper limit on \ion{C}{3} absorption, the $\zabs=0.06471$
\ion{O}{6} system could be reasonably modeled by either a photoionized
or collisionally-ionized medium. If the latter, the temperature is
constrained to be $T>2.2\times10^{5}\K$, which could be a probe of the
WHIM. The $\zabs=0.22555$ \ion{O}{6} system might be a single-phase,
collisionally-ionized absorber. No other metal lines were detected
with the tentative \ion{O}{6} absorber at $\zabs=0.22752$, and no
CLOUDY models were evaluated.

For the remaining two multiple metal-line systems, based primarily on
kinematic arguments, they are better modeled by a multi-phase
medium. In the case of the $\zabs=0.04222$ system, both \ion{C}{3} and
\ion{O}{6} are narrow but offset in by $50\kms$; this system could be
a two-phase photoionized medium with \Z\ $\approx-2$ to $-1$. On the other
hand, \ion{O}{6} in the $\zabs=0.09487$ system is broad, implying a
high temperature, while \ion{C}{3} and \ion{Si}{3} are narrow. Likely,
the broad feature is due to a collisionally-ionized phase, and the
narrow features are from a photoionized phase. The system has a
relatively low metallicity, \Z\ $\approx-2$, for the photoionized gas.

The \pks\ sightline has a galaxy survey complement.  The survey gives
compelling evidence that the metal-line absorption occurs in a diverse
set of galactic environments.  This includes a likely association with
individual galactic halos ($\zabs=0.09487,0.14253$), galaxy groups
($\zabs=0.094,0.192, 0.225$) and relatively poor environments ($\zabs
= 0.06471$).  The survey does not cover significant area at
$\zabs=0.00438$, the ninth metal-line system, but the sight line is
known to pass through the Virgo cluster at this redshift.

None of the four \ion{O}{6} absorbers detected in the \pks\ spectra
definitively trace the warm-hot intergalactic medium, which is defined
to be collisionally-ionized gas at $T\approx10^{5}$--$10^{7}\K$. The
systems at $\zabs=0.06471$ and 0.22555 may, though a firm conclusion
is difficult to draw.  In agreement with previous analysis
\citep{prochaskaetal04,richteretal04}, we find \ion{O}{6} absorption
in a multi-phase medium. The two systems with at least \ion{O}{6} and
\ion{C}{3} must be multi-phase since \ion{C}{4} is not detected and
the line profiles show different kinematic structure. However, these
systems do not necessarily probe the WHIM, as seen in hydrodynamic
simulations.  The metal-line systems appear to probe the predominantly
single-phase, photoionized intergalactic medium at low redshift.

From 28 intergalactic absorption systems in one sightline, simple
CLOUDY models, and a modest galaxy survey, we looked for qualitative
relationships between the various systems and between the systems'
environments. Roughly, one-third of \Lya\ absorbers also have
metal-line absorption. Two of five \ion{O}{6} absorbers are clearly in
multi-phase media. However, only one of the four has strong evidence
for being collisionally ionized and a potential WHIM candidate. The
strong \logHI\ $>14$ systems tend to be near galaxies
($\vert\dvg\vert<1000\kms$ and \ImPar\ $<500\hinv\kpc$). The nearest
galaxy distance tends to be correlated with \ion{H}{1} absorption.


\acknowledgements
Based on observations made with the NASA-CNES-CSA \emph{Far Ultraviolet
Spectroscopic Explorer}. \fuse\ is operated for NASA by the Johns Hopkins
University under NASA contract NAS5-32985. 

Based on observations made with the NASA/ESA \emph{Hubble Space
  Telescope} Space Telescope Imaging Spectrograph, obtained from the
  data archive at the Space Telescope Institute. STScI is operated by
  the association of Universities for Research in Astronomy,
  Inc. under the NASA contract NAS 5-26555.

The current study was funded by FUSE grant NAG5-12496.

{\it Facilities:} \facility{FUSE},
\facility{HST (STIS)},
\facility{Las Campanas: Dupont} 

\bibliographystyle{apj}
\bibliography{o6absorbers}

\end{document}